\documentclass{article}
\usepackage{epsfig}
\usepackage{amsmath}
\usepackage{amsfonts}\tolerance=10000
\usepackage{graphicx}
\usepackage[german, english]{babel}
\usepackage{a4wide}
\usepackage{amsmath}
\usepackage{amssymb}
\usepackage{ifthen}

\date{}

\newcommand{\la}{\lambda}
\newcommand{\ka}{\kappa}
\newcommand{\al}{\alpha}
\newcommand{\ta}{\theta}
\newcommand{\ga}{\gamma}
\newcommand{\Ga}{\Gamma}

\newcommand{\si}{\sigma}
\newcommand{\Si}{\Sigma}
\newcommand{\f}{\phi}
\newcommand{\vf}{\varphi}
\newcommand{\F}{\Phi}

\newcommand{\bnabla}{\mbox{\boldmath $\nabla$}}

\newcommand{\bOmega}{\mbox{\boldmath $\Omega$}}

\newcommand{\ee}{\end{equation}}
\newcommand{\eea}{\end{eqnarray}}
\newcommand{\be}{\begin{equation}}
\newcommand{\bea}{\begin{eqnarray}}

\newcommand{\pa}{\partial}

\newcommand{\vep}{\varepsilon}

\newcommand{\re}[1]{(\ref{#1})}
\newcommand{\R}{{\rm I \hspace{-0.52ex} R}}

\newcommand{\eins}{1\hspace{-0.56ex}{\rm I}}

\numberwithin{equation}{section}
\begin{document} 

\title{Notes on Yang-Mills--Higgs monopoles and dyons on $\R^D$,\\ and Chern-Simons--Higgs
solitons on $\R^{D-2}$:\\
 Dimensional reduction of Chern-Pontryagin densities} 
 
\author{
{\large D.~H.~Tchrakian}$^{\ddagger \star}$ \\ \\
\\
$^{\ddagger}${\small  School of Theoretical Physics -- DIAS, 10 Burlington
Road, Dublin 4, Ireland}
\\
$^{\star}${\small Department of Computer Science,
National University of Ireland Maynooth,
Maynooth,
Ireland}
  }
 
\maketitle 
\begin{abstract}
We review work on construction of Monopoles in higher dimensions.
These are solutions to a particular class of models descending from
Yang--Mills systems on even dimensional bulk, with Spheres as codimensions.
The topological lower bounds on the Yang--Mills action translate to
Bogomol'nyi lower bounds on the residual Yang-Mills--Higgs systems. Mostly,
consideration is restricted to 8 dimensional bulk systems, but extension to
the arbitrary case follows systematically. After presenting the monopoles,
the corresponding dyons are also constructed. Finally, new Chern--Simons
densities expressed in terms of Yang-Mills and Higgs fields are presented.
These are defined in all dimensions, including in even dimensional
spacetimes. They are constructed by subjecting the dimensionally reduced
Chern--Pontryagin densites to further descent by two steps.
\end{abstract}

\section{Introduction and Summary}

The central task of these notes is to explain how to subject $n-$th Chern--Pontryagin (CP)
density ${\cal C}^{(n)}$,
\be
\label{CP}
{\cal C}^{(n)}=\frac{1}{\omega(\pi)}\vep_{M_1M_2M_3M_4
\dots M_{2n-1}M_{2n}}\mbox{Tr}\,F_{M_1M_2}F_{M_3M_4}\dots F_{M_{2n-1}M_{2n}} 
\ee
defined on on the $2n-$dimensional space, to dimensional descent to $\R^D$ by considering \re{CP} on
the direct product space $\R^D\times S^{2n-D}$. The resulting residual density on $\R^D$ will be denoted as
$\Omega_{D}^{(n)}$.

The density ${\cal C}^{(n)}$ is by construction, a total divergence
\be
\label{totdivn}
{\cal C}^{(n)}=\bnabla\cdot\bOmega^{(n)}\,,
\ee
and it turns out that under certain retrictions, the dimensional descendant of \re{totdivn} is also a total divergence.
After demonstrating this result, it will be applied to the construction of monopoles, dyons and Chern-Simons solitons
in all dimensions.

Some special choices, or restrictions, are made for practical reasons.
Firstly, we have restricted to the codimension $S^{2n-D}$, the $(2n-D)-$sphere, as this is the most symmetric
compact coset space that is defined both in even and in odd dimensions. It can of course be replaced by any other
symmetric and compact coset space.

Secondly, the gauge field of the bulk gauge theory is chosen to be a $2^{n-1}\times 2^{n-1}$ array with complex valued
entries. Given our choice of spheres for the codimension, this leads to residual gauge fields on which take their
values in the Dirac matrix representation of the residual gauge group $SO(D)$.

As a result of the above two choices, it is possible to make the symmetry imposition (namely the dimensional reduction)
such that the residual Higgs field is described by a $D-$component isovector
multiplet. This choice is made specifically, with the requirement that the
resulting Higgs models support topologically stable, finite energy, solitons ("monopoles") whose asymptotic gauge
fields describe a Dirac-Yang~\cite{Dirac:1931kp,Yang:1977qv,Tchrakian:2008zz} monopoles.
These are $SO(D)$ monopoles defined on all $\R^D$
($D\ge 3$), generalising the usual $SO(3)$ 't~Hooft--Polyakov~\cite{'tHooft:1974qc,Polyakov:1974ek} monopole on $\R^3$.

The third and last restriction is to limit our concrete calculations to the case of
codimensions $S^{2n-D}$, for $n=2,3$ and $4$
only, $i.e.$, to the descents of the $2-$nd, $3-$rd and $4-$th CP densities only. The resulting
residual CP densities capture all qualitative features of the generic $n-$th case, the
calculus for $n\ge 3$ being inordinately more compicated, without yielding new qualitative
insight.

Inspite of the title, alluding to applications to the construction of various solitons, the
central result of these notes is the demonstration that the CP density ${\cal C}_{D}^{(n)}$
on $\R^D$ descended from the $2n-$dimensional bulk CP density ${\cal C}^{(n)}$ is a $total$ $divergence$
\be
\label{totdiv}
{\cal C}_{D}^{(n)}=\bnabla\cdot\bOmega^{(n,D)}
\ee
like ${\cal C}^{(n)}$ formally is on the bulk.

The various applications ensuing from this central result will be covered subsequently, rather
briefly since these present an open ended list of exercises. These fall under the following
broad headings:
\begin{itemize}
\item
When $D\ge 3$, the reduced density $\Omega_{D}^{(n)}$ can be interpreted as the $monopole$
charge density of a static YM--Higgs (YMH) theory on $\R^D$, and when $D=2$ it can be
interpreted as the $vortex$ charge of the YMH system on $\R^2$.
In the $D\ge 3$ case, these are the $D-$dimensional generalisations of the 't~Hooft-Polyakov
monopole on $\R^3$.
In the $D=2$ case, these are the generlisations of the Abrikosov--Nielsen-Olesen (ANO)
vortices~\cite{Abrikosov:1956sx,Nielsen:1973cs} of  Abelian Higgs models featuring higher order Higgs dependent functions.

In this application of the main result, a hierarchy of YMH models supporting
$monopoles$~\footnote{We are not much concerned with the construction of generalised
$vortices$ on $\R^2$, since such Abelian models cannot support dyons, that being the second
main application to be discussed.}
in all dimensions can be constructed systematically by subjecting the topological inequality
of YM systems in higher (even) dimensions, to dimensional reduction. The higher (even) dimensional
YM hierarchy~\cite{Tchrakian:1984gq} in question will be described in Section \ref{YM}, and this will be followed by a
brief description of the ensuing Higgs models after dimensional reduction.

It must be stressed, that this procedure is not unique, and one can alternatively follow a
scheme where the topological charge is defined directly as the $winding$ $number$~\cite{Tchrakian:2002ti} of the
Higgs field, which is suitably covariantised. We eschew this alternative because in that case the ensuing
Bogomol'nyi lower bounds result in energy densities that are described by
kinetic terms which are not exclusively quadratic in the velocity fields.
\item
The construction of monopoles is followed in a natural way by the construction of dyons and dyon like solutions
in $D+1$ dimensional Minkowski spacetime. This involves the formal extension of the models supporting
monopoles on $\R^D$ to Minkowskian theories on $D+1$ dimensional flat spacetime, supporting static electric Yang--Mills
fields in addition to monopoles. In this respect these solutions are the $D-$dimensional generalisations
of the Julia--Zee~\cite{Julia:1975ff} dyon~\footnote{For $D=2$ the resulting solitons are not $monopoles$, but are
$vortices$. In those cases the subsequest construction of a dyon is obstructed
by the Julia-Zee Theorem, $unless$ if the model is augmented with a suitable
Chern-Simons term.} on $\R^3$. However, only those defined on $\R^3$ describe an electric flux like the latter,
and they are referred to as dyons. The electric field carrying static solutions on $\R^D$, ($D\ge 4$) are referred to
as $pseudo-dyons$. Also, unlike the Julia--Zee dyon which does have a BPS limit,
solutions in dimensions other than $\R^{4p-1}$ do not have a BPS limit. These last dyonic configurations are presented
as a byproduct of monopolic configurations, presented in \cite{Radu:2005rf}. Apart from these, none of the dyonic
configurations presented below in Section {\bf 8} have been studied quantitatively (numerically) to date.
\item
From the reduced density $\bOmega^{(n,D)}\equiv\Omega^{(n,D)}_i$, where $i$ is the index of the spacelike
 coordinate $x_i$ with $i=1,2,\dots,D$,
one can formally identify a Chern--Simons
(CS) density as the $D-$th component of $\bOmega^{(n,D)}$. This quantity can then be 
interpreted as a CS term on $(D-1)-$ dimensional Minkowski space, $i.e.$,
on the spacetime $(t,\R^{D-2})$. The solitons of the
corresponding CS--Higgs (CSH) theory can be constructed systematically.
Note, that this is not the usual CS term defined in terms of a pure Yang--Mills field on
$odd$ dimensional spacetime, but rather these new CS terms are defined by both the YM and,
the Higgs fields. Most importantly, the definition of these new CS terms is not restricted to
$odd$ dimensional spacetimes, but covers also $even$ dimensional spacetimes. 
To date, such CSH solutions have not been studied concretely.
\end{itemize}

\bigskip
\noindent
{\bf\large Acknowledgements}

\medskip
\noindent
Most of the results covered in these notes have appeared in numerous publications, and the aim here
has been to collect all of these together in a coherent framework. Work on various aspects of monopoles in higher
dimensions, inclding generalisations of three dimension, were carried out in collaboration with Amithabha Chakrabarti,
Burkhard Kleihaus, Zhong--Qi Ma, Grainne O'Brien, Eugen Radu, Tom Sherry, Yisong Yang and Frank Zimmerschied.
Related work on vortices of generalised Abelian Higgs models in two dimensions, and their gauge decoupled versions,
was carried out in collaboration with Kieran Arthur, Yves Brihaye, Jurgen Burzlaff and Harald M\"uller-Kirsten.
Throughout, discussions with Lochlainn O'Raifeartaigh were invaluable.
Thanks are due to Neil Lambert, Valery Rubakov and Paul Townsend
for helpful comments. Special thanks are due to Eugen Radu for help and discussions throughout the preparation of
these notes.

\section{The Yang--Mills hierarchy}
\label{YM}

We seek finite energy solutions of Yang--Mills--Higgs (YMH) systems in arbitrary spacelike dimensions
$\R^D$. The construction of instantons of Yang--Mills (YM) instantons and YMH and monopoles in higher dimensions was first
suggested in \cite{Tchrakian:1978sf}. Yang--Mills--Higgs systems can be derived from the dimensional descent
of a suitable member of the Yang--Mills (YM)
hierarchy on the Euclidean space $\R^D\times K^{N}$ such that $K^{N}$ is a compact coset space. Here,
$D+N$ is even since the YM hierarchy introduced in \cite{Tchrakian:1984gq}, is defined in even
dimensions since Chern--Pontryagin densities are defined in even dimensions only.

The YM hierarchy of $SO(4p)$
gauge fields in the chiral
(Dirac matrix) representations consisting only of the $p$-YM term in \re{pYM} was
introduced in \cite{Tchrakian:1984gq} to construct selfdual instantons
in $4p$ dimensions. (The selfduality equation for the $p=2$ case was solved
indepenently in \cite{Grossman:1984pi}, whose authors subsequently stated in their {\it Erratum},
that this solution was the instanton of the
$p=2$ member of the hierarchy introduced earlier in~\cite{Tchrakian:1984gq}.) The instantons of the
generic system consisting of the sum of many terms \re{pYM} with different $p$, while stable, are not selfdual and cannot
be evaluated in closed form and are constructed numerically~\cite{Burzlaff:1993kf}. Restricting ourselves here to
finite action (instanton) solutions only, it is worth mentioning an alternative hierarchy which
supports selfdual instantons in $4p+2$ dimensions~\cite{Saclioglu:1986qn,Fujii:1986ty}.
While it is straightforward to construct spherically
symmetric solutions with gauge group $SO(4p+2)$ in the chiral Dirac representations, these
selfduality equations are even more overdetermined than those of
the $4p$ dimensional hierarchy. The action densities of
these systems are not positive definite so that, while the selfduality equations do solve
the second order field equations, they do not saturate a Bogomol'nyi bound and hence are
not necessarily stable. The selfduality equations employed in the works so far mentioned are nonlinear in the
Yang-Mills curvature. There have been other formulations higher dimensional
selfduality~\cite{Corrigan:1982th,Fairlie:1984mp,Fubini:1985jm} which differ from the
former in that the selfduality equations proposed there are linear in the
Yang-Mills curvature. These suffer from the same lack of stablity of the solutions of
\cite{Saclioglu:1986qn,Fujii:1986ty}, and some of them do not even have finite energy.

We start from the definition of the YM hierarchy.
Since we will be concerned with dimensional reduction over the $N-$dimensional codimension $K^N$, we will denote
the connection and the curvature in the bulk with $({\cal A},{\cal F})$, to distinguish these from the corresponding
quantities $(A,F)$ on the residual space $\R^D$. Using the notation ${\cal F}(2)={\cal F}_{\mu\nu}$ for the $2$-form
YM curvature, the $2p$-form YM tensor
\be
\label{2p}
{\cal F}(2p)={\cal F}(2)\wedge {\cal F}(2)\wedge...\wedge {\cal F}(2)\ ,\quad p-{\rm times}
\ee
is a $p$ fold totally antisymmetrised product of the $2$-form curvature.

The $p-$YM system of the YM hierarchy is defined, on $\R^{4p}$, by the Hamiltonian~\footnote{We call this a Hamiltonian
rather than a Largangian because by definition it is positive definite. The corresponding Largangian with
a given Minkowskian signature can then be defined systematically.}
density~\footnote{The canonical definition is
${\cal H}_{p}=\frac{1}{(2p)!^2}\,\mbox{Tr}\,{\cal F}(2p)^2$,
but here we use a simpler, unconvential, normalisation for convenience.}
\be
\label{pYM}
{\cal H}_{{\rm YM}_p}=\mbox{Tr}\,{\cal F}(2p)^2\,.
\ee

In $2n$ dimensions, partitioning $n$ as $n=p+q$, the Hodge dual of the
$2q$-form field $F(2q)$, namely $(^{\star}{\cal F}(2q))(2p)$, is a $2p$-form.

Starting from the inequality
\be
\label{ineq}
\mbox{Tr}[{\cal F}(2p)-\ka^{(p-q)}\ \ ^{\star}{\cal F}(2q)]^2\ge 0\ ,
\ee
it follows that
\be
\label{top-lb1}
\mbox{Tr}[{\cal F}(2p)^2+\ka^{2(p-q)}\ {\cal F}(2q)^2]\ge 2\ka^{(p-q)}\ {\cal C}^{(n)}\ ,
\ee
where ${\cal C}^{(n)}\equiv{\cal C}^{(n=p+q)}$ is the $n$-th Chern-Pontryagin density. In \re{ineq}
and \re{top-lb1}, the constant $\ka$ has the dimension of length if $p>q$ and the inverse if $p<q$.

The element of the YM systems labeled by $(p,q)$ in (even) $2(p+q)$ dimensions are defined
by Lagrangians defined by the densities on the {\it left hand side} of
\re{top-lb1}. When in particular $p=q$, then these systems are
conformally invariant and we refer to them as the $p-$YM members of the YM hierarchy.

The inequality \re{top-lb1} presents a topological lower bound which
guarantees that finite action solutions to the Euler--Lagrange equations
exist. Of particular interest are solutions to first order self-duality
equations which solve the second order Euler--Lagrange equations, when
\re{top-lb1} can be saturated.

For ${\bf M}^{2n}={\R}^{2n}$, the self--duality equations support
nontrivial solutions only if $q=p$,
\be
\label{sd-4p}
{\cal F}(2p)\ =\ ^{\star}{\cal F}(2p)\ .
\ee
For $p=1$, i.e. in four Euclidean dimensions, \re{sd-4p} is the usual
YM selfduality equation supporting instanton solutions. Of these, the
spherically symmetric~\cite{Belavin:1975fg,Tchrakian:1984gq} and axially
symmetric~\cite{Witten:1976ck,Chakrabarti:1985qj,Spruck:1997eb} instantons on $\R^{4p}$
are the known. For $p\ge 2$, i.e. in dimensions eight and higher, only
sphericaly symmetric~\cite{Tchrakian:1984gq} and axially symmetric~\cite{Chakrabarti:1985qj,Spruck:1997eb} solutions
can be constructed, because in these dimensions \re{sd-4p} are
overdetermined~\cite{Tchrakian:1990gc}.

In the $r\gg 1$ region, all these 'instanton' fields on ${\bf R}^{2n}$,
whether self--dual or not, asymptotically behave as pure--gauge
\[
{\cal A}\rightarrow g^{-1}dg\,.
\]
For ${\bf M}^{2n}=G/H$, namely on compact coset spaces, the self--duality
equations support nontrivial solutions for all $p$ and $q$,
\be
\label{sd-g/h}
{\cal F}(2p)\ =\ \ka\ ^{\star}{\cal F}(2q)\ 
\ee
where the constant $\ka$ is some power of the 'radius' of the (compact)
space. The simplest examples are ${\bf M}^{2n}=S^{2n}$, the
$2n$-spheres~\cite{O'Se:1987fx,O'Brien:1988rs,Kihara:2007di,Radu:2007az}, and ${\bf M}^{2n}={\bf CP}^n$, the complex
projective spaces~\cite{Ma:1990ja,Kihara:2008zg}.

Gravitating members of the YM hierarchy \cite{Kihara:2007vz} were studied and applicatied to dynamical compactification in
\cite{Kihara:2009ea,Chingangbam:2009jy}. (In \cite{Chingangbam:2009jy} in particular, the dimensional descent from $10$
dimensional spacetime yielding a Yang-Mills--Higgs system was considered, but unlike in the present notes this was not
done with a view to the construction of solitons.)

The above definitions of the YM systems can be formally extended to all dimensions, including all odd dimensions.
The only difference this makes is that in odd dimernsions topological lower bounds enabling the construction of
instantons are lost since Chern--Pontryagin charges are defined only on even dimensions.

Incidentally, on the subject of solutions to higher dimensional
Yang-Mills systems consisting of higher order curvature terms,
one might mention in passing that Meron~\cite{de Alfaro:1976qz} solutions in all even dimensions~\cite{O'Brien:1987jy}
can also be constructed. 

\section{Higgs models on ${\bf R}^D$}
\label{Higgs}

Higgs fields have the same dimensions as gauge connections and appear
as the extra components of the latter under dimensional reduction, when
the extra dimension is a compact symmetric space. Dimensional reduction of gauge fields over a compact
codimension is implemented by the imposition of the symmetry of the compact coset space on the
coordinates of the codimensions. In this respect, the calculus of dimensional reduction does not
differ from that of imposition of symmetries generally.

The calculus of imposition of symmetry on gauge fields that has been used in the works being reviewed here
is that of Schwarz~\cite{Schwarz:1977ix,Romanov:1977rr,Schwarz:1981mb}. This formalism was adapted to
the dimensional reduction over arbitrary codimensions in \cite{Ma:1986pu,Ma:1988um,O'Brien:1988xr}.
An alternative formalism
for the dimensional reduction of gauge fields, \cite{Forgacs:1979zs}, is familiar in the literature, but the calculus of
\cite{Schwarz:1977ix} was found to be more convenient for extension to codimensions of arbitrary dimensions.

In the following, we shall denote the codimension-$N$ by the index $I=1,2,\dots,N$ and the residual dimension-$D$ by the
index $i=1,2,\dots,D$. The bulk dimension-$(D+N)=2n$, which we will take to be even, permits the definition of the
bulk $n-$th Chern--Pontryagin (CP) density ${\cal C}^{(n)}$.
The dimensional descent of the densities ${\cal C}^{(n)}$ is the
first task of these notes. This ammounts to the imposition of symmetry appropriate to the codimension, followed by the
integration over this codimension. Integrating inequality \re{top-lb1} over this compact volume
\be
\label{descent}
\int_{{\R}^D\times K^{2(p+q)-D}}\mbox{Tr}[{\cal F}(2p)^2+\ka^2\ {\cal F}(2q)^2]\ge 2\ka\
\int_{{\R}^D\times K^{2(p+q)-D}}{\cal C}^{(n)}\ ,
\ee
results in the reduced YM-Higgs (YMH) energy density functional on the left hand side, and on the right hand side the
required residual CP density. In \re{descent}, ${\cal F}(2p)$ is the $2p-$form curvature of the $1-$form connection
${\cal A}$ on the higher dimensional space ${\bf R}^D\times K^{2(p+q)-D}$.

Imposing the symmetry appropriate to $K^{2(p+q)-D}$ on the gauge fields
results in the breaking of the original gauge group to the
residual gauge group $g$ for the fields on ${\bf R}^D$. This residual gauge group $g$ depends on the precise mode
of dimensional reduction, namely on the choice of the codimension $K^{2(p+q)-D}$. The details will be specified below
in Section \ref{YMdescent}. Performing then the integration over the compact space $K^{4p-D}$ leads to the (static)
Hamiltonian ${\cal H}[A,\f]$, of the residual YMH model on ${\bf R}^D$. In \re{descent}, the residual connection
and its curvature are denoted by
$(A,F),$ taking their values in the algebra of the residual gauge group $g$,
and $\phi$ is the Higgs multiplet whose structure under $g$ depends
on the detailed choice of $K^{2(p+q)-D}$, implying the following gauge
transformations
\[
{A}\rightarrow g^{-1}{A}\, g+g^{-1}d\,g 
\]
and depending on the choice of $K^{4p-D}$,
\[
\phi\rightarrow g\phi\ g^{-1}\quad ,\quad{\rm or}\quad ,\quad
\phi\rightarrow g\phi\quad ,\quad{\rm etc.}
\]
The inequality \re{descent} leads to
\bea
\int_{{\R}^D}{\cal H}[A,\f]&\ge&\int_{{\R}^D}{\cal C}^{(n)}_D=\int_{{\R}^D}\nabla\cdot
{\bf \Omega}^{(n,D)}[{A},\phi]=\int_{\bf\Sigma^{D-1}}{\bf\Omega}^{(n,D)}[{A},\phi]\ ,\label{top-lb2b}
\eea
where ${\cal H}[{A},\phi]=
{\cal H}[{F},{D}\phi,\vert\phi\vert^2,\eta^2]$ is the residual
Hamiltonian in terms of the residual gauge connection ${A}$ and its
curvature $F$, the Higgs fields $\phi$ and its covariant derivative
$D\phi$ and the inverse of the compactification 'radius' $\eta$.
The latter is simply the VEV of the Higgs field, seen clearly from the
typical form of the components of the curvature $F$ on the extra (compact)
space $K^{4p-D}$
\be
\label{VEV}
F\vert_{K^{2(p+q)-D}}\ \sim\ (\eta^2-\vert\phi\vert^2)
\otimes\Lambda^{(N)}\quad \Rightarrow
\quad\lim_{r\to\infty}\vert\phi\vert^2=\eta^2
\ee
where $\Lambda^{(N)}$ are, symbolically, spin-matrices/Clebsch-Gordan coefficients. Specifically, $\Lambda^{(N)}$
are representation
matrices of the stability group of the symmetry group of $K^N$. The precise definition of these matrices
will be given below in Section \ref{YMdescent}, for the special cases consireded here, namely for
$K^{2(p+q)p-D}=S^{2(p+q)p-D}$.

While the precise overall numerical factor in the density ${\bf\Omega}^{(n,D)}[{A},\phi]$ in \re{top-lb2b} can be
evaluated, in practice these quantities will be evaluated without regard to such factors, since in each case we will
normalise the monopole charge by requiring that in the spherically symmetric case this be the $unit$ charge.

Of course, the simplest choice for these (topological) inequalities is when $p=q$, $i.e.$, when $D+N=4p$, in which
cases these inequalities lead to Bogomol'nyi equations that do not feature a dimensional constant $\ka$. This is the
case with the familiar examples of $(D=3,N=1)$ for the BPS monopole and the $(D=2,N=2)$ of the critical Abelian
Higgs vortices, which are $selfdual$ solutions.
It should be noted at this stage that subjecting the selfduality equations \re{sd-4p}, for $p=q$, to this dimensional
descent results in Bogomol'nyi equations on $\R^D$, which for $p\ge 2$ in ${\R}^D\times K^{4p-D}$ ($cf.$ \re{descent})
turn out to be overdetermined~\cite{Tchrakian:1990gc} with only two exceptions, these being when $D=4p-1$ and $D=2$.

The residual gauge connection here is defined on the Euclidean space $\R^D$, so we will refer to it as the
$magnetic$ component. Subsequently, we will define the $electric$ component also, with reference to the Higgs
multiplet $\f$.

\subsection{Choice of condimension and residual gauge group}
In the particular plan of dimensional descent of gauge fields pursued here,
one is guided by the choice of residual gauge group. Here, we are
guided by the requirement that our residual gauge connections be described by
Dirac-Yang~\cite{Dirac:1931kp,Yang:1977qv,Tchrakian:2008zz}
monopoles asymptotically. This leads immediately to the choice of $SO(D)$ for the residual gauge field on $\R^D$.

As stated at the outset, our explicit considerations are restricted to the codimensions $K^N=S^N$, $i.e.$, to
$N-$spheres. The symmetry group of $S^N$ is $SO(N+1)$ with stability group $SO(N)$, and it is the representation
matrices of these stability groups which feature in in Schwarz's calculus of dimensional reduction, as the latter
exploits the symmetry imposition equations at a fixed point (say North pole) of $S^N$. In particular we have chosen
to employ the gamma (Dirac) matrix representations of $SO(N)$, so there arises a distinction between
{\bf odd} and {\bf even} $N$
due to existence of a chirality operator of the Clifford algebras in {\bf even} dimensions. Thus the concrete examples
of $\Lambda^{(N)}$ in \re{VEV} now are
\be
\label{soN}
\Gamma_{IJ}=-\frac14[\Gamma_{I},\Gamma_{J}]
\ee
in terms of the $N-$dimensional (Dirac) gamma matrices $\Gamma_I$.

In what follows, we will use a uniform index notation to label the coordinate $x_M=(x_i,x_I)$, with $x_i$
(lower case Latin $i=1,2,\dots,D$) the coordinate on the residual space $\R^D$ and with $x_I$
(upper case Latin $I=1,2,\dots,N$) the coordinate on the codimension $S^N$.
(We reserve the Greek letters $\mu,\nu,...$ for the Minkowskian index $\mu=(0,i)$ for later use in describing dyons.)

Since the dimensionality
of the bulk is $2n=D+N$, this distinction between {\bf odd} and {\bf even} $N$ will be refelcted in distinct features
of the residual $D-$dimensional fields for {\bf odd} and {\bf even} $D$, respectively.

The gamma matrices, $\Gamma_I$, used to represent the $SO(N)$ algebra are:

\begin{center}
for {\bf even} $N$, \ \ $2^{\frac{N}{2}}\times{2^{\frac{N}{2}}}$ \ \ complex valued arrays, and,

for {\bf odd} $N$, \ \ $2^{\frac{N-1}{2}}\times{2^{\frac{N-1}{2}}}$ \ \ complex valued arrays\ .
\end{center}

But in the fixed--point calculus of Schwarz, the
bulk gauge field ${\cal A}$ is a direct product of the residual gauge field $A$ times an element of the $N-$dimensional
Clifford algebra.
Thus if we choose the bulk gauge connections ${\cal A}$, irrespective of whether $N$ is odd or even, to be
$2^{(n-1)}\times{2^{(n-1)}}$ \ \ anti-Hermitian arrays, then the residual gauge connections $A$ will be:

\begin{center}
for {\bf even} $N$, \ $2^{\frac{D}{2}-1}\times{2^{\frac{D}{2}-1}}$ complex valued arrays, and,

for {\bf odd} $N$,\ \  $2^{\frac{D-1}{2}}\times{2^{\frac{D-1}{2}}}$ complex valued arrays.
\end{center}

Let us consider first the case of odd $N$ (and hence odd $D$).
Here, the residual connection $A$ is a $2^{\frac{D-1}{2}}\times{2^{\frac{D-1}{2}}}$ anti-Hermitian matrix
(not necessarily traceless). This allows the option of choosing the residual gauge
group to be $SO(D)$~\footnote{Had we chosen a larger rank gauge connection in the
bulk, we would have ended up with a larger residual gauge group, which however would include the $convenient$
gauge group $SO(D)$.}, such that $A$  takes its values in one or other of the two chiral (Dirac) matrix
representations of the algebra of $SO(D)$. With this choice of residual gauge group, the asymptotic configurations of the
residual YM connections are described by Dirac-Yang~\cite{Dirac:1931kp,Yang:1977qv,Tchrakian:2008zz}
monopoles, as required.

The case of even $N$ (and hence even $D$) is more subtle and restrictive. Here, the residual connection $A$ consists of
two $2^{\frac{D}{2}-1}\times{2^{\frac{D}{2}-1}}$ complex valued arrays, each being anti-Hermitian. However,
as we shall see below in the explicit dimensional reduction equations, these two matrices are "doubled up" $via$ the
chiral operator $\Gamma_{D+1}$. The resulting residual connection $A$, is the
$2^{\frac{D}{2}}\times{2^{\frac{D}{2}}}$ direct sum of the two ($left$ and $right$) chiral components.
This allows one to ascribe to $A$ the residual gauge group $SO(D)$, as in the case of odd $D$, but now with $A$
in the chirally symmetric (Dirac) matrix representation of $SO(D)$. Again, in the case of even
$D$, the (restricted) choice of $SO(D)$ is the natural one, describing asymptotically a Dirac--Yang monopole.

Since the options exercised above are specific to odd, $resp.$ even, dimensions, these two cases will be treated
separately when the concrete dimensional descent is presented. However, the scheme just described presents a unified
framework for both odd and even $D$, with $SO(D)$ being the residual gauge group for both.

The above described prescription for dimensional descent pertains to the ${\cal A}_i(x_i,x_I)$ components of the bulk
gauge fields, resulting in $SO(D)$ residual gauge fields $A$ on $\R^D$. The descent of the
components ${\cal A}_I(x_i,x_I)$ in turn result in Higgs multiplets in which, as is well
known~\cite{Arafune:1974uy,Tchrakian:2002ti}, the topology of the monopole is encoded. Again, the guiding feature
is that of having the asymptotic connection on $\R^D$ described by a
Dirac-Yang~\cite{Dirac:1931kp,Yang:1977qv,Tchrakian:2008zz}
monopole, which for the asymptotic Higgs field means gauging it away to the constant (trivial) configuration oriented
along the $x_D-$axis, resulting in the vanishing of its covariant derivative. There will be a Dirac line singularity
along the positive or negative $x_D-$axis, which is a gauge artefact. As will be seen in the next below, there is a
striking difference in the odd and even $D$ cases. In both
cases the Higgs field $\F$ is $not$ restricted to take its values in the $algebra$ of $SO(D)$ (except in the special case
of $D=3$). In even, (resp.) odd $D$, $\F$ consists of a $2^{\frac{D-1}{2}}\times{2^{\frac{D-1}{2}}}$, (resp.)
$2^{\frac{D}{2}}\times{2^{\frac{D}{2}}}$ compex valued array. As such it can be described by elements of the $left$ or
$right$ chiral (resp.), chirally symmetric $left$ plus $right$ (resp.), representations of the $SO(D+1)$ algerbra,
for odd and even $D$. A unified expression for the asymptotic Higgs field, taylored to present a Dirac-Yang field, is
\bea
\F|_{r\to\infty}&\simeq&\hat x_i\,\Si_{i,\,D+1}\label{Hodd}\\
\F|_{r\to\infty}&\simeq&\hat x_i\,\Gamma_{i,\,D+1}\label{Heven}
\eea
where $\hat x_i$ is the unit radius vector in $\R^D$ and $\Si_{ab}$ are the $left$ or $right$ $SO(D)$ chiral
representation matrices
\be
\label{Si}
\Si_{ab}^{(\pm)}=-\frac14\left(\frac{1\pm\Gamma_{D+1}}{2}\right)\Gamma_{ab}\ ,\quad a,b=1,2,\dots,D+1
\ee
It should be stressed here that although the Higgs field is not
restricted to take its values in the algebra of $SO(D)$, the gauge field does take its values in the algebra of $SO(D)$.
Thus the residual gauge group indeed remains $SO(D)$.

The above described prescription of dimensional reduction is fairly general, within the context of the retrictions
opted for. The exception, which we have eschewed here, is when the descent is over the codimension $S^2$, $i.e.$,
when $N=2$. In this case, the symmetry group of $S^2$, namely $SO(3)$, has the stability group
$SO(2)$. The latter, being Abelian, it affords a very much richer family of solutions to the symmetry equations imposed
for the descent~\cite{Schwarz:1981mb}, allowing for Higgs multiplets not $necessarily$ restricted as above. The
more general Higgs multiplets that result are superfluous for our purposes and are not consiedered here.

The YM field in the YMH models on $R^D$ discussed thus far, is purely magnetic supporting a 'magnetic' monopole. But 
when it comes to YMH models, as stated earlier, the presence of the Higgs field enables the support of the
electric component of the YM connection $A_0$ and hence enables the description of a Julia--Zee type dyon
in $d=D+1$ dimensional spacetime. In the usual~\cite{Julia:1975ff} sense as the dyon in $3+1$ dimensions, the Higgs
field partners the newly introduced 'electric' YM potential $A_0$. Thus we can describe $SO(d)$ dyons~\footnote{While
both the Higgs field and $A_0$ take their values in the Dirac matrix basis $\Gamma_{i,D+1}$, in the
in the familiar $3+1$ dimensional case the 'enlarged' algebra $SO(4)$ splits in the two $SU(2)$ subalgebras, whence the
$magnetic$ component of the connection $A_i$ and the $electric$ component of the connection $A_0$, are both described by
$SU(2)$ matrices. In all higher dimensions, this is not the case and the
full algebra employed is that is $SO(D+1)$, where $d=D+1$ is the dimension of the spacetime.} in $d-$dimensional
spacetime.

When the dimension of the spacetime $d=D+1$ is even, then the gauge field $A_{\mu}=(A_i,A_0)$ and Higgs field
$\F$ in the residual space are expanded in the basis of
the chiral $SO(d)$ matrices $(\Sigma_{ij}^{(\pm)},\Sigma_{i,d}^{(\pm)})$, $e.g.$, in $d=3+1$
spacetime.
By contrast, when the dimension of the spacetime $d$ is odd, then the chirally symmetric $SO(d)$ matrices
$\Gamma_{\mu\nu}=(\Gamma_{ij},\Gamma_{i,d})$ are used in the same situation, $e.g.$, in $d=4+1$
spacetime~\cite{Kleihaus:1998kd} when $D=4$.

In this enlarged context, namely that allowing the construction of dyons in addition to monopoles, the case of
$D=2$ is excluded, due to the well known obstruction of the Julia-Zee theorm~\cite{Julia:1975ff} to
the existence of dyons in $d=2+1$ spacetime.

\section{Dimensional reduction of gauge fields}
\label{YMdescent}
In this Section, we present the formalism of the dimensional reduction of Yang--Mills fields employed in the Sections
following it. We list the results of the calculus of dimensional reduction for the classes of descents considered
in a unified notation. This calculus is based on the formalism of A.~S.~Schwarz~\cite{Schwarz:1977ix},
which is specially transparent due to the choice of displaying the results only at a fixed point of the compact
symmetric codimensional space $K^N$ (the North or South pole for $S^N$). Our formalism is a straightforward extension
of \cite{Schwarz:1977ix,Romanov:1977rr,Schwarz:1981mb}.

The criterion of constructing monopoles that are asymptotically Dirac-Yang, restricts our calculus to the framework
described above. This does not include the restricting to the codimensions $K^N=S^N$, but we do this anyway, for the sake
of simplicity. (It is also the case that employing $K^{N}=CP^{N}$~\cite{ST}, for example, does not lead to any
qualitatively new results.)

\subsection{Descent over $S^N$: $N$ odd}
\label{YMdescentodd}
For the descent from the bulk dimension $2n=D+N$ down to {\bf odd} $D$ (over odd $N$),
the components of the residual connection evaluated at the Noth pole of $S^N$ are given by
\bea
{\cal A}_i&=&A_i(\vec x)\otimes\eins\label{aiodd}\\
{\cal A}_I&=&\F(\vec x)\otimes\frac12\Gamma_I\,.\label{aIodd}
\eea
The unit matrix in \re{aiodd}, like the $N-$dimensional gamma matrix in \re{aIodd}, are
$2^{\frac12(N-1)}\times 2^{\frac12(N-1)}$ arrays. Choosing the $2^{n-1}\times 2^{n-1}$ bulk gauge group to be, say,
$SU(n-1)$, allows the choice of $SO(D)$ as the gauge group of the residual connection $A_i(x)$. This choice is
made such that the asymptotic connections describe a Dirac--Yang monopole.

For the same reason, the choice for the multiplet structure of the Higgs field is made to be less restrictive.
The (anti-Hermitian) field $\F$, which is not necessarily traceless~\footnote{In practice, when constructing soliton
solutions, $\F$ is taken to be traceless
without loss of generality.}, can be and $is$
taken to be in the algebra of $SO(D+1)$, in particular, in one or other of the chiral
reprentations of $SO(D+1)$, $D+1$ here being even.
\be
\label{PHYodd}
\F=\f^{ab}\,\Si_{ab}\ ,\quad a=i,\,D+1\ ,\quad i=1,2,\dots, D\,.
\ee
(Only in the $D=3$ case does the Higgs field take its values
in the algebra of $SO(3)$, since the representations $SO(3)$ coincide with those of chiral $SO(4)$.)

In anticipation of the corresponding situation of even $D$ in the next Subsection, one can specialise \re{PHYodd}
to a $D-$component $isovector$ expression of the Higgs field
\be
\label{Hisoodd}
\F=\f^i\,\Si_{i,D+1}\,,
\ee
with the purpose of having a unified notation for both even and odd $D$, where the Higgs field takes its values
in the components $\Si_{i,D+1}$ orthogonal to elements $\Si_{ij}$ of the algebra of $SO(D+1)$.
This specialisation is not necessary, and
is in fact inappropriate should one consider, $e.g.$, axially symmetric fields. It is however adequate for the
presentation here, being
consistent with the asymptotic expressions \re{Hodd}, and is sufficiently general to describe spherically symmetric
monopoles~\footnote{While all concrete considerations in the following are restricted to spherically symmetric fields,
it should be emphasised that relaxing spherical symmetry results in the Higgs multiplet getting out of the orthogonal
complement $\Si_{i,D+1}$ to $\Si_{i,j}$. Indeed, subject to axial symmetry one has
\be
\label{ax}
\F=f_1(\rho,z)\Si_{\al\beta}\hat x_{\beta}+f_2(\rho,z)\Si_{\beta,D+1}\hat x_{\beta}+f_3(\rho,z)\Si_{D,D+1}\,,
\ee
where $x_i=(x_{\al},z)$, $|x_{\al}|^2=\rho^2$ and with $\hat x_{\al}=x_{\al}/\rho$. Clearly, the term in \re{ax}
multiplying the basis $\Si_{\al\beta}$ does not occur in \re{Hisoodd}.}.

In \re{aiodd} and \re{aIodd}, and everywhere henceforth,
we have denoted the components of the residual coordinates as $x_i=\vec x$.
The dependence on the codimension coordinate $x_I$ is suppressed since all fields are evaluated at a fixed point
(North or South pole) of the codimension space.

The resulting components of the curvature are
\bea
{\cal F}_{ij}&=&F_{ij}(\vec x)\otimes\eins\label{fijodd}\\
{\cal F}_{iI}&=&D_{i}\F(\vec x)\otimes\frac12\Gamma_I\label{fiIodd}\\
{\cal F}_{IJ}&=&S(\vec x)\,\otimes\Gamma_{IJ}\,,\label{fIJodd}
\eea
where $\Gamma_{IJ}=-\frac14[\Gamma_{I},\Gamma_{J}]$ are the Dirac representation matrices of $SO(N)$, the stability
group of the symmetry group of the $N-$sphere. In \re{fiIodd}, $D_{i}\F$ is the covariant derivative of the
Higgs field $\F$
\be
\label{covodd}
D_{i}\F=\pa_i\F+[A_i,\F]
\ee
and $S$ is the quantity
\be
\label{Sodd}
S=-(\eta^2\,\eins+\F^2)\,,
\ee
where $\eta$ is the inverse of the radius of the $N-$sphere.

\subsection{Descent over $S^N$:  $N$ even}
\label{YMdescenteven}
The formulae corresponding to \re{aiodd}-\re{fIJodd} for the case of {\bf even} $D$ are somewhat more complex.
The reason is the existence of a chiral matrix $\Gamma_{N+1}$, in addition to the Dirac matrices $\Gamma_{I}$,
$I=1,2,\dots,N$. Instead of \re{aiodd}-\re{aIodd} we now have
\bea
{\cal A}_i&=&A_i(\vec x)\otimes\eins+B_i(\vec x)\otimes\Gamma_{N+1}\nonumber\\
{\cal A}_I&=&\f(\vec x)\otimes\frac12\Gamma_I+\psi(\vec x)\otimes\frac12\Gamma_{N+1}\Gamma_I\nonumber\,,
\eea
where $A_i$, $B_i$, $\f$, and $\psi$ are again antihermitian matrices, but with only $A_i$ being traceless.
The fact that $B_i$ is not traceless here results in an Abelian gauge field in the reduced system.

Anticipating what follows, it is much more transparent to re-express these formulas in the form
\bea
{\cal A}_i&=&A_i^{(+)}(\vec x)\otimes\,P_++A_i^{(-)}(\vec x)\otimes\,P_-
+\frac{i}{2}\,a_i(\vec x)\,\Gamma_{N+1}\label{aieven}\\
{\cal A}_I&=&\vf(\vec x)\otimes\frac12\,P_+\,\Gamma_I-\vf(\vec x)^{\dagger}\otimes\frac12\,P_-\,\Gamma_I\label{aIeven}\,,
\eea
where now $P_{\pm}$ are the $2^{\frac{N}{2}}\times 2^{\frac{N}{2}}$ projection operators
\be
\label{proj}
P_{\pm}=\frac12\left(\eins\pm\Gamma_{N+1}\right)\,.
\ee
In \re{aieven}, the residual gauge connections $A_i^{(\pm)}$ are anti-Hermitian and traceless
$2^{\frac{D}{2}}\times 2^{\frac{D}{2}}$ arrays, and the Abelian connection $a_i$ results directly from the trace of the
field $B_i$. The $2^{\frac{D}{2}}\times 2^{\frac{D}{2}}$ "Higgs'' field $\vf$ in \re{aIeven} is neither Hermitian nor
anti-Hermitian. Again, to achieve the desired breaking of the gauge group, to lead eventually to the requisite Higgs
$isomultiplet$, we choose the gauge group in the bulk to be $SU(n-1)$, where $2n=D+N$.

The components of the curvaturs are readily calculated to give
\bea
{\cal F}_{ij}&=&F_{ij}^{(+)}(\vec x)\otimes\,P_++F_{ij}^{(-)}(\vec x)\otimes\,P_-
+\frac{i}{2}\ f_{ij}(\vec x)\,\Gamma_{N+1}\label{fijeven}\\
{\cal F}_{iI}&=&D_i\vf(\vec x)\otimes\,\frac12\,P_+\Gamma_I
-D_i\vf^{\dagger}(\vec x)\otimes\,\frac12\,P_-\Gamma_I\label{fiIeven}\\
{\cal F}_{IJ}&=&S^{(+)}(\vec x)\otimes\,P_+\Gamma_{IJ}+S^{(-)}(\vec x)\otimes\,P_-\Gamma_{IJ}\,,\label{fIJeven}
\eea
the curvatures in \re{fijeven} being defined by
\bea
F_{ij}^{(\pm)}&=&\pa_iA_j^{(\pm)}-\pa_jA_i^{(\pm)}+[A_i^{(\pm)},A_j^{(\pm)}]\label{curvpm}\\
f_{ij}&=&\pa_ia_j-\pa_ja_i\,.\label{curvabel}
\eea
The covariant derivative in \re{fiIeven} now is defined as
\bea
D_i\vf&=&\pa_i\vf+A_i^{(+)}\,\vf-\vf\,A_i^{(-)}+i\,a_i\,\vf\label{coveven}\\
D_i\vf^{\dagger}&=&\pa_i\vf^{\dagger}+A_i^{(-)}\,\vf^{\dagger}-\vf^{\dagger}\,A_i^{(+)}
-i\,a_i\,\vf^{\dagger}\,,\label{covevendag}
\eea
and the quantities $S^{(\pm)}$ in \re{fIJeven} are
\be
\label{Spm}
S^{(+)}=\vf\,\vf^{\dagger}-\eta^2\quad,\quad S^{(-)}=\vf^{\dagger}\,\vf-\eta^2\,.
\ee

In what follows, we will suppress the Abelian field $a_i$, since only when less stringent symmetry than spherical is
imposed is it that it would contribute. In any case, using the formal replacement
\[
A_i^{(\pm)}\leftrightarrow A_i^{(\pm)}\pm\frac{i}{2}\,a_i\,\eins
\]
yields the algebraic results to be derived below, in the general case.

We now refine our calculus of descent over even codimensions further. We see from \re{aieven} that $A_i^{(\pm)}$ being
$2^{\frac{D}{2}}\times 2^{\frac{D}{2}}$ arrays, that they can take their values in the two chiral representations,
repectively, of the algebra of $SO(D)$. It is therefore natural to introduce the full $SO(D)$ connection
\be
A_{i}=\left[
\begin{array}{cc}
A_{i}^{(+)} & 0\\
0 & A_{\mu}^{(-)}
\end{array}
\right]\,.\label{ASOD}
\ee
Next, we define the $D-$component $isovector$ Higgs field
\be
\Phi=\left[
\begin{array}{cc}
0 & \varphi\\
-\varphi^{\dagger} & 0
\end{array}
\right]=\f^i\,\Gamma_{i,D+1}\label{PHY}
\ee
in terms of the Dirac matrix representation of the algebra of $SO(D+1)$,
with $\Gamma_{i,D+1}=-\frac12\Gamma_{D+1}\Gamma_i$.

Note here the formal equivalence between the Higgs multiplet \re{PHY} in even $D$, to the corresponding one \re{Hisoodd}
in odd $D$. This formal equivalence turns out to be very useful in the calulus employed in following Sections.
In contrast with the former case of odd $D$ however, the form \re{PHY} for even $D$ is much more restrictive.
This is because in this case the Higgs multiplet is restricted to take its values in the components $\Gamma_{i,D+1}$
orthogonal to the elements $\Gamma_{ij}$ of $SO(D)$ by definition, irrespective of what symmetry is imposed.
Referring to footnote $8$, it is clear that relaxing the spherical symmetry here, does not result in $\F$ getting out
of the orthogonal complement of $\Gamma_{ij}$, when $D$ is even.

From \re{ASOD}, follows the $SO(D)$ curvature
\be
F_{ij}=\pa_iA_j-\pa_jA_i+[A_i,A_j]=\left[
\begin{array}{cc}
F_{ij}^{(+)} & 0\\
0 & F_{ij}^{(-)}
\end{array}
\right]\label{FSOD}
\ee
and from \re{ASOD} and \re{PHY} follows the covariant derivative
\be
D_{\mu}\Phi=\pa_i\F+[A_i,\F]=\left[
\begin{array}{cc}
0 & D_{i}\varphi\\
-D_{i}\varphi^{\dagger} & 0
\end{array}
\right]\,.
\label{covPHY}
\ee
From \re{PHY} there simply follows the definition of $S$ for even $D$
\be
\label{SPHY}
S=-(\eta^2\,\eins+\F^2)=\left[
\begin{array}{cc}
S^{(+)} & 0 \\
0 & S^{(-)}
\end{array}
\right]\,.
\ee

\medskip
\noindent
{\bf Note 1}: Given that $D+N$ is even, an Abelian gauge field Higgs system with $SO(D)=SO(2)$ results only from descents
over even dimensional codimensions. The finite energy solutions in these cases are Abelian $vortices$, which are
qualitatively different from all other cases describing $monopoles$. Unlike monpoles,
which asymptotically are (singular) Dirac--Yang monopoles, vortices do not support an asymptotic curvature.
Likewise, models supporting vortices cannot be adapted to support Julia--Zee type
dyons. Another such difference between monopoles and vortices is that in the former case the
boundary conditions (on the large sphere -- not circle!) can be adjusted to support monopole--antimonopole pairs and
chains, while this option is not open for the latter case with an asymptotic circle.

The dimensional reduction formulae for $D=2$ can in principle be read from the even--descent
formulae \re{aieven}-\re{SPHY}, but it is convenient to display them distinctly in a more
transparent notation, which is given in Section {\bf 4.2.1} below.

\medskip
\noindent
{\bf Note 2}: Descent by one dimension only is special and the calculus involved is rather trivial. In that case the
symmetry group of the YM field in the bulk does not break. In the (odd) $N=1$ case the matrices in \re{aiodd}-\re{aIodd}
contract to real numbers, hence $(A_i,A_I)$ take their values in the algebra of the bulk gauge group like
$({\cal A}_i,{\cal A}_I)$.

In addition, the components of the bulk gauge curvature on the codimension space, \re{fIJodd},
vanishes and the residual YMH model does not feature a symmetry breaking Higgs potential.

\medskip
\noindent
{\bf Note 3}: In the extreme case where $D=1$, the residual connection ${\cal A_i}$ in \re{aiodd} has only
one component and hence its curvature vanishes. It follows that this connection is gauge equivalent to zero, hence one
ends up with a residual system described by a scalar field $\F$ in \re{aIodd}. The covariant derivative in \re{fiIodd}
then becomes a partial derivative, and the components \re{fIJodd} on the codimension lead to symmetry breaking potentials.
The resulting systems are more nonlinear versions of the $\f^4$ model on $\R^1$. These will henceforth be ignored.

\subsubsection{Descent over $S^N$:  $N$ even and $D=2$}
Dimensional descendants of the connection $({\cal A}_i,{\cal A}_I)$ on $\R^2\times S^{N}$, expressed at the North pole of
$S^N$ are given by
\bea
{\cal A}_i&=&\frac{i}{2}\,A_i(\vec x)\,\Gamma_{N+1}\label{aiD=2}\\
{\cal A}_I&=&\frac{i}{2}\left[\vf(\vec x)\ P_+\Gamma_I+
\vf^{\star}(\vec x)\ P_-\Gamma_I\right]\label{aID=2}\,,
\eea
where $A_i$ is now the residual Abelian gauge field and $\vf$ is a complex valued scalar field.

The components of the curvaturs are readily calculated to give
\bea
{\cal F}_{ij}&=&\frac{i}{2}\,F_{ij}(\vec x)\,\Gamma_{N+1}\label{fijD2}\\
{\cal F}_{iI}&=&\frac{i}{2}\left[D_i\vf(\vec x)\ P_+\Gamma_I+
D_i\vf^*(\vec x)\ P_-\Gamma_I\right]\label{fiID2}\\
{\cal F}_{IJ}&=&S(\vec x)\,\Gamma_{IJ}\label{fIJD2}
\eea
where $F_{ij}=\pa_iA_j-\pa_jA_i$ is the residual Abelian curvature, the covariant derivative
$D_I\vf$ is
\be
\label{covD=2}
D_i\vf=\pa_i\vf+i\,A_i\,\vf
\ee
and $S$ now is
\be
\label{SD=2}
S=-\eta^2+|\vf|^2\,.
\ee

\section{Dimensional reduction of Chern-Pontryagin densities}
\label{CPdescent}
The crucial step in producing a YMH model on $\R^D$ that can support finite energy topologically stable monopoles,
$i.e.$, that one can establish a
topological lower bound on the energy, is to find the monopole charge density which is a dimensional
decsendant of the the relevant Chern--Pontryagin density. In other words, to show that the density on the right
hand side of \re{top-lb2b} is a total divergence. Denoting the CP density on $\R^D$ by ${\cal C}_D^{(n)}$ descended
from ${\cal C}^{(n)}$ on $\R^D\times S^{2n-D}$, the result to be demonstrated is
\be
\label{descCP}
{\cal C}_D^{(n)}={\bf\nabla}\cdot{\bf\Omega}^{(n,D)}\,.
\ee
The actual proof that the residual Chern--Pontryagin density is a
total divergence proceeds directly by taking arbitrary variations $\delta A_i$ and $\delta \F$ of it, and to show that
these lead to $trivial$ variational equations. On the other hand, should one wish to exploit the densities
${\bf\Omega}^{(n,D)}$ in the role of Chern--Simons densities in a $(D-1\ ,\ 1)$ Minkowskian theory, then one would need
the explicit expressions of ${\bf\Omega}[A_{\mu},\F]$. This is the task carried out in the present Section.

For practical reasons we restrict ourselves to the second, third and fourth Chern-Pontryagin (CP) densities, hence
it becoming clear that this result holds for arbitrary CP density. In each case we will consider all possible descents,
yielding the topological charges of the YMH models on the residual $\R^D$. Thus from the second CP density one
arrives at the vortex number of the Abelian Higgs model~\cite{Abrikosov:1956sx,Nielsen:1973cs} on $\R^2$ and
the ('t~Hooft-Polyakov) monopole charge of the
Georgi--Glashow model on $\R^3$ . From the third CP density one arrives at the topological charges of (one of the)
gereralised Abelian Higgs models~\cite{Burzlaff:1994tf,Arthur:1998nh} on $\R^2$, and, the monopole charges of generalised YMH models
on $\R^3$, $\R^4$ and $\R^5$ descended from the bulk YM model
\be
\label{3rd}
{\cal H}=\mbox{Tr}\,{\cal F}(2)^2+\ka^2\,\mbox{Tr}\,{\cal F}(4)^2\,.
\ee
From the fourth CP density one arrives at the topological charges of (one of the other)
gereralised Abelian Higgs models \cite{Burzlaff:1994tf,Arthur:1998nh} on $\R^2$, and, the monopole charges of generalised
YMH models on $\R^3$, $\R^4$, $\R^5$, $\R^6$  and $\R^7$ descended from the two bulk YM models
\bea
{\cal H}_1&=&\mbox{Tr}\,{\cal F}(4)^2\,,\label{4th1}\\
{\cal H}_2&=&\mbox{Tr}\,{\cal F}(2)^2+\ka^2\,\mbox{Tr}\,{\cal F}(6)^2\,.\label{4th2}
\eea

There is an important distinction between the "vortex numbers" and "monopiole charges" stated above. The former are the
topological charges of Abelian gauge field systems on $R^2$, hence the curvature on the large circle has
no curvature so that there exists no Dirac--Yang~\cite{Tchrakian:2008zz} fields asymptotically. By contrast,
all the non-Abelian gauge field systems on $R^D$, $D\ge 3$ do have asymptotic Dirac--Yang fields, with the important
consequence that the asymptotic gauge connection is $half\ pure\ gauge$, $i.e.$, decays as $r^{-1}$, typical of
monopoles rather than instantons.

\subsection{Topological densities on $\R^D$ from second Chern--Pontryagin density ${\cal C}^{(2)}$}
\label{CP2descent}
In this case $D+N=4$ and hence there are only two possible descents, with $N=1$ and $2$ (to $D=3$ and $2$
respectively), since we exclude descent to $D=1$.

\subsubsection{$N=1$,\ \ $D=3$}
\label{CP2descent1}
Using \re{fijodd} and \re{fiIodd}, one has
\bea
{\cal C}_3^{(2)}&=&\vep_{ijk4}\mbox{Tr}\ {\cal F}_{ij}\,{\cal F}_{k4}\nonumber\\
&\equiv&\vep_{ijk}\mbox{Tr}\ F_{ij}\,D_k\F\nonumber\\
&\stackrel{{\rm def.}}=&\bnabla\cdot\bOmega^{(2,3)}\nonumber
\eea
which is manifestly a total derivative defining the CS density
\be
\label{2,3}
\Omega^{(2,3)}_k=\vep_{ijk}\,\mbox{Tr}\ F_{ij}\,\F\,.
\ee

\subsubsection{$N=2$,\ \ $D=2$}
\label{CP2descent2}
The CP term
\bea
\frac12\,
{\cal C}_2^{(2)}&=&\vep_{ij}\vep_{IJ}\mbox{Tr}
\left({\cal F}_{ij}{\cal F}_{IJ}-2\,{\cal F}_{iI}{\cal F}_{jJ}\right)\label{CP22}
\eea
is subjected to dimensional reduction using the symmetry contraints \re{fijD2}, \re{fiID2}
and \re{fIJD2}, and performing the traces over the codimension indices $I,J$.

The result is
\bea
\frac12\,
{\cal C}_2^{(2)}&=&\vep_{ij}\left(S\,F_{ij}-2i\,D_i\vf^*\,D_j\vf
\right)\label{2,2}\\
&\stackrel{{\rm def.}}=&\bnabla\cdot\bOmega^{(2,2)}\nonumber
\eea
which is manifestly a total divergence defining the CS density
\be
\label{CS22}
\Omega^{(2,2)}_i\simeq\vep_{ij}\left(\eta^2\,A_j+i\,\vf^*D_j\vf\right)\,.
\ee
\subsection{Topological densities on $\R^D$ from third Chern--Pontryagin density ${\cal C}^{(3)}$}
\label{CP3descent}
In this case $D+N=6$ and hence there are four possible descents, with $N=1,2,3,$ and $4$ (to $D=5,4,3$ and $2$
respectively).

\subsubsection{$N=1$,\ \ $D=5$}
\label{CP3descent1}
Using \re{fijodd} and \re{fiIodd}, one has
\bea
{\cal C}_5^{(3)}&=&\vep_{ijklm6}\mbox{Tr}\ {\cal F}_{ijkl}\,{\cal F}_{m6}\nonumber\\
&\equiv&\vep_{ijklm}\mbox{Tr}\ F_{ij}\,F_{kl}\,D_m\F\nonumber\\
&\stackrel{{\rm def.}}=&3!\,\bnabla\cdot\bOmega^{(3,5)}\,,\label{35}
\eea
which is manifestly a total derivative defining the CS density
\be
\label{3,5}
\Omega^{(3,5)}_m=\vep_{ijklm}\,\mbox{Tr}\ F_{ij}\,F_{kl}\,\F\,.
\ee

\subsubsection{$N=2$,\ \ $D=4$}
\label{CP3descent2}
The CP term
\bea
{\cal C}_4^{(3)}&=&\vep_{ijkl}\,\vep_{IJ}\mbox{Tr}\left(
{\cal F}_{ijkl}\,{\cal F}_{IJ}-8\,{\cal F}_{ijkI}\,{\cal F}_{lJ}
+6\,{\cal F}_{ijIJ}\,{\cal F}_{kl}\right)\label{CP34}\\
&=&18\,\vep_{ijkl}\,\vep_{IJ}\mbox{Tr}
\left(\ {\cal F}_{ij}\,{\cal F}_{kl}\,{\cal F}_{IJ}-
4\,{\cal F}_{iI}\,{\cal F}_{jJ}\,{\cal F}_{kl}\right)\nonumber
\eea
is subjected to dimensional reduction using the symmetry contraints \re{fijeven}, \re{fiIeven}
and \re{fIJeven}, and performing the traces over the codimension indices $I,J$.

The result can be recast in transparent form by further using \re{ASOD}-\re{PHY} and
\re{FSOD}, \re{covPHY}, \re{SPHY}. This leads to the
compact~\footnote{In the ensuing manipulations,
note that $A_i$ and $F_{ij}$ commute with $\Gamma_5$, while $\F$ and $D_i\F$
anticommute with it.} expression
\bea
{\cal C}_4^{(3)}&=&18\,\vep_{ijkl}\,\mbox{Tr}\,\Gamma_5\,\left(
S\,F_{ij}F_{kl}+2\,D_i\F\,D_j\F\,F_{kl}\right)\label{3,4}\\
&\stackrel{{\rm def.}}=&\bnabla\cdot\bOmega^{(3,4)}\,.\nonumber
\eea
From the definition \re{SPHY} of $S$, it is clear that, the leading Higgs independent term
\be
\label{lead43}
\eta^2\,\vep_{ijkl}\,F_{ij}F_{kl}=\eta^2\,\bnabla\cdot\bOmega^{(4)}
\ee
is manifestly a total divergence in terms of the usual CS density of the pure YM field on $\R^4$
\bea
\Omega_{i}^{(4)}&=&\vep_{ijkl}\,\mbox{Tr}\,\Gamma_5\,A_{j}\left[F_{kl}-\frac23A_{k}A_{l}
\right]\label{CS4}
\eea
is manifestly a total divergence in terms of the usual CS density of the pure YM field on $\R^4$

The rest of the (Higgs dependent) terms in \re{3,4} can be shown to be also a total divergence,
such that
\bea
\Omega^{(3,4)}_i&=&3!^2\,\vep_{ijkl}\,\mbox{Tr}\,\Gamma_5\,\left[-2\eta^2A_{j}\left(F_{kl}
-\frac23A_{k}A_{l}\right)+\left(\F\,D_j\F-D_j\F\,\F\right)\,F_{kl}
\right]\,.\label{CS34}
\eea

\subsubsection{$N=3$,\ \ $D=3$}
\label{CP3descent3}
The CP density
\bea
\frac14
{\cal C}_3^{(3)}&=&\vep_{ijk}\,\vep_{IJK}\mbox{Tr}\left({\cal F}_{iIjk}\,{\cal F}_{JK}
+{\cal F}_{iIJK}\,{\cal F}_{jk}+3{\cal F}_{jkJK}\,{\cal F}_{iI}\right)\label{CP33}\\
&=&3\,\vep_{ijk}\,\vep_{IJK}\mbox{Tr}\left(3{\cal F}_{jk}\,{\cal F}_{iI}\,{\cal F}_{JK}+
3{\cal F}_{jk}\,{\cal F}_{JK}\,{\cal F}_{iI}-4{\cal F}_{iI}\,{\cal F}_{jJ}\,{\cal F}_{kK}\right)\nonumber
\eea
is subjected to dimensional reduction using the symmetry constraints
\re{fijodd}, \re{fiIodd} and \re{fIJodd}.

This results directly in
\bea
{\cal C}_3^{(3)}&=&3!^2\vep_{ijk}\,\mbox{Tr}\left[3F_{jk}\left(D_i\F\,S+S\,D_i\F\right)+2
D_i\F\,D_j\F\,D_k\F\right]\label{43,3}\\
&\stackrel{{\rm def.}}=&\bnabla\cdot\bOmega^{(4,3)}\nonumber
\eea
which is manifestly a total divergence defining the CS density
\bea
\Omega^{(3,3)}_k&=&(18)^2\,\vep_{ijk}\,\mbox{Tr}\left[-(3\eta^2-\F^2)\,\F\,F_{ij}+
\F\,D_i\F\,D_j\F\right]\,.\label{CS33}
\eea

The calculus described in this subsection was first developed in \cite{Sherry:1982fd}.

\subsubsection{$N=4$,\ \ $D=2$}
\label{CP3descent4}
This example wqas first considered in \cite{Ma:1986pu}.
The CP term
\bea
\frac18\,
{\cal C}_2^{(3)}&=&\vep_{ij}\vep_{IJKL}\mbox{Tr}
\left(6\,{\cal F}_{ijIJ}{\cal F}_{KL}-8\,{\cal F}_{iIKL}{\cal F}_{jJ}+
{\cal F}_{IJKL}{\cal F}_{ij}\right)\nonumber\\
&=&18\,\vep_{ij}\vep_{IJKL}\mbox{Tr}\left({\cal F}_{ij}{\cal F}_{IJ}{\cal F}_{KL}
-4{\cal F}_{iI}{\cal F}_{jJ}{\cal F}_{KL}\right)\label{CP32}
\eea
is subjected to dimensional reduction using the symmetry contraints \re{fijD2}, \re{fiID2}
and \re{fIJD2}, and performing the traces over the codimension indices $I,J,K,L$.

The result is
\bea
\frac18\,
{\cal C}_2^{(3)}&=&9i\vep_{ij}\left(S^2\,F_{ij}-4i\,S\,D_i\vf^*\,D_j\vf
\right)\label{3,2}\\
&\stackrel{{\rm def.}}=&\bnabla\cdot\bOmega^{(3,2)}\nonumber
\eea
which is manifestly a total divergence defining the CS density
\be
\label{CS32}
\Omega^{(3,2)}_i\simeq\vep_{ij}\left[\eta^4\,A_j-i(2\eta^2-|\vf|^2)
\,\vf^*D_j\vf\right]\,.
\ee

\subsection{Topological densities on $\R^D$ from fourth Chern--Pontryagin density ${\cal C}^{(4)}$}
\label{CP4descent}
In this case $D+N=8$ and hence there are six possible descents, with $N=1,2,3,4,5$ and $6$ (to $D=7,6,5,4,3$ and $2$
respectively).
For future convenience, namely when studying the Bogomol'nyi lower bounds on descendants of the energy density
\re{4th1}, we express the $4-$th CP density in the form
\[
{\cal C}_D^{(4)}=\mbox{Tr}\ {\cal F}(4)\wedge{\cal F}(4) 
\]
rather than as $\mbox{Tr}\ {\cal F}(2)\wedge{\cal F}(2)\wedge{\cal F}(2)\wedge{\cal F}(2)$. This is just a practical
option which enables the treatment of all the above listed cases in a uniform manner.

\subsubsection{$N=1$,\ \ $D=7$}
\label{CP4descent1}
Using \re{fijodd} and \re{fiIodd}, one has
\bea
{\cal C}_7^{(4)}&=&\vep_{ijklmnp}\mbox{Tr}\ {\cal F}_{ij}\,{\cal F}_{kl}\,{\cal F}_{mn}\,{\cal F}_{p6}\nonumber\\
&=&\vep_{ijklmnp}\mbox{Tr}\ F_{ij}\,F_{kl}\,F_{mn}\,D_p\F\nonumber\\
&\stackrel{{\rm def.}}=&3!^2\,\bnabla\cdot\bOmega^{(4,7)}\nonumber
\eea
which is manifestly a total derivative defining the CS density
\be
\label{4,7}
\Omega^{(4,7)}_p=\vep_{ijklmnp}\,\mbox{Tr}\ F_{ij}\,F_{kl}\,F_{mn}\,\F\,.
\ee

\subsubsection{$N=2$,\ \ $D=6$}
\label{CP4descent2}
The CP term
\bea
\frac14
{\cal C}_6^{(4)}&=&\vep_{ijklmn}\,\vep_{IJ}\mbox{Tr}\left(3{\cal F}_{ijkl}\,{\cal F}_{mnIJ}-4{\cal F}_{ijkI}\,{\cal F}_{lmnJ}\right)
\nonumber\\
&=&36\,\vep_{ijklmn}\,\vep_{IJ}\mbox{Tr}\left(
{\cal F}_{ij}\,{\cal F}_{kl}\,{\cal F}_{mn}\,{\cal F}_{IJ}-
4{\cal F}_{ij}\,{\cal F}_{kl}\,{\cal F}_{mI}\,{\cal F}_{nJ}
-2\,{\cal F}_{ij}\,{\cal F}_{mI}\,{\cal F}_{kl}\,{\cal F}_{nJ}\right)\label{CP46}
\eea
is subjected to dimensional reduction using the symmetry contraints \re{fijeven}, \re{fiIeven}
and \re{fIJeven}, and performing the traces over the codimension indices $I,J$.

The result can be recast in transparent form by further using \re{ASOD}-\re{PHY} and
\re{FSOD}, \re{covPHY}, \re{SPHY}. This leads to the compact
expression~\footnote{See footnote 9.}
\bea
\frac{1}{144}
{\cal C}_6^{(4)}&=&\vep_{ijklmn}\,\mbox{Tr}\,\Gamma_7\,
\left[S\,F_{ij}F_{kl}F_{mn}+2\,F_{ij}F_{kl}D_m\F D_n\F+F_{ij}D_m\F F_{kl}D_n\F\right]
\label{4,6}\\
&\stackrel{{\rm def.}}=&\bnabla\cdot\bOmega^{(4,6)}\,.\nonumber
\eea
From the definition \re{SPHY} of $S$, it is clear that, the leading Higgs independent term
\be
\label{lead6}
\eta^2\,\vep_{ijklmn}\,F_{ij}F_{kl}F_{mn}=\eta^2\,\bnabla\cdot\bOmega^{(6)}
\ee
is manifestly a total divergence in terms of the usual CS density of the pure YM field on $\R^6$
\bea
\Omega_{i}^{(6)}&=&2\,\vep_{ijklmn}\,\mbox{Tr}\,\Gamma_7\,A_{j}\left[F_{kl}F_{mn}-
F_{kl}A_{m}A_{n}+\frac25A_{k}A_{l}A_{m}A_{n}\right]\,.\label{CS6}
\eea
The rest of the (Higgs dependent) terms in \re{4,6} can be shown to be also a total divergence,
such that
\bea
-\frac{1}{144}\Omega^{(4,6)}_i&=&\Omega_{i}^{(6)}+\vep_{ijklmn}\,\mbox{Tr}\,\Gamma_7\,
D_j\F\left(\F F_{kl}F_{mn}+F_{kl}\F F_{mn}+F_{kl}F_{mn}\F\right)\label{CS46}
\eea

\subsubsection{$N=3$,\ \ $D=5$}
\label{CP4descent3}
The CP density
\bea
\frac18
{\cal C}_5^{(4)}&=&
\vep_{ijklm}\,\vep_{IJK}\mbox{Tr}\left({\cal F}_{ijkl}\,{\cal F}_{mIJK}+6{\cal F}_{ijIJ}\,{\cal F}_{klmK}\right)
\label{CP45}\\
&=&36\,\vep_{ijklm}\,\vep_{IJK}\mbox{Tr}\left(\ {\cal F}_{ij}\,{\cal F}_{kl}\,{\cal F}_{IJ}\,{\cal F}_{mK}+
\ {\cal F}_{ij}\,{\cal F}_{kl}\,{\cal F}_{mK}\,{\cal F}_{IJ}+\ {\cal F}_{ij}\,{\cal F}_{mK}\,{\cal F}_{kl}\,{\cal F}_{IJ}-
4\,{\cal F}_{iI}\,{\cal F}_{jJ}\,{\cal F}_{kK}\,{\cal F}_{mn}\right)\nonumber
\eea
is subjected to dimensional reduction using the symmetry constraints
\re{fijodd}, \re{fiIodd} and \re{fIJodd}.

This results directly in
\bea
\frac18
{\cal C}_5^{(4)}&=&3\,\vep_{ijklm}\,\mbox{Tr}
\left[ D_m\F\,(3\eta^2\,F_{ij}F_{kl}+F_{ij}F_{kl}\,\F^2+\F^2\,F_{ij}F_{kl}+F_{ij}\,\F^2 F_{kl})
-2F_{ij}D_k\F D_l\F D_m\F\right]\nonumber\\
&\stackrel{{\rm def.}}=&\bnabla\cdot\bOmega^{(4,5)}\label{4,5}
\eea
which is manifestly a total divergence defining the CS density
\bea
\frac18\Omega^{(4,5)}_m&=&(18)^2\,\vep_{ijklm}\,\mbox{Tr}\bigg[
\F\left(\eta^2\,F_{ij}F_{kl}+\frac29\,\F^2\,F_{ij}F_{kl}+\frac19\,F_{ij}\F^2F_{kl}\right)
\nonumber\\
&&\qquad\qquad\qquad\qquad-\frac29\left(\F D_i\F D_j\F-D_i\F\F D_j\F+D_i\F D_j\F\F\right)F_{kl}\bigg]\,.\label{CS45}
\eea
This example was considered in \cite{Kleihaus:1998kd}.

\subsubsection{$N=4$,\ \ $D=4$}
\label{CP4descent4}
The CP term
\bea
\frac12
{\cal C}_4^{(4)}&=&\vep_{ijkl}\,\vep_{IJKL}\mbox{Tr}\left(
{\cal F}_{ijkl}\,{\cal F}_{IJKL}-16\,{\cal F}_{ijkK}\,{\cal F}_{lLIJ}
+18\,{\cal F}_{ijIJ}\,{\cal F}_{klKL}\right)\nonumber\\
&=&36\,\vep_{ijkl}\,\vep_{IJKL}\mbox{Tr}\bigg[
2{\cal F}_{ij}\,{\cal F}_{kl}\,{\cal F}_{IJ}\,{\cal F}_{KL}+
{\cal F}_{ij}\,{\cal F}_{IJ}\,{\cal F}_{kl}\,{\cal F}_{KL}\nonumber\\
&&\qquad\qquad\qquad\qquad-8\left({\cal F}_{iI}{\cal F}_{jJ}{\cal F}_{kl}{\cal F}_{KL}+
{\cal F}_{iI}{\cal F}_{jJ}{\cal F}_{KL}{\cal F}_{kl}
+{\cal F}_{iI}{\cal F}_{kl}{\cal F}_{jJ}\,{\cal F}_{KL}\right)\nonumber\\
&&\qquad\qquad\qquad\qquad\qquad+8{\cal F}_{iI}{\cal F}_{jJ}{\cal F}_{kK}{\cal F}_{lL}\bigg]\label{CP44}
\eea
is subjected to dimensional reduction using the symmetry contraints \re{fijeven}, \re{fiIeven}
and \re{fIJeven}, and performing the traces over the codimension indices $I,J,K,L$.

The result can be recast in transparent form by further using \re{ASOD}-\re{PHY} and
\re{FSOD}, \re{covPHY}, \re{SPHY}. This leads to the
compact~\footnote{See footnote 9.} expression
\bea
\frac12
{\cal C}_4^{(4)}&=&18\,\vep_{ijkl}\,\mbox{Tr}\,\Gamma_5\,\bigg[
2\,S^2\,F_{ij}F_{kl}+F_{ij}\,S\,F_{kl}\,S\nonumber\\
&&\qquad\qquad\qquad+4\left(D_i\F\,D_j\F\,\{S,F_{kl}\}+D_i\F\,F_{kl}\,D_j\F\,S\right)\nonumber\\
&&\qquad\qquad\qquad+2\,D_i\F\,D_j\F\,D_k\F\,D_l\F\bigg]\label{4,4}\\
&\stackrel{{\rm def.}}=&\bnabla\cdot\bOmega^{(4,4)}\,.\nonumber
\eea
From the definition \re{SPHY} of $S$, it is clear that, the leading Higgs independent term
\be
\label{lead6-n}
\eta^4\,\vep_{ijkl}\,F_{ij}F_{kl}=\eta^4\,\bnabla\cdot\bOmega^{(4
)}
\ee
is manifestly a total divergence in terms of the usual CS density of the pure YM field on $\R^4$
\bea
\Omega_{i}^{(4)}&=&\vep_{ijkl}\,\mbox{Tr}\,\Gamma_5\,A_{j}\left[F_{kl}-\frac23A_{k}A_{l}
\right]\,.\label{CS4-n}
\eea
The rest of the (Higgs dependent) terms in \re{4,4} can be shown to be also a total divergence,
such that
\bea
\Omega^{(4,4)}_i&=&\vep_{ijkl}\,\mbox{Tr}\,\Gamma_5\,\bigg\{6\eta^4\,A_j\left(F_{kl}-\frac23\,A_k\,A_l
\right)\nonumber\\
&&\qquad\quad-6\,\eta^2\left(\F\,D_j\F-D_j\F\,\F\right)\,F_{kl}\nonumber\\
&&\qquad\quad+\left[\left(\F^2\,D_j\F\,\F-\F\,D_j\F\,\F^2\right)-2\left(\F^3\,D_j\F-D_j\F\,\F^3\right)\right]F_{kl}
\bigg\}\label{CS44}
\eea
\subsubsection{$N=5$,\ \ $D=3$}
\label{CP4descent5}
The CP density
\bea
\frac14
{\cal C}_3^{(4)}&=&2\vep_{ijk}\,\vep_{MIJKL}\mbox{Tr}\left({\cal F}_{ijkM}\,{\cal F}_{IJKL}
+6{\cal F}_{ijIJ}\,{\cal F}_{kMKL}\right)\label{CP43}\\
&=&4!^2\,\vep_{ijk}\,\vep_{MIJKL}\mbox{Tr}\left(\ {\cal F}_{ij}\,{\cal F}_{kM}\,{\cal F}_{IJ}\,{\cal F}_{KL}+
\ {\cal F}_{ij}\,{\cal F}_{IJ}\,{\cal F}_{kM}\,{\cal F}_{KL}+\ {\cal F}_{ij}\,{\cal F}_{IJ}\,{\cal F}_{kM}\,{\cal F}_{KL}-
4\,{\cal F}_{iI}\,{\cal F}_{jJ}\,{\cal F}_{kM}\,{\cal F}_{KL}\right)\nonumber
\eea
is subjected to dimensional reduction using the symmetry constraints
\re{fijodd}, \re{fiIodd} and \re{fIJodd}.

This results directly in
\bea
{\cal C}_3^{(4)}&=&4\,4!^2\,\vep_{ijk}\,\mbox{Tr}
\bigg(3\eta^4F_{ij}D_k\F+\eta^2\left[3F_{ij}\left(\F^2D_k\F+D_k\F\F^2\right)-2D_i\F D_j\F D_k\F\right]\nonumber\\
&&\qquad\qquad\qquad+\left[F_{ij}\left(\F^4D_k\F+D_k\F \F^4+\F^2D_k\F\F^2\right)-2\F^2D_i\F D_j\F D_k\F\right]\bigg)
\label{4,3}\\
&\stackrel{{\rm def.}}=&\bnabla\cdot\bOmega^{(4,3)}\nonumber
\eea
which is manifestly a total divergence defining the CS density
\bea
\Omega^{(4,3)}_k&=&(18)^2\,\vep_{ijk}\,\mbox{Tr}\bigg[3\eta^4\F F_{ij}+2\eta^2
\left(\F^3 F_{ij}-\F D_i\F D_j\F\right)\nonumber\\
&&\qquad\qquad\qquad+\frac15\left[3\F^5F_{ij}-2\left(2\F^3D_i\F D_j\F-\F^2D_i\F\,\F\,D_j\F\right)\right]
\bigg]\label{CS43}
\eea

\subsubsection{$N=6$,\ \ $D=2$}
\label{CP4descent6}
The CP term
\bea
\frac14\,
{\cal C}_2^{(4)}&=&\vep_{ij}\vep_{IJKLMN}\mbox{Tr}\left(3{\cal F}_{ijIJ}{\cal F}_{KLMN}-
4{\cal F}_{iIKL}{\cal F}_{jJMN}\right)\nonumber\\
&=&3!^2\vep_{ij}\vep_{IJKLMN}\mbox{Tr}\left({\cal F}_{ij}{\cal F}_{IJ}{\cal F}_{KL}{\cal F}_{MN}
-4{\cal F}_{iI}{\cal F}_{jJ}{\cal F}_{KL}{\cal F}_{MN}-2
{\cal F}_{iI}{\cal F}_{KL}{\cal F}_{jJ}{\cal F}_{MN}
\right)\label{CP42}
\eea
is subjected to dimensional reduction using the symmetry contraints \re{fijD2}, \re{fiID2}
and \re{fIJD2}, and performing the traces over the codimension indices $I,J,K,L$.

The result is
\bea
\frac14\,
{\cal C}_2^{(4)}&=&3!^2\vep_{ij}\bigg\{-\eta^6F_{ij}+3\eta^4\left(|\vf|^2F_{ij}
-2iD_i\vf^*D_j\vf\right)-3\eta^2|\vf|^2\left(|\vf|^2F_{ij}-4iD_i\vf^*D_j\vf\right)\nonumber\\
&&\qquad\qquad\qquad\qquad\qquad\qquad+(|\vf|^2)^2\left(|\vf|^2F_{ij}-6iD_i\vf^*D_j\vf\right)
\bigg\}\label{4,2}\\
&\stackrel{{\rm def.}}=&\bnabla\cdot\bOmega^{(4,2)}\nonumber
\eea
which is manifestly a total divergence defining the CS density~\footnote{The dimensional reduction of CP densities on
$\R^2\times S^{4p-2}$ down to $\R^2$ can be carried out systematically.}
\be
\label{CS42}
\Omega^{(4,2)}_i\simeq-\vep_{ij}\left(\eta^6A_j+i\left[3\eta^4-3\eta^2|\vf|^2+(|\vf|^2)^2\right]
\,\vf^*D_j\vf\right)\,.
\ee

\subsection{Gauge transformation properties of ${\bf\Omega}^{(n,D)}$}
The descended Chern--Pontryagin densities ${\cal C}_D^{(n)}$ on $\R^D$ presented above are
total divergences
\[
{\cal C}_D^{(n)}={\bf\nabla}\cdot{\bf\Omega}^{(n,D)}
\]
where the densities ${\bf\Omega}^{(n,D)}$ are the descended Chern--Simons densities on $\R^D$.
The most remarkable feature of the descended Chern--Simons densities ${\bf\Omega}^{(n,D)}$ on $\R^D$
is that for $odd\ D$ they are $gauge\ invariant$ functions of the Yang--Mills--Higgs fields, while
those on $even\ D$ are $gauge\ variant$ densities. The best known examples featuring this property are the magnetic
field ${\bf B}=\mbox{Tr}\F\,\vec B$ of the '~Hooft--Polyakov monopole on $\R^3$, which is $gauge\  invariant$, and, the
magnetic Maxwell potential $A_i$ yielding the vortex number
of the Abelian Higgs model on $\R^2$ which is $gauge\ variant$.

It is remarkable that the CS densities in even dimensional residual space $\R^D$ are cast
in two distinct parts; one {\it gauge variant} and the other {\it gauge invariant}.
The leading term is the gauge variant part which
does not feature the Higgs field and contributes to the surface integral over $S^{D-1}$,
yielding the monopole (or vortex) charge. The
rest, which is the Higgs dependent part, is gauge invariant and does not contribute to the topological charge.
The reason for this in the case of monopoles, which is the subject of interest here, is that both the Yang--Mills
curvature and the
covariant derivative of the Higgs field decay as $r^{-2}$. This property is explained by the
fact that monopoles are Dirac-Yang fields
as exposed in Section {\bf 6} below, or otherwise stated the monopole solutions must obey the
{\it finite energy conditions.}

What is perhaps more remarkable, if not unexpected, is the fact these
gauge variant leading terms are formally identical to the generic Chern--Simons
densities defined in terms of the Yang--Mills connection and curvature in $D+1$ dimensional spacetimes.
The only difference between
these types of densities is that for monopoles on $\R^D$ the gauge connection takes its values in the
Dirac representation of $SO(D)$,
while the connection of the generic Chern--Simons densities takes its values in the {\bf chiral}
representation of $SO(D)$.

The gauge invariant densities ${\bf\Omega}^{(n,D)}$ for odd $D$ also split up in a leading terms which
contributes to the topological
charge surface integral, and another part which decays too fast and has vanishing contribution. In this case,
both parts feature the Higgs field.

\subsection{Normalisation of the topological charge}
At this stage, it is in order to state the normalisation constants of the higher dimensional monopole charges, even
though the simplest presentation involves the spherically symmetric field configurations and hence anticipates results
in Section {\bf 6} below. Normalisation involves setting the charge of the {\it spherically symmetric} monopole equal
to $unity$. The topological charge (volume) integral over $\R^D$ then reduces to a one dimensional integral whose
integrand now is a total derivative and is integrated
trivially. This will not be carried out explicitly here since it is easily carried out in each case.

What is important to consider here is the normalisation of the monopole charge on $\R^D$ when $D$ is even, relative to the
normalisation of the corresponding instanton charge. These instantons are those alluded to in Section {\bf 2}. Since the
decay properties of the (asymptotically pure gauge) instantons are quite different from the milder decay of the
monopoles in the same dimensions, it might be expected that the respective normalisations will be different. In fact,
they are identical.

For pedagogical simplicity, it is sufficient to consider the $D=4$ case, and in particular the respective spherically
symmetric field configurations. The spherically symmetric Ansatz for the connection of the $D=4$ monopole is given by
\re{spheven} below, and the finite energy asymptotics of the function $w(r)$ there is that given by \re{asymw}.  The
corresponding Ansatz for the $D=4$ instanton is formally identical to that stated in \re{sphodd},
with no Higgs field, and where now $\Si_{ab}^{\pm}$ is that given by \re{Si} and not by \re{chiraloddmono}.
The asymptotics for finite energy/action in this case are
\be
\label{asyminst}
\lim_{r\to 0}\,w=\pm 1\qquad\qquad\lim_{r\to\infty}\,w=\mp 1\,,
\ee
which are quite different from \re{asymw}, since unlike the latter the instanton is pure gauge at infinity.

Now in both these cases the reduced Pontryagin density is proportional to the same total derivative and its integral is
\[
\int_{w(0)}^{w(\infty)}\frac{d}{dr}\left(w-\frac13w^3\right)\,dr\,.
\]
Thus in the case of the instanton the limits in \re{asyminst} result in double the integral of the monopole with limits
\re{asymw}. However, the Trace in the second CP density in the
case of the monopole is twice as large as that for the instanton, since in
the former case the matrices involved are the $4\times 4$ Dirac matrices
while in the latter case they are the $2\times 2$ chiral matrices. This
counting repeats in every $4p$ dimensions. Hence, the normaisation of the
$D=4$ monopole and the $D=4$ instanton are equal. This holds for all even $D$.

\noindent

\section{Spherical symmetry and Dirac--Yang monopoles}
\label{Dirac-Yang}
It is natural to introduce the Dirac--Yang (DY) monopoles at this stage, after having stated the spherically
symmetric Ans\"atze for
the $SO(D)$ gauge fields on $\R^D$, because the Dirac--Yang~\cite{Dirac:1931kp,Yang:1977qv,Tchrakian:2008zz}
fields are the asymptotic 
gauge fields of the spherically symmetric monopole solutions themselves. In that context, the
asymptotics result from the $finite\ energy\ boundary\ conditions$. These boundary conditions result
also from requiring that the monopole charge be a $topological\ charge$, $i.e.$, that for the monopole on $\R^D$,
the surface integral on the large spheres $S^{D-1}$ be convergent and normalised to an integer. Imposing spherical
symmetry on the monopole charge densities presented in the previous section, which are all manifestly total
divergences, yields total deriviative expressions.

The Dirac~\cite{Dirac:1931kp} monopole can be constructed by gauge transforming the asymptotic
't~Hooft-Polyakov monopole~\cite{'tHooft:1974qc,Polyakov:1974ek} in $D=3$, which can be taken to be spherically
symmetric~\footnote{Strictly speaking, it is not necessary to restrict to spherically symmetric
fields only. By choosing to start with the asymptotic axially symmetric fields
characterised with vorticity $n$, the gauge tranformed connection is just $n$
times the usual Dirac monopole field.}, such that the $SO(3)$ isovector
Higgs field is gauged to a (trivial) constant, and the $SU(2)\sim SO(3)$
gauge group of the Yang-Mills (YM) connection breaks down to
$U(1)\sim SO(2)$, the resulting Abelian connection developing a line
singularity on the positive or negative $(x_3=)z$-axis.

In exactly the same way, the Yang~\cite{Yang:1977qv} monopole can be constructed by transforming the
$D=5$ dimensional monopole
such that the $SO(5)$ isovector Higgs field is gauged to a (trivial) constant, and
the $SO(5)$ gauge group of the YM connection breaks down to
$SO(4)$, the resulting non Abelian connection developing a line
singularity on the positive or negative $x_5$-axis. In fact, the residual non
Abelian connection can take its values in one or other chiral representations of
$SU(2)$, as formulated by Yang~\cite{Yang:1977qv}, but this is a low dimensional accident which
does not apply to the higher diemnsional analogues to be defined below, all of which
are $SO(D-1)$ connctions. Just like the
't~Hooft-Polyakov monopole monopole is the regular counterpart of the Dirac monopole,
so is the $D=5$ dimensional 'monopole' the regular counterpart of the
Yang monopole.

The above two definitions of the Dirac and of the Yang monopoles will
be the template for our definition of what we will refer to as the
hierarchy of Dirac--Yang (DY) monopoles in all dimensions. The two
examples just given are both in odd ($D=3$ and $D=5$) dimensions, but
the DY hierarchy is in fact defined in all, including even, dimensions.

We start by stating the spherically symmetric Ansatz for the $SO(D)$ gauge connection $A_i$ and the $iso-D-vector$
Higgs field $\F$, for odd and even $D$ respectively, as
\bea
A_i^{(\pm)}=\frac{1}{r}\,(1-w(r))\,\Si_{ij}^{(\pm)}\,\hat x_j\quad&,&\quad
\F=2\,\eta\,h(r)\,\hat x_i\,\Si_{i,D+1}^{(\pm)}\qquad,\qquad{\rm for\ odd}\ \ D\label{sphodd}\\
A_i=\frac{1}{r}\,(1-w(r))\,\Ga_{ij}\,\hat x_j\quad&,&\quad
\F=2\,\eta\,h(r)\,\hat x_i\,\Ga_{i,D+1}\qquad,\qquad{\rm for\ even}\ \ D\,.\label{spheven}
\eea
In relations \re{sphodd}-\re{spheven}, $\hat x_i=\frac{x_i}{r}$, $i=1,2,..,D$, is the unit radius vector. $\Ga_i$ are the
Dirac gamma matrices in $D$ dimensions with the chiral matrix $\Ga_{D+1}$ for even
$D$, so that
\[
\Ga_{ij}=-\frac{1}{4}\left[\Ga_i,\Ga_j\right]
\]
are the Dirac representations of $SO(D)$. The matrices $\Si_{ij}$, employed only in
the odd $D$ case, are
\be
\label{chiraloddmono}
\Si_{ij}^{(\pm)}=-\frac{1}{4}\left(\frac{\eins\pm\Ga_{D+2}}{2}\right)
\left[\Ga_i,\Ga_j\right]\,,
\ee
$\Ga_{D+1}$ being the chiral matrix in $D+1$ dimensions, and $\Si_{ij}^{(\pm)}$
being one or other of the two possible chiral representations of the $SO(D)$
subgroup of $SO(D+1)$. 

Just as the Dirac monopole can be defined as a gauge transform of the
asymptotic spherically symmetric 't~Hooft-Polyakov monopole,
our definition for the DY fields in arbitrary $D$ dimensions starts from the asymptotic
(non Abelian) $SO(D)$ YM field $A_i$ and the $D$-tuplet Higgs field $\F$, with
\bea
&&\lim_{r\to\infty}w(r)=0\,,\qquad\lim_{r\to\infty}w(r)=1\label{asymw}\\
&&\lim_{r\to\infty}h(r)=1\,,\qquad\lim_{r\to\infty}h(r)=0\label{asymh}\\
\eea
leading to
\bea
A_i^{(\pm)}=\frac{1}{r}\,\Si_{ij}^{(\pm)}\,\hat x_j\quad&,&\quad
\F=2\,\eta\,\hat x_i\,\Si_{i,D+1}^{(\pm)}\qquad,\qquad{\rm for\ odd}\ \ D\label{YMHodd}\\
A_i=\frac{1}{r}\,\Ga_{ij}\,\hat x_j\quad&,&\quad
\F=2\,\eta\,\hat x_i\,\Ga_{i,D+1}\qquad,\qquad{\rm for\ even}\ \ D\,.\label{YMHeven}
\eea

The DY monopoles result from the action of the following $SO(D)$ gauge group element
\be
\label{g}
g_{\pm}=\frac{(1\pm\cos\ta_1)\eins\pm\Ga_D\Ga_{\al}\hat x_{\al}\sin\ta_1}
{\sqrt{2(1\pm\cos\ta_1)}}\,,
\ee
having parametrised the $\R^D$ coordinate $x_i=(x_{\al},x_D)$ in terms of the radial
variable $r$ and the polar angles
\be
\label{polar}
(\ta_1,\ta_2,..,\ta_{D-2},\vf)
\ee
with the index alpha running over
$\al=1,2,..,D-1$.  The meaning of the $\pm$ sign in \re{g} is as follows~\cite{Tchrakian:1999me}: Choosing
these signs the Dirac line singularity will be along the negative or positive
$x_D$--axis, respectively. (In the case of odd $D$ if we chose the opposite sign
on $\Si$ in \re{YMHodd} the situation will be reversed.) In other words the DY
field will be the $SO(D-1)$ connection on the upper or lower half $D-1$
sphere, $S^{D-1}$, respectively, the transition gauge transformation being given
by $g_+^{-1}g_-$. Notice that the dimensionality of the matrices $g$, \re{g}, and
those of both \re{YMHodd} and \re{YMHeven}, match in each case.

The result of the action of \re{g} on \re{YMHodd} or \re{YMHeven},
\bea
A_i\rightarrow &&g^{-1}\,A_ig+g^{-1}\pa_ig\nonumber\\
\F\rightarrow &&g^{-1} \F g\nonumber
\eea
yields the required DY fields
$\hat A_i^{(\pm)}=(\hat A_{\al}^{(\pm)},\hat A_D^{(\pm)})$
\bea
\hat A_{\al}^{(\pm)}&=&\frac{1}{r(1\pm\cos\ta_1)}\,\Si_{\al\beta}
\hat x_{\beta}\quad,\quad\hat A_{D}^{(\pm)}=0\quad,\quad{\rm for\ odd}\ D
\label{DYodd}\\
\hat A_{\al}^{(\pm)}&=&\frac{1}{r(1\pm\cos\ta_1)}\,\Ga_{\al\beta}
\hat x_{\beta}\quad,\quad\hat A_{D}^{(\pm)}=0\quad,\quad{\rm for\ even}\ D\,,
\label{DYeven}
\eea
and the Higgs field is gauged to a constant, i.e. it is trivialised.

The components of the DY curvature
$\hat F_{ij}^{(\pm)}=(\hat F_{\al\beta}^{(\pm)},\hat F_{\al D}^{(\pm)})$
follow from \re{DYodd}-\re{DYeven} straightforwardly. To save space we give only the
curvature corresponding to \re{DYodd}
\bea
\hat F_{\al\beta}^{(\pm)}&=&-\frac{1}{r^2}\left[\Ga_{\al\beta}+
\frac{1}{(1\pm\cos\ta_1)}\,\hat x_{[\al}\,\Ga_{\beta]\ga}\hat x_{[\ga}\right]
\label{fab}\\
\hat F_{\al D}^{(\pm)}&=&\pm\frac{1}{r^2}\,\Ga_{\al\ga}\hat x_{\ga}\,,\label{fad}
\eea
where the notation $[\al\beta]$ implies the antisymmetrisation of the indices, and
the components of the curvature for even $D$ corresponding to \re{DYeven} follows
by replacing $\Ga$ in \re{fab}-\re{fad} with $\Si^{(\pm)}$.
The parametrisation \re{DYodd}-\re{DYeven} and \re{fab}-\re{fad} for the DY field
appeared in \cite{Ma:1992zy} and \cite{Tchrakian:1999me}.

That the DY field \re{DYodd}-\re{DYeven} in $D$ dimensions, constructed by
gauge transforming the asymptotic fields \re{YMHodd}-\re{YMHeven} of a $SO(D)$ EYM
system, is a $SO(D-1)$ YM field is obvious. For $D=3$ and $D=5$, these are the
Dirac~\cite{Dirac:1931kp} and Yang~\cite{Yang:1977qv} monopoles, respectively.

In retrospect, we point out that to construct DY monopoles it is not even necessary
to start from a YMH system, but ignoring the Higgs field and simply applying the
gauge transformation \re{g} to the YM members of \re{YMHodd}-\re{YMHeven} results
in the DY monopoles \re{DYodd}-\re{DYeven}. In other words the only function of the
Higgs fields in \re{YMHodd}-\re{YMHeven} is the definition of the gauge group
element \re{g} designed to gauge it away.

It is perhaps reasonable to emphasise there that DY monopoles exist is all dimensions $D$, whether $D$ is odd
or even, as presented above. The most prominent difference between these cases is that for odd $D$, one has a
$gauge\ covariant$, $D-1$ form generalisation~\cite{Tchrakian:1999me} for the definition of a 't~Hooft electromagnetic
tensor, while for even $D$ this is higher form is $gauge\ variant$.

\section{Monopoles on $\R^D$: $D\ge 3$}
\label{monopoles}

These are topologically stable static finite energy solutions generalising the 't~Hooft--Polyakov monopole on $\R^3$.
The fundamental inequality yielding the appropriate Bogomol'nyi lower bound is \re{top-lb2b}, which is descended
from the inequality \re{top-lb1}. The simplest option here is to take $p=q$ in \re{top-lb1}, in which case the density on
the right hand side will consist of a single term \re{pYM} scale invariant in $4p$ dimensions. For the purposes of the
present notes, attention will be restricted to bulk systems in 6 and 8 dimensions, since these two suffice to expose
all quatitative features of monopoles in higher dimensions. In the first case, $p+q=3$ the only choice we have is
$p=1$ and $q=2$, $i.e.,$ $p\neq q$, while in the second case $p+q=8$ allows one to opt for the simpler choice of
$p=q=4$, which is what will be done here.

The Bogomol'nyi bounds \re{descent}, descending
from \re{top-lb1}, are in general not saturated. When $p\neq q$, the presence of the dimensionful
constant $\ka$ obstructs the construction of self-dual solutions exactly as is the case in
Skyrme~\cite{Skyrme:1962vh} theory.
When $p=q$ on the other hand, the inequality
can be saturated in the bulk but its descendants on the residual space result in Bogomol'nyi
equations that are in general overdetermined~\cite{Tchrakian:1990gc} and do not
support any nontrivial $self$-$dual$ monopole solutions. Such monopoles are solutions of second order Euler-Lagrange
equations. The exceptions~\footnote{The only other exceptions are the vortices supported by the Abelian
Higgs models on $\R^2$ descended from the $p-$YM systems on $\R^{4p}$, given in \cite{Burzlaff:1994tf,Arthur:1998nh}.
These do have a BPS limit, which solutions might be referred to as $p-$BPS vortices. An analytic proof of existence for
the $p=2$ BPS vortices can be found in \cite{Tchrakian:1996bc},
which can readily be adapted to the arbitrary $p$ case.} are 
the monopoles supported by YMH models on $\R^{4p-1}$ descended from the $p-$YM systems on $\R^{4p}$, given in
\cite{Radu:2005rf}, which might be referred to as $p-$BPS monopoles. An analytic proof of existence for the $p-$BPS
monopoles is given in \cite{Yisong1,Yisong2,Yisong3}. The $p=1$ case is the BPS monopole which is known in closed form,
all other $p-$BPS monopoles being constructed only numerically.

The presentation here is restricted to YMH systems on $\R^D\ , D\ge 3$, supporting monopoles. The vortices on $\R^2$
supported by models descending from higher dimensional YM systems \cite{Burzlaff:1994tf,Arthur:1998nh} are not included
here since consideration is restricted to monopoles only.

In any dimension $\R^D$ there is an infinite tower of YM-Higgs (YMH) models, each descending from
the $p-$YM member \re{pYM} of the Yang--Mills hierarchy on the bulk of dimension $D+N$, for all $N$, $i.e.$, such
that $4p\ge D+N$. In these notes, only the first nontrivial elements of these towers are presented explicitly, namely
that consideration is restricted to monopoles of YMH models descending from the $p=1$ and $p=2$ members of the
Yang--Mills hierarchy defined on the bulk dimensions $D+N=4$, $D+N=6$ and $D+N=8$.

In the $D+N=4$ case, the monopole charge density is the descendant of the second Chern--Pontryagin density, and
the YMH models in question are the descendants of the usual ($p=1$) YM system. There
are the two possibilities, $(D=3\ ,\ N=1)$ and ($D=2\ ,\ N=2$), the first of which is the usual 't~Hooft--Polyakov
monopole in the BPS limit and the second is the usual Abelian Higgs model supporting the ANO vortex, consideration
of which is excluded.

In the $D+N=6$ case, the monopole charge density is the descendant of the third Chern--Pontryagin density, and
the YMH models in question are the descendants of the sum of the $p=1$ and $p=2$ members of the YM hierarchy.
There are the four possible monopoles, $(D=5\ ,\ N=1)$, $(D=4\ ,\ N=2)$ and $(D=3\ ,\ N=3)$, the vortex case
$(D=2\ ,\ N=4)$ again being excluded. In fact, the $(D=5\ ,\ N=1)$ and $(D=4\ ,\ N=2)$ are also excluded.  The
reason is that the energy of the usual ($p=1$) YM term diverges in both $D=5$ and $D=4$ dimensions,
since the asymptotic connection is a Dirac--Yang (DY) and decays as $r^{-1}$.

In the $D+N=8$ case, the monopole charge density is the descendant of the fourth Chern--Pontryagin density, and
the YMH models are those descending from the $p=2$ members of the YM hierarchy. In this case one does not have the option
of employing the descendants of the sum of the $p=1$ and $p=2$ members of the YM hierarchy, since the only possibility
is the (generalised~\cite{Burzlaff:1994tf,Arthur:1998nh}) ANO vortex~\cite{Abrikosov:1956sx,Nielsen:1973cs}
on $\R^2$, outside of interest here. This results in
the possibilities $(D=7\ ,N=1)$, $(D=6\ ,N=2)$, $(D=5\ ,N=3)$, $(D=4\ ,N=4)$ and $(D=3\ ,N=5)$.  

To date, only two of the above mentioned models resulting from a descent over $S^N$ with $N\ge 2$ have been studied
quantitatively. These are the $(D=3\ ,N=5)$ \cite{Kleihaus:1998kd} and the $(D=4\ ,N=4)$ \cite{O'Brien:1988xr} monopoles,
respectively. In addition, the monopoles resulting from a descent over $S^1$ with $N\ge 1$, $i.e.$, those on
$R^3$, $R^5$ and $\R^7$ are readily constructed as special cases of the monopoles on arbitrary dimensions $\R^{2n+1}$
given in \cite{Tchrakian:1978sf,Kihara:2007di,Radu:2005rf,Breitenlohner:2009zi}.

Since all descents are perfomed from compact bulk dimensions to Euclidean residual dimensions, the resulting
residual systems are described as static Hamiltonians~\footnote{The description of 'static
Hamiltonian' for the descended system is used here, in anticipation of employing the name 'static Lagrangian' in the
next section, where the corresponding towers of Julia--Zee dyons will be described.}.
In what follows, the (candidate static) Hamiltonian densities with $(D,N)$ are denoted as ${\cal H}^{(N)}_D$.

\subsection{Monopole on $\R^3$  with descended second Chern--Pontryagin charge}
\label{2monopole3}
This is the usual 't~Hooft--Polyakov monopole on $\R^3$ in the BPS limit. It is also the first one in the hierarchy of
monopoles~\cite{Tchrakian:1978sf,Kihara:2007di,Radu:2005rf,Breitenlohner:2009zi} on $\R^{2n+1}$.

\subsection{Vortex on $\R^2$ with descended second Chern--Pontryagin charge}
Since the density whose surface integral yielding winding number \re{CS22} pertaining to this case was given in
Section {\bf 5.1.2}, it is sufficient here to state that the model supporting these vortices is the usual Abelian Higgs
model~\cite{Abrikosov:1956sx,Nielsen:1973cs}.

\subsection{Monopole on $\R^3$  with descended third Chern--Pontryagin charge}
\label{3monopole3}
This is the monopole whose topological lower bound is given by the volume integral of the topological charge density
\re{CP33}, or the surface integral of \re{43,3}. It is the $only$ monopole in this class, since the energy integrals
of those on $\R^5$ and $\R^4$ are divergent due to the presence of the usual ($p=1$) YM term. Another special feature
of this model is that it features a dimensionful constant $\ka$, exactly like the usual Skyrme model. In this sense,
the question of saturating the topological lower bound does not arise. It is of course likewise, possible to estimate
the amount by which the lowest energy monopole exceeds this lower bound numerically.

What distinguishes the model employed here from the rest of the examples given later~\footnote{The models whose monopole
charge density is the descendant of the 4th Chern-Pontryagin density, are descended from 8 dimensional bulk space
and hence one has the option of employing a bulk action density with $p=q=2$ in \re{descent},
featuring only ${\cal F}(4)$, or, one with $p=2$ and $q=3$, featuring both ${\cal F}(2)$ and ${\cal F}(6)$. The latter
choice is eschewed since the added technical complexity does not bring any new qualitative features of the
monopole whose Hamiltonian
includes the $6-$form term $\mbox{Tr}{\cal F}_{ijkIJK}^2\to\mbox{Tr}\,S^2\{F_{[ij},D_{k]}\F\}^2$.},
is that the energy density functional in 6
dimensions that is bounded from below by the 3rd Chern-Pontryagin density, must feature both ${\cal F}_{ij}$ and
${\cal F}_{ijkl}$ curvature terms. This means that there will appear a dimensionful constant $\ka$, whose function is
to compensate the difference in the dimensions of these two curvature terms. In this respect, the resulting model is
akin to the Skyrme model.

The static Hamiltonian can be labelled by $(D,N)=(3,3)$, which is the dimensional descendant of the $p=1$ and the
$p=2$ members of the YM hierarchy on $\R^3\times S^3$, appearing in \re{3rd}.

The static Hamiltonian is readily calculated using \re{fijodd}-\re{fIJodd},
\bea
\label{H33}
{\cal H}^{(3)}_3&=&\mbox{Tr}\left({\cal F}_{ij}^2+2{\cal F}_{iI}^2+{\cal F}_{IJ}^2\right)-\frac{\ka^4}{12}\mbox{Tr}
\left(4{\cal F}_{ijkI}^2+6{\cal F}_{ijIJ}^2+4{\cal F}_{iIJK}^2\right)\nonumber\\
&=&\mbox{Tr}\left(2F_{ij}^2+\frac32 D_i\F^2-3S^2\right)\nonumber\\
&-&\frac{\ka^4}{4}\mbox{Tr}\left(\{F_{[ij},D_{k]}\F\}^2
-3\left(\{S,F_{ij}\}+[D_i\F,D_j\F]\right)^2-\frac92\{S,D_i\F\}^2\right)\,,
\eea
where the constant $\ka$ has the dimension of $length$. \re{H33}, which is positive/negative definite
(employing antihermtian fields) is bounded from below by the topological charge density
\be
\label{rho33}
\varrho^{(3)}_3=\ka^2\,\vep_{ijk}\mbox{Tr}\left(3F_{ij}\{S,D_k\F\}+2D_i\F D_j\F D_k\F\right)\,,
\ee
read from \re{43,3}.

\subsubsection{Spherical symmetry}
\label{3monopole3sph}
Subject to spherical symmetry \re{sphodd}, the density \re {H33}, viewed as a static Hamiltonian, reduces to the one
dimensional subsystem
\bea
\label{H33sph}
H^{(3)}_3&\simeq&\left([2w'^2+r^{-2}(1-w^2)^2]+\la_1\eta^2[r^2\,h'^2+2w^2h^2]+\la_2\eta^4r^2(1-h^2)^2\right)\nonumber\\
&+&\ka^4\bigg(\la_3\eta^2([(1-w^2)h]')^2+\la_4\eta^4\left(2([(1-h^2)w]')^2+r^{-2}[(1-h^2)(1-w^2)+2w^2h^2]^2\right)
\nonumber\\
&&\qquad+\la_5\eta^6(1-h^2)^2\left(r^2h'^2+2w^2h^2\right)\bigg)\,,
\eea
whose final normalisation will be fixed after the monopole charge of the hedgehog solution is fixed to $unity$. The
constants $(\la_1,\la_2,\la_3,\la_4,\la_5)$ must be positive for the topological lower bound to remain valid.

Subject to the same symmetry, the monopole charge density \re{rho33} bounding $H^{(3)}_3$ from below reduces to
\be
\label{rho33red}
\rho^{(3)}_3=\ka^2\,\eta^3\,\frac{d}{dr}\{(h-\frac13h^3)-(1-h^2)w^2\,h\}
\ee
which as expected, is a total derivative. It is clear that the only contribution to the
integral of \re{rho33red} comes from the first term, since the second term yields $nil$, according to the
boundary values \re{asymw}-\re{asymh} resulting from the finite energy conditions. This was of course known in advance
since the $monopole$ gauge fields are asymptotically Dirac-Yang fields.

The integral of \re{rho33red} is equal to $\frac23$, which means that both it and the static Hamiltonian \re{H33sph}
must each be multiplied by $\frac32$ for the monopole charge of the Hedgehog to be $unity$.

In this and all subsequent monopoles considered, the number of Bogomol'nyi equations if greater than the number of
functions parametrising the fields, so the systems are overdetermined. But in the case in hand the overdetermination
is even more pronounced. Here, the situation is more similar to the usual Skyrme\cite{Skyrme:1962vh} model which
likewise features a dimensionful constant, in addition to the Bogomol'nyi
equations numbering $two$, for $one$ Hedgehog function $f(r)$. One would expect therefore that the Bogomol'nyi
bound is, like in the Skyrme model, violated more severely than in the case of the corresponding first order equations
in the following, which pertain to systems descended from a single member of the Yang--Mills hierarchy, and which
consequently do not feature any additional dimensional constant other than $\eta$, the inverse raduius of the
codimensional compact space.

\subsection{Vortex on $\R^2$ with descended third Chern--Pontryagin charge}
Since the density whose surface integral yielding the winding number \re{CS32} pertaining to this case was given in
Section {\bf 5.2.4}, we state the generalised~\cite{Burzlaff:1994tf,Arthur:1998nh} Abelian Higgs model in this case.
This consists of the usual Abelian Higgs model, {\bf plus} the density descended from the $p=2$ YM density
displayed in \re{p=2vortex} below in Section {\bf 7.10}.

\subsection{Monopole on $\R^7$  with descended fourth Chern--Pontryagin charge}
\label{4monopole7}
 This is the monopole whose topological lower bound is given by the volume integral of the topological charge density
\re{4,7}, or the surface integral of \re{4,7}. The static Hamiltonian can be labelled by $(D,N)=(7,1)$, which is the
dimensional descendant of the $p=2$ member of the YM hierarchy on $\R^7\times S^1$, appearing in \re{3rd}.
The static Hamiltonian is readily calculated using \re{fijodd}-\re{fIJodd},
\bea
\label{H71}
{\cal H}^{(1)}_7&=&\mbox{Tr} \left({\cal F}_{ijkl}^2+4{\cal F}_{ijk8}^2\right)\nonumber\\
&=&\mbox{Tr}\left(F_{ijkl}^2
+4\,\{F_{[ij},D_{k]}\F\}^2 \right)\,,
\eea
which is bounded from below by the topological charge density
\be
\label{varrho71}
\varrho^{(1)}_7=8\,\vep_{ijklmnp}\mbox{Tr}\,F_{ij}\,F_{kl}\,F_{np}\,D_m\F\,.
\ee
The topological lower bound in the class of models on $\R^{4p-1}$ descended from the $p-$YM system on $\R^{4p-1}\times S^1$
is very special in that the Bogomol'nyi equations
\[
F_{ijkl}=\frac{1}{3!}\vep_{ijklmnp}\,\{F_{[mn},D_{p]}\F\}=\frac{1}{2}\vep_{ijklmnp}\,\{F_{mn},D_{p}\F\}\,.
\]
which saturate the topological lower bound are not
overdetermined. The general case on $\R^{4p-1}$ is given in \cite{Radu:2005rf}, and here we present the
monopole in this class on $\R^7$.

\subsubsection{Spherical symmetry}
\label{4monopole7sph}
Subject to spherical symmetry \re{sphodd}, the density \re {H71} reduces to the one dimensional subsystem
\footnote{In the case of arbitrary $p$, the corresponding expression for the static energy density of the monopole
on $\R^{4p-1}$ is~\cite{Radu:2005rf}
\bea
H^{(1)}_{4p-1}&=&\frac12(1-w^2)^{2(p-1)}\left[2p\,w'^2+(2p-1)
\frac{(1-w^2)^2}{r^2}\right]\nonumber\\
&&\,+\frac12\eta^2
\left[\frac{r^2}{2p-1}\left(\left[(1-w^2)^{p-1}\,h\right]'\right)^2
+2p(1-w^2)^{2(p-1)}\,w^2h^2\right]\,,\nonumber
\eea
which is bounded from below by the density
\[
\rho^{(1)}_7=\eta\,\frac{d}{dr}
\left[(1-w^2)^{2p-1}\,h\right]\,,
\]
the lower bound being saturated by the Bogomol'nyi equations
\bea
w'&\mp&\eta\,wh=0,\nonumber\\
\eta\,r\left[(1-w^2)^{p-1}\,h\right]'&\pm&
\frac{(2p-1)}{r}\,(1-w^2)^p=0\nonumber\,.
\eea
}
\bea
H^{(1)}_7&=&\frac12(1-w^2)^{2}\left[4\,w'^2+3\frac{(1-w^2)^2}{r^2}\right]\nonumber\\
&&+\frac12\eta^2
\left[\frac{r^2}{3}\left(\left[(1-w^2)\,h\right]'\right)^2
+4(1-w^2)^{2}\,w^2h^2\right]\,,\label{H17}
\eea
which is bounded from below by the density
\be
\label{rho71}
\rho^{(1)}_7=\eta\,\frac{d}{dr}
\left[(1-w^2)^{2p-1}\,h\right]\,,
\ee
which as expected is a total derivative. Both \re{H71} and \re{rho71} are normalised such that the hedgehog
has $unit$ monopole charge.

The Bogomol'nyi equations for this hedgehog configuration are
\bea
w'&\mp&\eta\,wh=0,\label{Bog17}\\
\eta\,r\left[(1-w^2)\,h\right]'&\pm&
\frac{3}{r}\,(1-w^2)^2=0\nonumber\,.
\eea
which are not overdetermined and can be solved numerically~\cite{Radu:2005rf}. These are obviously the direct
generalisations of the BPS monopoles on $\R^3$, so they will be referred to as the tower of $p-$BPS monopoles on
$\R^{4p-1}$, in these notes.

An analytic proof of existence to the BPS equations \re{Bog17} on $\R^7$ was given in \cite{Yisong1,Yisong2}. Clearly, that
proof can be readily adapted to the solutions of the BPS equations on $\R^{4p-1}$ of \cite{Radu:2005rf}.

\subsection{Monopole on $\R^6$  with descended fourth Chern--Pontryagin charge}
\label{4monopole6}
This is the monopole whose topological lower bound is given by the volume integral of the topological charge density
\re{CP46}, or the surface integral of \re{4,6}.

The static Hamiltonian can be labelled by $(D,N)=(6,2)$, which is the dimensional descendant of the $p=2$ member of the YM
hierarchy on $\R^6\times S^2$, appearing in \re{4th1}.

The static Hamiltonian is readily calculated from
\re{fijeven}-\re{fIJeven}, in the compact notation of \re{FSOD}, \re{covPHY} and \re{SPHY},
\bea
\label{H26}
{\cal H}^{(2)}_6&=&\mbox{Tr}\left({\cal F}_{ijkl}^2+4{\cal F}_{ijkI}^2+6{\cal F}_{ijIJ}^2\right)\nonumber\\
&\simeq&\mbox{Tr}\left(F_{ijkl}^2+4\la_1\{F_{[ij},D_{k]}\F\}^2
-3\la_2\left(\{S,F_{ij}\}+[D_i\F,D_j\F]\right)^2\right)\,,
\eea
the second line of which is expressed up to an overall numerical factor, since the final normalisation will be made
by requiring the Hedgehog to have $unit$ monopole charge. Also in the second line of \re{H44}, the fictitious dimensioless
and $positive$ constants $(\la_1,\,\la_2)$ are inserted since the Bogomol'nyi inequality remains valid
as long as these constants are all positive. The Bogomol'nyi inequalities in question can be saturated $only$ when each
of these constants is equal to $1$. \re{H44} is a positive definite Hamiltonian density, in which
negative signs appear under the trace since we have used an antihermitian connection.

The Hamiltonian density \re{H26} is bounded from below by the topological charge density
\bea
\label{rho26}
\varrho^{(2)}_6&=&\vep_{ijklmn}\,\mbox{Tr}\,\gamma_7\,
\left[S\,F_{ij}F_{kl}F_{mn}+2\,F_{ij}F_{kl}D_m\F D_n\F+F_{ij}D_m\F F_{kl}D_n\F\right]
\label{2,6}
\eea
read from \re{4,6}.

\subsubsection{Spherical symmetry}
\label{4monopole6sph}
Subject to spherical symmetry, \re{spheven}, the density \re{H26} reduces to the one
dimensional subsystem
\bea
H^{(2)}_6&\simeq&r^{-1}\left(2\,[(1-w^2)\,w']^2+r^{-2}(1-w^2)^4\right)\nonumber\\
&+&\frac29\,\la_1\,\eta^2\,r\left[([(1-w^2)h]')^2+3^2\,r^{-2}(1-w^2)^2w^2h^2\right]\nonumber\\
&+&\frac{1}{36}\,\la\,_2\eta^4r^{3}\left[([(1-h^2)w]')^2+2r^{-2}[(1-w^2)(1-h^2)+2w^2h^2]^2\right]\label{H26sph}
\eea
whose overall normalisation will be chosen such, that the monopole charge of the hedgehog solution is fixed to $unity$,
and for convenience each of the $\la$'s is rescaled.

We now rewrite \re{H26sph} with a given choice of the constants $\la_1=\la_2=1$,
\bea
H^{(2)}_6&\simeq&\left(2r^{-1}\{(1-w^2)w'-\frac16\eta^2r[(1-h^2)(1-w^2)+2w^2h^2]\}^2+
\frac13\eta^2[(1-h^2)(1-w^2)+2w^2h^2](1-w^2)w'
\right)\nonumber\\
&&\qquad+\left(\frac29\eta^2r\{[(1-w^2)h]'+3r^{-1}(1-w^2)wh\}^2-\frac43\eta^2(1-w^2)wh[(1-w^2)h]'\right)\nonumber\\
&&\qquad+\frac16\left(r^3\{\eta^2[(1-h^2)w]'-r^{-3}(1-w^2)^2\}^2+\frac13\eta^2(1-w^2)^2[(1-h^2)w]'\right)\,,
\label{H26quad}
\eea
such that the Bogomol'nyi lower bound is exposed.
This density is bounded from below by
\be
\label{rho6red}
\rho^{(2)}_6=\eta^2\,\frac{d}{dr}\left\{(w-\frac23w^3+\frac15w^5)-\left[(1-w^2)\,w\,h^2\right]\right\}
\ee
which is a total derivative descending from \re{4,6}, or \re{CS46}. It is clear that the only contribution to the
integral of \re{rho6red} comes from the first term only, since the second term yields $nil$, according to the
boundary values \re{asymw}-\re{asymh} resulting from the finite energy conditions. This was of course known in advance
since the $monopole$ gauge fields are asymptotically Dirac-Yang fields.

The integral of \re{rho6red} is equal to $\frac{8}{15}$, which means that both it and the static Hamiltonian \re{H26sph}
must each be multiplied by $\frac{15}{8}$ for the monopole charge of the Hedgehog to be $unity$.

Finally we state the Bogomol'nyi equations following from \re{H26quad}
\bea
(1-w^2)w'&=&\pm\frac16\eta^2r[(1-h^2)(1-w^2)+2w^2h^2]\nonumber\\
r[(1-w^2)h]'&=&{\mp}3(1-w^2)wh\label{Bog26} \\
\eta^{2}[(1-h^2)w]'&=&{\pm}\,r^{-3}(1-w^2)^2\nonumber
\eea
which are overdetermined~\cite{Tchrakian:1990gc}.

\subsection{Monopole on $\R^5$ with descended fourth Chern--Pontryagin charge}
\label{4monopole5}
This is the monopole whose topological lower bound is given by the volume integral of the topological charge density
\re{CP45}, or the surface integral of \re{4,5}.

The static Hamiltonian can be labelled by $(D,N)=(5,3)$, which is the dimensional descendant of the $p=2$ member of the YM
hierarchy on $\R^5\times S^3$, appearing in \re{4th1}.

The static Hamiltonian is readily calculated from
\re{fijodd}-\re{fIJodd},
\bea
\label{H35}
{\cal H}^{(3)}_5&=&\mbox{Tr}\left({\cal F}_{ijkl}^2+4{\cal F}_{ijkI}^2+6{\cal F}_{ijIJ}^2+4{\cal F}_{iIJK}^2
\right)\nonumber\\
&\simeq&\mbox{Tr}\left(F_{ijkl}^2+4\la_1\{F_{[ij},D_{k]}\F\}^2
-18\la_2\left(\{S,F_{ij}\}+[D_i\F,D_j\F]\right)^2-54\la_3\{S,D_i\F\}^2\right)\,,
\eea
the second line of which is expressed up to an overall numerical factor, since the final normalisation will be made
by requiring the Hedgehog to have $unit$ monopole charge. Also in the second line of \re{H44}, the fictitious
dimensionless
and $positive$ constants $(\la_1,\,\la_2,\,\la_3)$ are inserted since the Bogomol'nyi inequality remains valid
as long as these constants are all positive. The Bogomol'nyi inequalities in question can be saturated $only$ when each
of these constants is equal to $1$. \re{H35} is a positive definite Hamiltonian density, in which
negative signs appear under the trace since we have used an antihermitian connection.

The Hamiltonian density \re{H44} is bounded from below by the topological charge density
\bea
\label{varrho35}
\varrho^{(3)}_5&=&\vep_{ijklm}\,\mbox{Tr}\left[\left(S\,F_{ij}F_{kl}+F_{ij}\,S\,F_{kl}+F_{ij}F_{kl}\,S\right)D_m\F
+2F_{ij}D_k\F D_l\F D_m\F\right]\,,
\eea
read from \re{4,5}.

\subsubsection{Spherical symmetry}
\label{4monopole5sph}
Subject to spherical symmetry, \re{sphodd}, the density \re{H35} reduces to the one
dimensional subsystem
\bea
H^{(3)}_5&\simeq&r^{-2}(1-w^2)^2\left[4\,w'^2+r^{-2}(1-w^2)^2\right]\nonumber\\
&+&\la_1\eta^2\left[([(1-w^2)h]')^2+6r^{-2}(1-w^2)^2w^2h^2\right]\nonumber\\
&+&\frac13\la_2\eta^4r^{2}\left[2\,([(1-h^2)w]')^2+3\,r^{-2}[(1-w^2)(1-h^2)+2w^2h^2]^2\right]\label{H35sph}\\
&+&\frac14\la_3\eta^6r^4\,(1-h^2)^2\left[h'^2+4\,r^{-2}w^2h^2\right]\nonumber
\eea
whose normalisation is chosen such, that the monopole charge of the hedgehog solution is fixed to $unity$,
and for convenience each of the $\la$'s is rescaled.

We now rewrite \re{H35sph}, with $\la_1=\la_2=\la_3=1$ in the following way
\bea
\label{H35quad}
H^{(3)}_5&\simeq&\bigg(4\{r^{-1}(1-w^2)w'+\frac12\,\eta^3\,r(1-h^2)\,wh\}^2-4\,\eta^3(1-h^2)(1-w^2)\,h\,w\,w'\bigg)
\nonumber\\
&+&\bigg(\{\eta[(1-w^2)h]'-\eta^2[(1-h^2)(1-w^2)+2w^2h^2]\}^2+2\eta^3[(1-h^2)(1-w^2)+2w^2h^2]\,[(1-w^2)h]'\bigg)
\nonumber\\
&+&\bigg(\frac23\{\eta^2r[(1-h^2)w]'+3\eta r^{-1}(1-w^2)wh\}^2-4\eta^3(1-w^2)wh\,[(1-h^2)w]'\bigg)\nonumber\\
&+&\bigg(\frac14\{\eta^3r^2(1-h^2)h'-2r^{-2}(1-w^2)^2\}^2+\eta^3(1-h^2)(1-w^2)^2h'\bigg)\,,
\eea
such that the Bogomol'nyi bound is exposed.
This density is bounded from below by
\be
\label{rho5red}
\rho^{(3)}_5=\eta^3\,\frac{d}{dr}\left\{(3h-h^3)-\left[3(2-w^2)-(6-5w^2)h^2\right]\,w^2\,h\right\}
\ee
which is a total derivative descending from \re{4,5}, or \re{CS45}. It is clear that the only contribution to the
integral of \re{rho5red} comes from the first term only, since the second term yields $nil$, according to the
boundary values \re{asymw}-\re{asymh} resulting from the finite energy conditions. This was of course known in advance
since the $monopole$ gauge fields are asymptotically Dirac-Yang fields.

The integral of \re{rho5red} is equal to $2$, which means that both it and the static Hamiltonian \re{H26sph}
must each be multiplied by $\frac12$ for the monopole charge of the Hedgehog to be $unity$.

Finally we state the Bogomol'nyi equations following from \re{H35quad}
\bea
(1-w^2)\,w' &=&\mp\,\frac12\eta^3\,r^2\,(1-h^2)\,wh\nonumber\\
{[(1-w^2)\,h]}'&=&\pm\,\eta\,[(1-h^2)(1-w^2)+w^2\,h^2]\nonumber\\
\eta\,r^{2}\,{[(1-h^2)\,w]}'&=&\mp\,3\,(1-w^2)\,wh\nonumber\\
\eta^{3}\,r^{4}(1-h^2)\,h' &=&\pm\,2\,(1-w^2)^2\label{Bog35}
\eea
which are overdetermined~\cite{Tchrakian:1990gc}.

\subsection{Monopole on $\R^4$  with descended fourth Chern--Pontryagin charge}
\label{4monopole4}
This is the monopole whose topological lower bound is given by the volume integral of the topological charge density
\re{CP44}, or the surface integral of \re{4,4}.

The static Hamiltonian can be labelled by $(D,N)=(4,4)$, which is the dimensional descendant of the $p=2$ member of the YM
hierarchy on $\R^4\times S^4$, appearing in \re{4th1}.

The static Hamiltonian is readily calculated from
\re{fijeven}-\re{fIJeven}, in the compact notation of \re{FSOD}, \re{covPHY} and \re{SPHY},
\bea
\label{H44}
{\cal H}^{(4)}_4&=&\mbox{Tr}\left({\cal F}_{ijkl}^2+4{\cal F}_{ijkI}^2+6{\cal F}_{ijIJ}^2+4{\cal F}_{iIJK}^2
+{\cal F}_{IJKL}^2\right)\nonumber\\
&\simeq&\mbox{Tr}\bigg(F_{ijkl}^2+4\la_1\{F_{[ij},D_{k]}\F\}^2
-18\la_2\left(\{S,F_{ij}\}+[D_i\F,D_j\F]\right)^2\nonumber\\
&&\qquad\qquad\qquad\qquad\qquad\qquad\qquad\qquad-54\la_3\{S,D_i\F\}^2+54\la_4S^4\bigg)\,,
\eea
the second line of which is expressed up to an overall numerical factor, since the final normalisation will be made
by requiring the Hedgehog to have $unit$ monopole charge. Also in the second line of \re{H44}, the fictitious
dimensionless and $positive$ constants $(\la_1,\,\la_2,\,\la_3,\,\la_4)$ are inserted since the Bogomol'nyi inequality
remains valid as long as these constants are all positive. The Bogomol'nyi inequalities in question can be saturated
$only$ when each of these constants is equal to $1$. \re{H44} is a positive definite Hamiltonian density, in which
negative signs appear under the Trace since we have used an antihermitian connection.

The Hamiltonian density \re{H44} is bounded from below by the topological charge density
\bea
\label{rho44}
\varrho^{(4)}_4&=&\vep_{ijkl}\,\mbox{Tr}\,\gamma_5\,\bigg[
2\,S^2\,F_{ij}F_{kl}+F_{ij}\,S\,F_{kl}\,S\nonumber\\
&&\qquad\qquad\qquad+4\left(D_i\F\,D_j\F\,S\,F_{kl}+D_i\F\,D_j\F\,F_{kl}\,S+D_i\F\,F_{kl}\,D_j\F\,S\right)\nonumber\\
&&\qquad\qquad\qquad+2\,D_i\F\,D_j\F\,D_k\F\,D_l\F\bigg]\,,
\eea
read from \re{4,4}.

\subsubsection{Spherical symmetry}
\label{4monopole4sph}
Subject to spherical symmetry, \re{spheven}, the density \re{H44} reduces to the one
dimensional subsystem
\bea
\label{H44sph}
H^{(4)}_4&\simeq&r^{-3}[(1-w^2)\,w']^2\nonumber\\
&+&\frac13\la_1\eta^2r^{-1}\left[([(1-w^2)h]')^2+3r^{-2}(1-w^2)^2w^2h^2\right]\nonumber\\
&+&\frac12\la_2\eta^4r^{-2}\left[([(1-h^2)w]')^2+r^{-2}[(1-w^2)(1-h^2)+2w^2h^2]^2\right]\nonumber\\
&+&\la_3\eta^6r^3\left[[(1-h^2)h']^2+3r^{-2}(1-h^2)^2w^2h^2\right]\nonumber\\
&+&\la_4\eta^8r^3(1-h^2)^4\,.
\eea
whose normalisation is chosen such, that the monopole charge of the hedgehog solution is fixed to $unity$,
and for convenience each of the $\la$'s is rescaled.

We now rewrite \re{H44sph}, with $\la_1=\la_2=\la_3=\la_4=1$ in the following way
\bea
\label{H44quad}
H^{(4)}_4&\simeq&\bigg(r^{-3}\{(1-w^2)w'-\eta^4r^3(1-h^2)^2\}^2+4\eta^4(1-h^2)^2(1-w^2)w'\bigg)\nonumber\\
&+&\frac{1}{3}\bigg(\eta^2r^{-1}\{[(1-w^2)h]'+3\eta^2r(1-h^2)wh\}^2-6\eta^4(1-h^2)wh[(1-w^2)h]'\bigg)\nonumber\\
&+&\frac12\bigg(\eta^4r^{-2}\{[(1-h^2)w]'-r^{-1}[(1-h^2)(1-w^2)+2w^2h^2]\}^2+\nonumber\\
&&\qquad\qquad\qquad\qquad\qquad\qquad+2\eta^4[(1-h^2)(1-w^2)+2w^2h^2][(1-h^2)w]'\bigg)\nonumber\\
&+&\bigg(r^3\{\eta^2(1-h^2)h'+\eta r^{-3}(1-w^2)wh\}^2-2\eta^4(1-h^2)(1-w^2)whh'\bigg)\,,
\eea
such that the Bogomol'nyi bound is exposed. This density is bounded from below by
\be
\label{rho4red}
\rho^{(4)}_4=\frac{d}{dr}\left\{(3w-w^3)-\left[(2-h^2)(3-w^2)-4(1-h^2)w^2\right]\,h^2\,w\right\}
\ee
which is a total derivative descending from \re{4,4}, or \re{CS44}. It is clear that the only contribution to the
integral of \re{rho4red} comes from the first term only, since the second term yields $nil$, according to the
boundary values \re{asymw}-\re{asymh} resulting from the finite energy conditions. This was of course known in advance
since the $monopole$ gauge fields are asymptotically Dirac-Yang fields.

The integral of \re{rho4red} is equal to $2$, which means that both it and the static Hamiltonian \re{H44sph}
must each be multiplied by $\frac12$ for the monopole charge of the Hedgehog to be $unity$.

Finally we state the Bogomol'nyi equations following from \re{H44quad}
\bea
(1-w^2)\,w' &=&\pm\frac12\,\eta^4\,r^3\,(1-h^2)^2\nonumber\\
{[(1-w^2)\,h]}'&=&\mp3\,\eta^2\,r\,(1-h^2)\,w\,h\nonumber\\
{[(1-h^2)\,w]}'&=&\pm\, r^{-1}[(1-h^2)(1-w^2)+w^2\,h^2]\nonumber\\
(1-h^2)\,h' &=&\mp\eta^{-2}\,r^{-3}\,(1-w^2)\,w\,h\label{Bog44}
\eea
which are overdetermined~\cite{Tchrakian:1990gc}.

\subsection{Monopole on $\R^3$  with descended fourth Chern--Pontryagin charge}
\label{4monopole3}
This is the monopole whose topological lower bound is given by the volume integral of the topological charge density
\re{CP43}, or the surface integral of \re{4,3}.

The static Hamiltonian can be labelled by $(D,N)=(3,5)$, which is the dimensional descendant of the $p=2$ member of the YM
hierarchy on $\R^5\times S^3$, appearing in \re{4th1}.

The static Hamiltonian is readily calculated using again \re{fijodd}-\re{fIJodd},
\bea
\label{H53}
{\cal H}^{(5)}_3&=&\left(4{\cal F}_{ijkI}^2+6{\cal F}_{ijIJ}^2+4{\cal F}_{iIJK}^2
+{\cal F}_{IJKL}^2\right)\nonumber\\
&\simeq&\mbox{Tr}\left(\{F_{[ij},D_{k]}\F\}^2
-6\la_1\left(\{S,F_{ij}\}+[D_i\F,D_j\F]\right)^2-27\la_1\{S,D_i\F\}^2+54\la_3\,S^4\right)\,,
\eea
again inserting dimensionless non-negative coefficients $(\la_1,\la_2,\la_3)$. \re{H53} is bounded from below by the
topological charge density
\be
\label{rho53}
\varrho^{(5)}_3=\vep_{ijk}\mbox{Tr}\left(F_{ij}\,S^2\,D_k\F
+F_{ij}\,D_k\F\,S^2+F_{ij}\,S\,D_k\F\,S+2S\,D_i\F D_j\F D_k\F\right)\,,
\ee
read from \re{4,3}.

\subsubsection{Spherical symmetry}
\label{3monopole3sph-n}
Subject to spherical symmetry \re{sphodd}, the static Hamiltonian density \re{H53} reduces to the one
dimensional subsystem
\bea
\label{H53sph}
H^{(5)}_3&\simeq&\quad\ \eta^2\,r^{-2}\,\left([(1-w^2)h]'\right)^2\nonumber\\
&+&\la_1\,\eta^4\left[2\left([(1-h^2)w]'\right)^2+r^2\left[(1-w^2)(1-h^2)+w^2h^2\right]^2\right]\nonumber\\
&+&\frac89\la_2\,\eta^6\,r^{2}\left[(1-h^2)^2\,h'^2+2r^{-2}(1-h^2)^2w^2h^2\right]\nonumber\\
&+&\frac19\la_3\,\eta^8\,r^{2}(1-h^2)^4\,.
\eea
whose final normalisation will be fixed after the monopole charge of the hedgehog solution is fixed to $unity$,
and for convenience each of the $\la$'s is rescaled.

Rewriting \re{H53sph} with $\la_1=\la_2=\la_3=1$
\bea
\label{H53quad}
H^{(5)}_3&\simeq&\bigg(r^{-2}\{\eta\,[(1-w^2)h]'-\frac13\eta^4\,r^2\,(1-h^2)^2\}^2
+\frac23\,\eta^5\,(1-h^2)^2\,[(1-w^2)h]'\bigg)\nonumber\\
&+&2\bigg(\{\eta^2[(1-h^2)w]'+\frac23\eta^3(1-h^2)\,wh\}^2-\frac43\,\eta^5(1-h^2)\,wh\,[(1-h^2)w]'\bigg)\nonumber\\
&+&2\bigg(\{\frac23\eta^3(1-h^2)\,h'-\eta^2\,r^{-2}[(1-h^2)(1-w^2)+2\,w^2h^2]\}^2+\nonumber\\
&&\qquad\qquad\qquad\qquad\qquad+\frac43\eta^5[(1-h^2)(1-w^2)+2\,w^2h^2]\,(1-h^2)\,h'\bigg)
\eea
the Bogomol'nyi bound can be conveniently exposed.

Subject to the same symmetry, the monopole charge density \re{rho53} bounding $H^{(5)}_3$ from below reduces to
\bea
\label{rho3red}
\rho^{(5)}_3=\eta^5\,\frac{d}{dr}\{(h-\frac23h^3+\frac15h^5)-(1-h^2)^2\,w^2h\}
\eea
which is a total derivative. It is clear that the only contribution to the
integral of \re{rho3red} comes from the first term, since the second term yields $nil$, according to the
boundary values \re{asymw}-\re{asymh} resulting from the finite energy conditions. This was of course known in advance
since the $monopole$ gauge fields are asymptotically Dirac-Yang fields.

The integral of \re{rho3red} is equal to $\frac{8}{15}$, which means that both it and the static Hamiltonian \re{H53sph}
must each be multiplied by $\frac{15}{8}$ for the monopole charge of the Hedgehog to be $unity$.

Finally we state the Bogomol'nyi equations following from \re{H53quad}
\bea
[(1-w^2)h]'&=&\pm\frac13\eta^3r^2(1-h^2)^2\nonumber\\
{[(1-h^2)w]'}&=&{\mp}\frac23\eta^2(1-h^2)\,wh\label{Bog53} \\
\eta\,r^2(1-h^2)h'&=&{\pm}\frac32[(1-h^2)(1-w^2)+2w^2h^2]\nonumber
\eea
which are overdetermined~\cite{Tchrakian:1990gc}.

\subsection{Vortex on $\R^2$ with descended fourth Chern--Pontryagin charge}
The density whose surface integral yields the winding number \re{CS42} pertaining to this case was given in
Section {\bf 5.3.6}, so we state the generalised~\cite{Burzlaff:1994tf,Arthur:1998nh} Abelian Higgs model in this case.
This consists {\bf only} of the density descended from the $p=2$ YM
density~\footnote{The corresponding density for arbitrary $p$ can be expressed as
\[
{\cal H}_2^{(4p-1)}=(\eta^2-|\vf|^2)^{2(p-2)}\bigg([(\eta^2-|\vf|^2)F_{ij}+iD_{[i}\vf\,D_{j]}\vf^*]^2+4p(2p-1)(\eta^2
-|\vf|^2)^2|D_i\vf|^2+2(2p-1)(\eta^2-|\vf|^2)^4\bigg)\,.
\]} given by
\be
\label{p=2vortex}
{\cal H}_2^{(6)}=[(\eta^2-|\vf|^2)F_{ij}+iD_{[i}\vf\,D_{j]}\vf^*]^2+24(\eta^2-|\vf|^2)^2|D_i\vf|^2+6(\eta^2-|\vf|^2)^4\,.
\ee

\subsection{Bogomol'nyi bounds and bound states}
A noteworthy feature of monopoles on $\R^D$ discussed above is that with the exception of those pertaining to
Yang-Mills--Higgs models on $\R^{4p-1}$ descended from the $p-$th member of the Yang-Mills heirarchy, their energies
do not saturate the topological lower bound. The first member, that with $p=1$, of this exceptional class is the
't~Hooft--Polyakov~\cite{'tHooft:1974qc,Polyakov:1974ek} monopole itself. (In addition to these monopoles, all the
vortices\cite{Abrikosov:1956sx,Nielsen:1973cs,Burzlaff:1994tf,Arthur:1998nh} on $\R^2$ in models which are
descended from a $single$ member of the Yang-Mills heirarchy, do saturate the Bogomol'nyi lower bound.)

The first order Bogomol'nyi equations of the monopoles that do not saturate the topological lower bound, in common with
those of the skyrmion~\cite{Skyrme:1962vh}, are overdetermined~\cite{Tchrakian:1990gc}. There is however a marked
quanititative difference between the excess of the energy of the skyrmion above the topological lower bound, and the
corresponding excess in the case of monopoles in Yang-Mills--Higgs models descended from {\it one single} member of the
Yang-Mills heirarchy. In the latter case this excess is several orders of magnitude smaller than the excess in the former,
as pointed out in \cite{Kleihaus:1998kd}. In this sense, the higher dimensional monopoles are quantitatively closer to
the energy bound saturating vortices~\cite{Abrikosov:1956sx,Nielsen:1973cs,Burzlaff:1994tf,Arthur:1998nh} than the
skyrmion~\cite{Skyrme:1962vh}. Also, the Yang-Mills--Higgs models supporting higher dimensional monopoles $always$ feature
at least one dimensionful constant that cannot be scaled away, $like$ the Abelian Higgs models, and $unlike$ the usual
Skyrme model (without pion mass potential and sextic kinetic terms). In the latter case the (non-selfdual) skyrmions
have a nonzero interaction energy independently of the value of a dimensionful parameter, while in the case of Abelian
Higgs vortices it is well is known from the work of \cite{Jacobs:1978ch}
that the interaction energy depends on the value of a parameter, which
for a critical value results in BPS noninteracting configurations. This feature is repeated also for all generalised
vortices~\cite{Arthur:1998nh}.

The corresponding situation for higher dimensional monopoles is very close to that of vortices, whose YMH models
always feature at least one dimensionful coupling constant that cannot be scaled away. While there is (are) no value(s)
of this constant(s) for which the topological energy bound is saturated, in certain "almost selfdual" configurations
this bound is approached~\cite{Kleihaus:1998kd} quantitatively very closely. It is therefore not unreasonable to expect
that for various values of this parameter the interaction energy of the monopoles may exhibit bound states. It was
verified in \cite{Kleihaus:1998gy}, in the physically most relevant case on $\R^3$ describing the monopole presented
in Section {\bf 7.9} above, that some configurations do feature positive binding energy. It is not surprising also that
monopoles of the model presented in Section {\bf 7.3} should also feature positive binding energy, even though the
lowest energy in that case is not quantitavely close to the Bogomol'nyi lower bound.

\subsection{Gauge decoupling limits: Global monopoles}
The two best known solitons of gauged Higgs models are the vortices of the Abelian Higgs
model~\cite{Abrikosov:1956sx,Nielsen:1973cs} on $\R^2$
and the monopoles of the 't~Hooft--Polyakov~\cite{'tHooft:1974qc,Polyakov:1974ek} model on $\R^3$.
In both these cases the energy is divergent in the gauge decoupling limit as in that limit the Derrick scaling
requirement is not satisfied. As seen in Sections {\bf 7.} and {\bf 7.1} respectively, these models descend from the
usual $p=1$ Yang-Mills system on $\R^2\times S^2$ and $\R^3\times S^1$.
This situation changes drastically when considering the gauged Higgs models descending from $p\ge 2$ Yang-Mills systems,
simply because there these models feature high enough nonlinear terms in the covariant derivative of the Higgs field that
after gauge decoupling the model still contains the requisite terms to satisfy the Derrick scaling requirement
yielding finite energy solutions.

The gauge decoupled versions of all the non-Abelian and Abelian models described in this Section are immediately
found by suppressing the gauge connection and curvature. In the spherically or radially symmetric cases, this is
achieved by setting the $w(r)=1$. In particular, the topological charges in that case are the winding numbers resulting
from the substitution  $w(r)=1$ in the monopole, and, $w(r)=n$ in the vortex charge densities.
The resulting symmetry breaking models, featuring a scalar isovector field and no gauge fields, support
solitons. In the $2-$dimensional case such models were first employed in \cite{MullerKirsten:1990qw}, and in the
$3-$dimensional case in \cite{Tchrakian:1990qx}.
These were variously described as Goldstone models~\cite{Paturyan:2005ik,Radu:2007zz}.
In the three dimensions both spherically and axially symmetric solitons, as well as soliton--antisolitons, were
constructed in \cite{Paturyan:2005ik}. The solitons in arbitrary dimensions~\cite{Radu:2007zz} can also be described.
In the Abelian analogue on $\R^2$, such vortices were studied in detail in \cite{Arthur:1995ws}.

The $D=3$ Goldstone model is apparently similar to the Skyrme model~\cite{Skyrme:1962vh} but
unlike the latter, none of its higher topological charge solitons have a positive binding energy~\cite{Paturyan:2005ik},
unable to describe bound states of nucleons. The Goldstone model is not an alternative for the Skyrme model.

Solitons of these Goldstone models can be considered to be global
monopoles~\cite{Barriola:1989hx,Harari:1990cz,Maison:1999pi},
in the non-Abelian case, and global strings~\cite{Cohen:1988sg,Gregory:1988xc}
in the Abelian case. The gravitating versions of
such solitons may find application as topological defects in the context of phase transitions in the early Universe.
The difference
from those employed previously~\cite{Cohen:1988sg,Gregory:1988xc,Barriola:1989hx,Harari:1990cz,Maison:1999pi}
is that the ones proposed here are both topologically stable and have finite energy.

\section{Dyon and pseudo-dyon solutions on $\R^D$: $D\ge 3$}
\label{dyon}
The solutions described in this Section are static finite energy solutions of the Euler--Lagrange equations in $D+1$
dimensional Minkowski space, in the static limit. They are close analogues of the
Julia--Zee dyon~\cite{Julia:1975ff} in the sense
that the 'magnetic' components of the gauge field describe monopoles, namely that they
are asymptotically Dirac-Yang fields. In
this respect they are completetly different from the {\it dyonic instantons}~\cite{Lambert:1999ua} in $4+1$
dimensions. The
'magnetic' components of the gauge field of the latter describe instantons~\footnote{One might
think that in higher dimensions it
may be possible to exploit the $4p$ dimensional selfdual instantons described in Section {\bf 2} but no such
solution is identified to date.}, which by contrast are asymptotically pure gauge.

The Lagrangians from which the Euler--Lagrange equations are derived, are those
pertaining to the (static) Hamiltonians presented in Section {\bf 7}, which support topologically stable
monopoles on $\R^D$. The (static) solutions described here support both the magnetic components
$F_{ij}$ and the electric components $F_{i0}$ of the Yang--Mills curvature $F_{\mu\nu}=(F_{ij},F_{i0})$.
In this respect these solutions are similar to the Julia--Zee (JZ) dyon~\cite{Julia:1975ff},
the electric component $A_0$ of the Yang--Mills connection being introduced as a partner
of the Higgs field $\F$. This is the crux of the construction of the solutions presented here. However, there are some
clear departures between the electric-YM field carrying (static) solutions on $\R^D$ for $D\ge 4$, and the JZ dyon
on $\R^3$. Before proceeding to describe $pseudo-$dyons in higher dimensions and the $excited-$dyons in three (space)
dimensions, it is in order to comment on the JZ dyon itself since the analytic and stability features of this are not
at all on the same firm footing as that of the 't~Hooft-Polyakov monopole and the higher dimensional
monopoles described above in Section {\bf 7}.

Concerning existence, the Euler-Lagrange equations of the Julia-Zee dyon arise from the variation of a Lagrangian given
on a Minkowskian space, $i.e.$, an action density that is not positive definite. Thus the proofs of existence of monopoles
do not carry over. The existence proof~\cite{Schechter:1980cz}
of the Julia-Zee dyon is a much more involved problem. To date, the main description of the JZ dyon is through
numerical construction, except in the BPS limit.

Concerning topological stability, in the case of monopoles this is guaranteed by the (generalised) Bogomol'nyi lower
bounds demonstrated in Sections {\bf 4} and {\bf 7} above, while this is not the case with the Julia-Zee
dyon. This criterion is not applicable to the dyon, again since the Lagrangian density is not positive definite.

However, the situation is quite different, and definitive, in the case of
the Julia-Zee dyon in the BPS limit. The JZ dyon presented in Ref. \cite{Julia:1975ff} pertains to
the full Georgi--Glashow model, with nonvanishing
Higgs self-interaction potential. (For an in depth numerical analysis of this system see
\cite{Brihaye:1998vr}.) However it was soon realised that in the absence of the Higgs potential the dyon solutions
satisfy the first order Bogomol'nyi equations~\cite{Prasad:1975kr}. In this limit, the monopole
satisfies first order Bogomol'nyi equations saturating the topological lower bound. More importantly in this limit,
there is a complete symmetry between the Higgs field $\F$ and the (non-Abelian) electric potential $A_0$ in the
Lagrangian, and hence also in the field equations. As a result, the electric potential obeys the very same first order
Bogomol'nyi equations as the Higgs field does. Hence the existence of the JZ dyon in the BPS limit follows directly from
the existence of the BPS monopole. Like the latter, this dyon is topologically stable as the
topological lower bound is saturatd.

The new dyon-like configurations described in the present section, in addition to sharing some properties with
the JZ dyon (in and out of BPS limit), also differ from the latter. A qualitative discussion of these features is listed
here, before presenting the specific examples. The new dyon-like configurations fall in two main categories,
\begin{itemize}
\item
$excited-$dyons in $3+1$ Minkowskian dimensions,
\item
$pseudo-$dyons in $D+1$ Minkowskian dimensions, $D\ge 4$.
\end{itemize}
Qualitative properties of these two types of dyon-like configurations are listed here.
\begin{itemize}
\item
$excited-$dyons in $3+1$ dimensions:

These partner the monopoles descended from higher dimensional Yang-Mills, down to $\R^3$. Examples of these are the
monopole in section {\bf 7.2} descended from 6 dimensional YM, and the monopole in section {\bf 7.7} descended from 8
dimensional YM. They have the following qualitative features:

\begin{itemize}
\item
The gauge group of the residual (excited) monopole on $\R^3$ being $SO(3)$, the Higgs field $\F$ is an iso-triplet and
hence is also its partner $A_0$. Thus, both $A_0$ and the magnetic $A_i$ are both iso-triplets. This feature is in
common with the Julia--Zee dyon~\cite{Julia:1975ff}.
\item
In common with the Julia--Zee dyon, these excited dyons are ascribed an electric flux. This is the
surface integral of the electric
component, $E_i={\cal F}_{i0}$, of the 't~Hooft electromagnetic tensor ${\cal F}_{\mu\nu}=({\cal F}_{ij},{\cal F}_{i0})$,
\bea
Q&=&\frac{1}{4\pi}\int{\bf E}\cdot d{\bf S}\nonumber\\
&=&\frac{1}{4\pi}\int\mbox{Tr}\,(\F\,F_{i0})\,dS_i\,.\label{q}
\eea
\item
In contrast with the Julia--Zee dyon, these dyons do {\bf not} have a BPS limit. This is because their partner monopoles
{\it do not saturate their respective Bogomol'nyi bounds}.

In the first case, namely the monopole in section {\bf 7.1}, the presence of the dimensionful constant $\ka$ obstructs the
possibility of finding $self$-$dual$ solutions exactly in exactly the same way as it happens in Skyrme theory.

In the second case, namely the monopole in section {\bf 7.7}, the $self$-$duality$ equations are $overdetermined$ and
again there are no $self-dual$ solutions saturating the topological lower bound.

The nonexistence of $self$-$dual$ solutions is tantamount to the absence of complete symmetry between the functions
parametrising the Higgs field $\F$ and the functions parametrising the electric YM connection $A_0$. This will be seen
below when the examples in question are presented concretely, and is always the case when the dimensional descent
resulting in the monopole is over a codimension greater than one, which happens when a Higgs potential is present.

The result is that the excited dyons in both these (categories of) cases are solutions of second-order Euler-Lagrange
equations, and not first order self-duality equations symmetric between electric and magnetic functions. Consequently
both the existence and stability of such dyons is open to question.

\end{itemize}
\item
$pseudo-$dyons in $D+1$ dimensions:

These partner the monopoles descended from higher dimensional Yang-Mills, down to $\R^D$, $D\ge 4$. Examples of these
are the monopoles in sections {\bf 7.3}-{\bf 7.6} descended from 8 dimensions. The salient feature in which they
differ from the dyons and excited dyons on $\R^3$ is that here there is no simple or natural definition of a
(scalar) electric flux. In the context of models describing monopoles, the definition \re{q} cannot be
extended~\footnote{Indeed, the definition
of the 't~Hooft electromagnetic tensor ${\cal F}_{\mu\nu}$ can be extended to higher dimensions~\cite{Tchrakian:1999me}.
Just as ${\cal F}_{\mu\nu}$ is defined as the Abelian
density described by the dimensionally reduced Chern--Simons density of the 't~Hooft--Polyakov monopole, so can its
higher order generalisations be defined as the Chern--Simons densities described by the higher dimensional monopoles,
presented in Section \ref{CPdescent} above. A quick inspection of these leads to the conclusion that the resulting
Abelian curvature ${\cal F}_{\mu_1\mu_2\dots\mu_{D-1}}$ on $\R^D$
is an antisymmetric $D-1$ form. Hence its electric component ${\cal F}_{i_1i_2\dots i_{D-2},0}$ is an antisymmetric
$D-2$ form. On a space with spherical horizon, namely the flat space restricted to here, there is no natural definition
of a flux for this electric tensor.} to $\R^D$ for $D\ge 4$. This is because the magnetic field
$F_{ij}$ and the Higgs covariant derivative $D_i\F$ both decay like $r^{-2}$, and hence so does the electric field
$F_{i0}=D_iA_0$ which likewise decays as $D_i\F$ does. Hence when $D\ge 4$, the integral \re{q} diverges.

Another special feature of pseudo-dyons (on $\R^d,\ \ D\ge 4$) contrasting with dyons and and excited dyons (on $\R^3$)
is the fact that the electric and magnetic components for the YM connection have different multiplet structure.
Both $\F$ and $A_0$ take their values in the orthogonal complement $L_{i,D+1}$ of $L_{ij}$
($i=1,2,\dots,D$) of the $SO(D+1)$ algebra $L_{ab}=(L_{i,D+1},L_{ij})$ ($a=i,D+1$), while $A_i$ takes values in
$L_{ij}$, $i.e.,$ in $SO(D)$ algebra. Except when $D=3$, where the algebra of $SO(4)$ splits up in two $chiral$
$SU(2)$ pieces, the electric connection $A_0$ and the magnetic connection $A_i$
do not belong to the same isotopic multiplet, in cotrast to when
$D=3$, $A_i$, $A_0$ and $\F$ are all iso-triplets.

Pseudo-dyons fall in two very different categories: Those in $4p$ dimensional Minkowskian spaceimes, and, those in
the rest.

\begin{itemize}
\item
Pseudo-dyons in $4p$ dimensional spacetimes:

In common with the JZ dyon in the BPS limit, the pseudo-dyon in $7+1$ Minkowski space partnering the monopole on $\R^7$
in section {\bf 7.3} satisfies first-order self-duality equations that are symmetric between the electric and magnetic
functions. Likewise, the magnetic function(s) saturate the Bogomo'lnyi bound (see Footnote 16). Nontrivial
monopole solutions to these are constructed numerically~\cite{Radu:2005rf}. It follows that the corresponding
pseudo-dyons also exist~\cite{Yisong3}.

In fact, such pseudo-dyons exist in all $4p-$dimensional Minkowskian spacetimes, the first ($p=1$) member of that
hierarchy boing the Julia-Zee dyon itself, in the BPS limit. However, only the latter is a genuine dyon with a scalar
electric flux, while all the higher dimensional members are pseudo-dyons with no natural definition for an electric flux
in flat Minkowski space.
\item
Pseudo-dyons in all spacetime dimensions different from $4p$:

The monopoles partnering these dyon like configurations {\bf do not} saturate their respective Bogomol'nyi lower bounds,
even in the case of descents from higher dimensional YM models not featuring a dimensional constant. Examples of this
are the Bogomol'nyi equations \re{Bog26}, \re{Bog35} and \re{Bog44}, for $D=6,5,4$ respectively. These are all
overdetermined and are satisfied only by the trivial solution. Their static Hamiltonians, as well as Lagrangians,
feature a Higgs potential which destroys the symmetry between the electric and magnetic potentials in the latter.

These are solutions to the second order Euler-Lagrange equations and can be constructed numerically. Like the JZ dyon
(not the one in the BPS limit!), their existence and stability proofs are open. 
\end{itemize}
\end{itemize}
There remains to present the examples arising from the dimensional descent of YM in 6 and 8 dimensions. Since one is
dealing with static fields only, the component of the YM curvature $F_{i0}=D_i\,A_0$, is the covariant derivative of
the electric YM connection $A_0$, and it plays a similar role to the covariant derivative $D_i\F$ of the Higgs field.
Like the Higgs field, the asymptotic behaviour of $A_0$, consistent with finiteness of energy, will be such that its
magnitude tends to a constant at infinity. As long as this magnitude is not larger than that of the Higgs VEV $\eta$,
the solutions can support a soliton. If the asymptotic value of the magnitude of $A_0$ is larger than $\eta$,
the solutions become oscillatory.

The construction of the Lagrangian (for the static fields) on $D+1$ dimensional Minkowski space follows systematically
from the static Hamiltonians constructed in Section \ref{monopoles} above, as in \cite{Julia:1975ff}. In a symbolic way,
this construction can be achieved by the following replacements in
\re{H33}, \re{H71}, \re{H26}, \re{H35}, \re{H44} and \re{H53},
\bea
F_{ij}&\rightarrow&F_{\mu\nu}\nonumber\\
D_{i}\F&\rightarrow&D_{\mu}\F\label{replace}
\eea
taking account always to insert the correct sign in the Lagrangian according to the Minkowskian signature chosen.

While the existence of these dyons is not dependent on the degree of symmetry the full YMH systems are subjected to, it
is nevertheless convenient to demonstrate this in the spherically symmetric case. Especially so, since all discussion
in the present notes is restricted to spherically symmetric YMH monopoles (and dyons).

To this end, we extend the (static) spherically symmetric Ans\"atze \re{sphodd} and \re{spheven} for the $A_i$ and $\F$
fields on $\R^D$, to include also the Ansatz for the static spherically electric component $A_0$ of the YM
connection $A_{\mu}=(A_i,A_0)$ in $d=D+1$ dimensional Minkowski spacetime,
\bea
A_i^{(\pm)}&=&\frac{1}{r}\,(1-w(r))\,\Si_{ij}^{(\pm)}\,\hat x_j\ ,\quad A_0^{(\pm)}=u(r)\,\Si_{j,D+1}^{(\pm)}\,\hat x_j
\ \ ,\quad\F=2\,\eta\,h(r)\,\hat x_i\,\Si_{i,D+1}^{(\pm)}\ \,\ \ {\rm for\ odd}\ \ D\label{dysphodd}\\
A_i&=&\frac{1}{r}\,(1-w(r))\,\Ga_{ij}\,\hat x_j\quad , \quad A_0=u(r)\,\Ga_{j,D+1}\,\hat x_j
\ \ ,\quad\F=2\,\eta\,h(r)\,\hat x_i\,\Ga_{i,D+1}\  ,\ \ {\rm for\ even}\ \ D\,.\label{dyspheven}
\eea

The (static) Lagrangians on $d=D+1$ dimensional Minkowski spacetime, subject to spherical symmetry on $\R^D$, pertaining
to the monopoles discussed in the previous section will be presented here. These Lagrangian support dyons for the models
on $\R^3$, and pseudo-dyons for the models on $\R^D$, $D\ge 4$.

\subsection{Dyon in $d=3+1$ Minkowski space with second CP magnetic charge}
\label{2dyon3}
This the JZ dyon in the BPS limit. Its stablity follows immediately from the fact that the equations of motion of the
electric potential are identical to the equations of motion of the Higgs field, the latter being the Bogomol'nyi
equations of the absolutely stable BPS monopole.

By contrast, adding the symmetry breaking Higgs self interaction potential to this
system spoils the symmetry between the electric potential and the Higgs field, such that the ensuing equations of
motion are not solved by the first order Bogomol'nyi equations any more. In the absence of the electric potential,
the static monopole solutions of the resulting system are still topologically stable, but, in the presence of the
electric potential the stability of the resulting JZ dyon do not follow from the stability of the monopole.

\subsection{Excited-dyon in $d=3+1$ Minkowski space with third CP magnetic charge}
\label{3dyon3}
This solution is a genuine dyon in that it supports a nonvanishing electric flux,
but it is not $the$ JZ dyon, the latter being the dyon in $d=3+1$ Minkowski space with
$second$ CP magnetic charge. The magnetic charge of the present example is that descended from the $third$ CP charge,
and this dyon is referred to as an $excited$ dyon~\footnote{Clearly,
there is an infinite tower of such excited dyons, each pertaining to the $n-$th CP
magnetic monopole ($n\ge 4$, $n=4$ being the highest order considered here). Even within this remit, there is the excited
monopole descending from 8 diensional bulk for simplicity. See footnote 16.}.

In some convenient normalisation, the (static) Lagrangian corresponding to the Hamiltonian \re{H33} is,
\bea
\label{L33}
{\cal L}^{(3)}_3
&=&\mbox{Tr}\left(2F_{\mu\nu}^2-\frac32 D_{\mu}\F^2-3S^2\right)\nonumber\\
&-&\frac{\ka^4}{4}\mbox{Tr}\left(\frac14 F_{\mu\nu\rho\si}^2-\{F_{[\mu\nu},D_{\rho]}\F\}^2
-3\left(\{S,F_{\mu\nu}\}+[D_{\mu}\F,D_\nu\F]\right)^2+\frac92\{S,D_{\mu}\F\}^2\right)\,.
\eea
It should be noted here that \re{L33} does not result simply from the replacement \re{replace}, but in addition the
first term $\frac14 F_{\mu\nu\rho\si}^2$ in the second line is inserted by hand.
Introducing this term in the Lagrangian leads to enhanced symmetry between the doublet of functions ($u,h$),
already present in the $fourth$ line of \re{L33sph}, also in its $second$ line.
This is not necessary for the existence of the dyon~\footnote{If the symmetry
between the doublet of functions ($u,h$) were complete, then the system would be solved by first-order
(Bogomol'nyi) equations identical for $u$ and $h$. If in addition the partner monopole were self-dual, then the
existence of the dyon would be guaranteed. This is not the case here since the symmetry between ($u,h$) is in any case
violated in the first and last lines of \re{L33sph}. However, it may be reasonable to maximise this incomplete symmetry,
at least on aesthetic grounds.}.

Setting $A_0=0$ results in the Hamiltonian
\re{L33} supporting the partner monopole, but of course in the presence of this extra term
the energy of the resulting (excited-)dyon is affected.

Subject to spherical symmetry \re{sphodd}, and further redefining the numerical coefficients,
the density \re {L33}  reduces to the one dimensional subsystem
\bea
\label{L33sph}
L^{(3)}_3&\simeq&
\left([2w'^2+r^{-2}(1-w^2)^2]-[r^2\,u'^2+2w^2u^2]+\eta^2[r^2\,h'^2+2w^2h^2]+\eta^4r^2(1-h^2)^2\right)\nonumber\\
&+&\ka^4\bigg(-([(1-w^2)u]')^2+\eta^2([(1-w^2)h]')^2\nonumber\\
&&\qquad+\eta^4\left(2([(1-h^2)w]')^2+r^{-2}[(1-h^2)(1-w^2)+2w^2h^2]^2\right)
\nonumber\\
&&\qquad-\eta^4(1-h^2)^2\left(r^2u'^2+2w^2u^2\right)+\eta^6(1-h^2)^2\left(r^2h'^2+2w^2h^2\right)\bigg)\,.
\eea
The symmetry between the functions $u$ and $h$ in \re{L33sph},
is absent. These solutions do not satisfy first-order Bogomol'nyi equations
as is the case for the BPS limit of the JZ dyon in \re{2dyon3} above, and for the (pseudo) dyon on $\R^7$, in \re{4dyon7}
to be presented below. It means that the existence of these excited-dyons, like the Julia-Zee dyon (not in the BPS limit),
do not follow from the existence of the partner
monopole and do not inherit its topological stability.

The static energy density of the excited-dyon is given by the expression of \re{L33sph}, with all minus signs
replaced by plus signs.

\subsection{Pseudo-dyon in $d=7+1$ Minkowski space with fourth CP magnetic charge}
\label{4dyon7}
This is the dyon living on the $2-$BPS monopole on $\R^7$ explicitly discussed in Section {\bf (7.3)}.
It is in fact a $pseudo-$dyon in the nomenclature used above, in the sense that it describes a nonvanishing electric
YM potential $A_0$, but no scalar electric flux in addition to the magnetic charge. It is a solution to the
Euler-Lagrange equations of the corresponding Lagrangian in $d=7+1$ dimensional Minkowski spacetime.

The Lagrangian corresponding to the Hamiltonian \re{H71} is
\bea
\label{L17}
{\cal L}^{(1)}_7&=&\mbox{Tr} \left(F_{\mu\nu\rho\si}^2-4\,\{F_{[\mu\nu},D_{\rho]}\F\}^2\right)\ \ ,\qquad
\mu=i,0\ \ ,\quad i=1,2,\dots,7\nonumber\\
&=&\mbox{Tr}\left(\left(F_{ijkl}^2-4F_{ijk0}^2\right)
-4\left(-\{F_{[ij},D_{k]}\F\}^2+3\{F_{[ij},D_{0]}\F\}^2\right)\right)\,.
\eea

Subject to the static spherically symmetric Ansatz \re{dysphodd}, the reduced one dimensional (static) Lagrangian density
\re{L17} is
\bea
L^{(1)}_7&=&\frac12(1-w^2)^{2}\left[4\,w'^2+3\frac{(1-w^2)^2}{r^2}\right]\nonumber\\
&&+\frac12\eta^2
\left[\frac{r^2}{3}\left(\left[(1-w^2)\,h\right]'\right)^2
+4(1-w^2)^{2}\,w^2h^2\right]\label{L17sph}\\
&&-\frac12\left[\frac{r^2}{3}\left(\left[(1-w^2)\,u\right]'\right)^2
+4(1-w^2)^{2}\,w^2u^2\right]\,.\nonumber
\eea
In contrast with the previous example, the symmetry between the functions ($u,h$) in \re{L17sph} is complete, and as a
result the system is solved by the Bogomol'nyi equations \re{Bog17} of the partner monopole, and identical first-order
equations where the electric function $u$ replaces the Higgs function $h$, $via$ the following substitution
\be
\label{subst}
h(r)=f(r)\,\cosh\gamma\quad,\quad u(r)=\eta\,f(r)\,\sinh\gamma\,,
\ee
with a constant parameter $\gamma$. This hyperbolic rotation renders the action functional
\re{L17} identical to the energy functional \re{H17}, with $h(r)$
replaced by $f(r)$. Thus the solution~\cite{Radu:2005rf} of the Bogomol'nyi equations \re{Bog17}, augmented by the
same first-order equations with $u$ replacing $h$, yield the dyon field.

This situation can occur in every $4p-$dimensional spacetime, provided that the partner monopole pertains to a model
that is descended from the bulk Yang-Mills system $via$ the descent by {\bf one codimension only!} (In those cases,
the residual system does {\bf not} feature a Higgs potential.)
In the following four examples, the dyons are constructed as solutions to the second order Euler--Lagrange equations
arising from the respective Lagrangian, and their existence cannot be inferred $via$ a rotation like \re{subst}
from the existence of the corresponding (partner) monopole.

\subsection{Pseudo-dyon in $d=6+1$ Minkowski space with fourth CP magnetic charge}
\label{4dyon6}
This is a pseudo-dyon with no electric charge defined in Minkowski space.
The Lagrangian corresponding to the Hamiltonian \re{H26}, with convenient normalisations, is
\bea
\label{L26}
{\cal L}^{(2)}_6
&\simeq&\mbox{Tr}\left(F_{\mu\nu\rho\si}^2-\{F_{[\mu\nu},D_{\rho]}\F\}^2
-\left(\{S,F_{\mu\nu}\}+[D_{\mu}\F,D_{\nu}\F]\right)^2\right)\,.
\eea
Subject to spherical symmetry, \re{spheven}, the density \re{L26} reduces to the one
dimensional subsystem
\bea
L^{(2)}_6&\simeq&r^{-1}(1-w^2)^2\left(2\,w']^2+r^{-2}(1-w^2)^2\right)\nonumber\\
&&-r\left[([(1-w^2)u]')^2+9\,r^{-2}(1-w^2)^2w^2u^2\right]
+\eta^2\,r\left[([(1-w^2)h]')^2+9\,r^{-2}(1-w^2)^2w^2h^2\right]\nonumber\\
&&+\eta^4r^{3}\left[([(1-h^2)w]')^2+2r^{-2}[(1-w^2)(1-h^2)+2w^2h^2]^2\right]\,.\label{L26sph}
\eea
This model has no BPS limit and the consequent absence of symmetry between the functions $u$ and $h$ in \re{L26sph}
means that the existence of this (pseudo-)dyon does not follow from the existence of its partner monopole. 
\subsection{Pseudo-dyon in $d=5+1$ Minkowski space with fourth CP magnetic charge}
\label{4dyon5}
The Lagrangian of this pseudo-dyon corresponding to the Hamiltonian \re{H35}, is
\bea
\label{L35}
{\cal L}^{(3)}_5
&\simeq&\mbox{Tr}\left(F_{\mu\nu\rho\si}^2-\{F_{[\mu\nu},D_{\rho]}\F\}^2
-\left(\{S,F_{\mu\nu}\}+[D_{\mu}\F,D_{\nu}\F]\right)^2+\{S,D_{\mu}\F\}^2\right)\,.
\eea
Subject to spherical symmetry, \re{spheven}, the density \re{L35} reduces to the one
dimensional subsystem
\bea
L^{(3)}_5&\simeq&r^{-2}(1-w^2)^2\left[4\,w'^2+r^{-2}(1-w^2)^2\right]\nonumber\\
&-&\left[([(1-w^2)u]')^2+6r^{-2}(1-w^2)^2w^2u^2\right]
+\eta^2\left[([(1-w^2)h]')^2+6r^{-2}(1-w^2)^2w^2h^2\right]\nonumber\\
&+&\eta^4r^{2}\left[2\,([(1-h^2)w]')^2+3\,r^{-2}[(1-w^2)(1-h^2)+2w^2h^2]^2\right]\label{L35sph}\\
&-&\eta^4r^4\,(1-h^2)^2\left[u'^2+4\,r^{-2}w^2u^2\right]
+\eta^6r^4\,(1-h^2)^2\left[h'^2+4\,r^{-2}w^2h^2\right]\nonumber
\eea
Again, this model has no BPS limit and hence the existence of this $A_0$ carrying solution does not follow
from the existence of its partner monopole.
\subsection{Pseudo-dyon in $d=4+1$ Minkowski space with fourth CP magnetic charge}
\label{4dyon4}
The Lagrangian of this pseudo-dyon corresponding to the Hamiltonian \re{H44}, is
\bea
\label{L44}
{\cal L}^{(4)}_4
&\simeq&\mbox{Tr}\left(F_{\mu\nu\rho\si}^2+\{F_{[\mu\nu},D_{\rho]}\F\}^2
-\left(\{S,F_{\mu\nu}\}+[D_{\mu}\F,D_{\nu}\F]\right)^2-\{S,D_{\mu}\F\}^2\right)\,,
\eea
Subject to spherical symmetry, \re{spheven}, the density \re{L44} reduces to the one
dimensional subsystem
\bea
L^{(4)}_4&\simeq&r^{-2}(1-w^2)^2\left[4\,w'^2+r^{-2}(1-w^2)^2\right]\nonumber\\
&-&\left[([(1-w^2)u]')^2+6r^{-2}(1-w^2)^2w^2u^2\right]
+\eta^2\left[([(1-w^2)h]')^2+6r^{-2}(1-w^2)^2w^2h^2\right]\nonumber\\
&+&\eta^4r^{2}\left[2\,([(1-h^2)w]')^2+3\,r^{-2}[(1-w^2)(1-h^2)+2w^2h^2]^2\right]\label{L44sph}\\
&-&\eta^4r^4\,(1-h^2)^2\left[u'^2+4\,r^{-2}w^2u^2\right]
+\eta^6r^4\,(1-h^2)^2\left[h'^2+4\,r^{-2}w^2h^2\right]\nonumber
\eea
with no BPS limit, such that the existence of this psuodo-dyon does not follow from the existence of
its partner monopole. 
\subsection{Excited-Dyon in $d=3+1$ Minkowski space with fourth CP magnetic charge}
\label{4dyon3}
Like the excited-dyon in section {\bf 8.2}, this solution is also a genuine dyon as it supports a nonvanishing electric
flux. It differs from the latter however, since its Lagranian does not sit on top of that of the Georgi-Glashow model,
as a consequence of the distinct dimensional descent giving rise to its partner monopole which
in this case is stabilised by the 
$fourth$ CP magnetic charge. It can nonetheless be described as a(nother) $excited$ dyon.

In some convenient normalisation, the (static) Lagrangian corresponding to the Hamiltonian \re{H53} is,
\bea
\label{L53}
{\cal L}^{(5)}_3
&\simeq&\mbox{Tr}\left(\frac14 F_{\mu\nu\rho\si}^2-\{F_{[\mu\nu},D_{\rho]}\F\}^2
-\left(\{S,F_{\mu\nu}\}+[D_{\mu}\F,D_{\nu}\F]\right)^2+\{S,D_{\mu}\F\}^2+S^4\right)\,.
\eea
Again as in section {\bf 8.2} the term $\frac14 F_{\mu\nu\rho\si}^2$ is inserted by hand in \re{L53}, for the purpose of
maximising the symmetry between the functions $u$ and $h$ in \re{L53sph} below.

Subject to spherical symmetry \re{sphodd} and with some further redefinitions of the numerical coefficients,
the density \re{L53} reduces to the one dimensional subsystem
\bea
\label{L53sph}
H^{(5)}_3&\simeq&-r^{-2}\,\left([(1-w^2)u]'\right)^2+\eta^2\,r^{-2}\,\left([(1-w^2)h]'\right)^2\nonumber\\
&+&\eta^4\left[2\left([(1-h^2)w]'\right)^2+r^2\left[(1-w^2)(1-h^2)+w^2h^2\right]^2\right]\nonumber\\
&-&\,r^{2}(1-h^2)^2\left[u'^2+2r^{-2}w^2u^2\right]+\eta^6\,r^{2}(1-h^2)^2\left[h'^2+2r^{-2}w^2h^2\right]\nonumber\\
&+&\eta^8\,r^{2}(1-h^2)^4\,.
\eea
The static energy density of the excited-dyon is given by the expression of \re{L53sph}, with all minus signs
replaced by plus signs.

\section{New Chern--Simons terms}
\label{CS}
Chern-Simons densities can be defined in all odd dimensional spaces irrespective of the signature, $i.e.$, on
Minkowskian or Euclidean spaces. They are defined in terms of the gauge connection and curvature, both Abelian
or non-Abelian. In the Abelian case, they are very simple quantities and their properties can be easily analysed.
Here, we are concerned exclusively with non-Abelian Chern-Simons densities. We are not concerned at all with
Chern-Simons densities of (Abelian) antisymmetric potentials $A_{\mu_1\mu_2\dots\mu_n}$ and their field strengths
$F_{\mu_1\mu_2\dots\mu_n\mu_{n+1}}$.

The usual Chern--Simons densities in odd dimensions defined in terms of the non-Abelian
gauge connection are first recalled. This is
followed by the subsection in which the new Chern--Simons are introduced
in both odd and even dimensions. These are defined in terms of the
non-Abelian connection and its partner Higgs field which appears in the dimensionally descended Chern-Pontryagin terms
presented above in section {\bf 5}. As will be seen below, the new Chern--Simons densities are exclusively non-Abelian.
\subsection{Usual non-Abelian Chern--Simons terms in odd dimensions}
Topologically massive gauge field theories in $2+1$ dimensional spacetimes were first introduced in
\cite{Deser:1982vy,Deser:1981wh}. The salient feature of these theories is the presence of a Chern-Simons (CS)
dynamical term. To define a CS density one needs to have a gauge connection, and hence also a curvature. Thus,
CS densities can be defined both for Abelian (Maxwell) and non-Abelian (Yang--Mills) fields. They can also be defined
for the gravitational~\cite{Jackiw:2003pm} field since in that system too one has a (Levi-Civita or otherwise) connection,
akin to the Yang-Mills connection in that it carries frame indices analogous to the isotopic indices of the YM
connection. Here we are interested exclusively in the (non-Abelian) YM case, in the presence of an $isovector$ valued
Higgs field.

The definition of a Chern-Simons (CS) density follows from the definition of the corresponding Chern-Pontryagin (CP)
density \re{CP}. As stated by \re{totdivn}, this quantity is a total divergence and the density
$\bOmega^{(n)}=\Omega^{(n)}_M$ ($M=1,2,\dots,2n$)
in that case has $(2n)-$components. The Chern-Simons density is then defined as one fixed component of
$\bOmega^{(n)}$, say the $2n-$th component,
\be
\label{CS-n}
\Omega^{(n)}_{\rm{CS}}=\Omega_{2n}^{(n)}
\ee
which now is given in one dimension less, where $M=\mu,2n$ and $\mu=1,2,\dots (2n-1)$.

This definition of a (dynamical) CS term holds in all odd dimensional spacetimes
$(t,\R^D)$, with $x_{\mu}=(x_0,x_i)$, $i=1,2,\dots,D$, with $D$ being an even integer. That $D$ must be even is clear
since $D+2=2n$, the $2n$ dimensions in which the CP density \re{CP} is defined, is itself even.

The properties of CS densities are reviewed in \cite{Jackiw:1985}.
Most remarkably, CS densities are defined in odd (space or spacetime) dimensions and are $gauge$ $variant$. The context
here is that of a $(2n-1)-$dimensional Minkowskian space. It is important to realise that dynamical Chern-Simons theories
are defined on spacetimes with Minkowskian signature. The reason is that the usual CS densities appearing in the
Lagrangian are by construction $gauge$ $variant$, but in the definition of the energy densities the CS term itself does
not feature, resulting in a Hamiltonian (and hence energy) being $gauge$ $invariant$ as it should be~\footnote{Should one
employ a CS density on a space with Euclidean signature, with the CS density appearing in
the static Hamiltonian itself, then the energy would not be $gauge$ $invariant$. Hamiltonians of this
type have been considered in the literature, $e.g.$, in \cite{Rubakov:1986am}. Chern-Simons densities
on Euclidean spaces, defined in terms of the composite connection of a sigma model, find application as the
topological charge densities of Hopf solitons.}.

Of course, the CP densities and the resulting CS densities, can be defined in terms of both Abelian and non-Abelian
gauge connections and curvatures. The context of the present notes is the construction of soliton
solutions~\footnote{The term soliton solutions here is used rather loosely, implying only the construction of
regular and finite energy solutions, without insisting on topological stability in general.},
unlike in \cite{Deser:1982vy,Deser:1981wh}. Thus in any given dimension,
our choice of gauge group must be made with due regard to
regularity, and the models chosen must be consistent with the Derrick scaling requirement for the finiteness of
energy. Accordingly, in all but $2+1$ dimensions, our considerations are restricted to non-Abelian gauge fields.

Clearly, such constructions can be extended to all odd dimensional spacetimes systematically. We list
$\Omega_{\rm CS}$, defind by \re{CS}, for $D=2,4,6$, the familiar densities
\bea
\Omega_{\rm CS}^{(2)}&=&\vep_{\la\mu\nu}\mbox{Tr}\,
A_{\la}\left[F_{\mu\nu}-\frac23A_{\mu}A_{\nu}\right]\label{CS3}\\
\Omega_{\rm CS}^{(3)}&=&\vep_{\la\mu\nu\rho\si}\mbox{Tr}\,
A_{\la}\left[F_{\mu\nu}F_{\rho\si}-F_{\mu\nu}A_{\rho}A_{\si}+
\frac25A_{\mu}A_{\nu}A_{\rho}A_{\si}\right]\label{CS5}
\\
\Omega_{\rm CS}^{(4)}&=&\vep_{\la\mu\nu\rho\si\tau\ka}
\mbox{Tr}\,A_{\la}\bigg[F_{\mu\nu}F_{\rho\si}F_{\tau\ka}
-\frac45F_{\mu\nu}F_{\rho\si}A_{\tau}A_{\ka}-\frac25
F_{\mu\nu}A_{\rho}F_{\si\tau}A_{\ka}\nonumber\\
&&\qquad\qquad\qquad\qquad\qquad\qquad
+\frac45F_{\mu\nu}A_{\rho}A_{\si}A_{\tau}A_{\ka}-\frac{8}{35}
A_{\mu}A_{\nu}A_{\rho}A_{\si}A_{\tau}A_{\ka}\bigg]\,.\label{CS7}
\eea
Note that \re{CS3} and \re{CS5} coincide with the leading terms in \re{CS44} and \re{CS46} respectively, except for the
chiral matrices $\Ga_5$ and $\Ga_7$ in the latter.

Concerning the choice of gauge groups, one notes that the
CS term in $D+1$ dimensions features the product of $D$ powers of the (algebra valued) gauge field/connection in front
of the Trace, which would vanish if the gauge group $is\ not\ larger\ than$ $SO(D)$.  In that case, the YM connection
would describe only a 'magnetic' component, with the 'electric' component necessary for the the nonvanishing of the CS
density would be absent.As in \cite{Brihaye:2009cc}, the
most convenient choice is $SO(D+2)$. Since $D$ is always even, the representation of $SO(D+2)$ are the $chiral$
representation in terms of (Dirac) spin matrices. This completes the definition of the usual non-Abelian Chern-Simons
densities in $D+1$ spacetimes.

From \re{CS3}-\re{CS7}, it is clear that the CS density is $gauge$ $variant$. The Euler--Lagrange equations of the
CS density is nonetheless $gauge$ $invariant$, such that for the examples \re{CS3}-\re{CS7} the corresponding arbitarry
variations are
\bea
\delta_{A_{\la}}\Omega_{\rm CS}^{(2)}&=&\vep_{\la\mu\nu}F_{\mu\nu}\label{ELCS3}\\
\delta_{A_{\la}}\Omega_{\rm CS}^{(3)}&=&\vep_{\la\mu\nu\rho\si}F_{\mu\nu}F_{\rho\si}\label{ELCS5}\\
\delta_{A_{\la}}\Omega_{\rm CS}^{(4)}&=&\vep_{\la\mu\nu\rho\si\ka\eta}F_{\mu\nu}F_{\rho\si}F_{\ka\eta}\,.\label{ELCS7}
\eea
This, and other interesting properties of CS densities are given in \cite{Jackiw:1985}. A remarkable property of a CS
density is its transformation under the action of an element, $g$,
of the (non-Abelian) gauge group. We list these for the two examples \re{CS3}-\re{CS5},
\bea
\Omega_{\rm CS}^{(2)}\to\tilde\Omega_{\rm CS}^{(2)}&=&\Omega_{\rm CS}^{(2)}-
\frac23\vep_{\la\mu\nu}\mbox{Tr}\,\al_{\la}\al_{\mu}\al_{\nu}
-2\vep_{\la\mu\nu}\,\pa_{\la}\mbox{Tr}\,\al_{\mu}\,A_{\nu}\label{gaugeCS3}\\
\Omega_{\rm CS}^{(3)}\to\tilde\Omega_{\rm CS}^{(3)}&=&\Omega_{\rm CS}^{(3)}-\frac25\,\vep_{\la\mu\nu\rho\si}
\mbox{Tr}\,\al_{\la}\al_{\mu}\al_{\nu}\al_{\rho}\al_{\si}\nonumber\\
&&+
2\,\vep_{\la\mu\nu\rho\si}\,\pa_{\la}\mbox{Tr}\,\al_{\mu}\bigg[
A_{\nu}\left(F_{\rho\si}-\frac12A_{\rho}A_{\si}\right)+\left(F_{\rho\si}-\frac12A_{\rho}A_{\si}\right)A_{\nu}\nonumber\\
&&\qquad\qquad\qquad\qquad\qquad\qquad\qquad\qquad-\frac12\,A_{\nu}\,\al_{\rho}\,A_{\si}-\al_{\nu}\,\al_{\rho}\,A_{\si}
\bigg]\,,\label{gaugeCS5}
\eea
where $\al_{\mu}=\pa_{\mu}g\,g^{-1}$, as distinct from the algebra valued quantity $\beta_{\mu}=g^{-1}\,\pa_{\mu}g$ that
appears as the inhomogeneous term in the gauge transformation of the non-Abelian curvature (in our convention).

As seen from \re{gaugeCS3}-\re{gaugeCS5}, the gauge variation of $\Omega_{\rm CS}$ consists of a term which is
explicitly a total divergence, and, another term
\be
\label{om}
\omega^{(n)}\simeq\vep_{\mu_1\mu_2\dots\mu_{2n-1}}\mbox{Tr}\,\al_{\mu_1}\al_{\mu_2}\dots\al_{\mu_{2n-1}}\,,
\ee
which is {\it effectively total divergence}, and in a concrete group representation parametrisation becomes
{\it explicitly total divergence}. This can be seen by subjecting \re{om} to variations with respect to the
function $g$, and taking into account the Lagrange multiplier term resulting from the (unitarity) constraint
$g^{\dagger}\,g=g\,g^{\dagger}=\eins$.

The volume integaral of the CS density then transforms under a gauge transformation as follows.
Given the appropriate asymptotic decay of the connection (and hence also the curvature), the surface integrals in
\re{gaugeCS3}-\re{gaugeCS5} vanish. The only contribution to the gauge variation of the CS action/energy then comes from
the integral of the density \re{om}, which (in the case of Euclidean signature) for the appropriate
choice of gauge group yields an integer, up to the angular volume as a multiplicative factor.

All above stated properties of the Chern-Simons (CS) density hold irrespective of the signature of the space. Here, the
signature is taken to be Minkowskian, such that the CS density in the Lagrangian does not contribute to the energy density
directly. As a consequence the energy of the soliton is gauge invariant and does not suffer the gauge transformation
\re{gaugeCS3}-\re{gaugeCS5}. Should a CS density be part of a static Hamiltonian (on a space of Euclidean signature),
then the energy of the soliton would change by a multiple of an integer.

\subsection{New Chern--Simons terms in all dimensions}
The plan to introduce a completely new type of Chern-Simons term. The usual CS densities $\bOmega_{\rm{CS}}^{(n)}$,
\re{CS}, are defined with reference to the total divergence expression \re{totdivn} of the $n-$th
Chern-Pontryagin density \re{CP}, as the $2n-$th component $\Omega_{2n}^{(n)}$ of the density $\bOmega^{(n)}$.
Likewise, the new CS terms are defined with reference to the total divergence expression \re{descCP} of the
dimensionally reduced $n-$th CP density, with the dimension $D$ of the residual space replaced
formally by $\bar{D}$
\be
\label{totdivbar}
{\cal C}_{\bar{D}}^{(n)}={\bf\nabla}\cdot{\bf\Omega}^{(n,\bar{D})}\,.
\ee
The densities ${\bf\Omega}^{(n,\bar{D})}$ can be read off from ${\bf\Omega}^{(n,D)}$ given in Section {\bf 5}, with
the formal replacement $D\to\bar{D}$.
The new CS term is now identified as the $\bar{D}-$th component of ${\bf\Omega}^{(n,\bar{D})}$. The final step
in this identification is to assign the value $\bar{D}=D+2$, where $D$ is the spacelike dimension of the $D+1$
dimensional Minkowski space, with the new Chern-Simons term defined as
\be
\label{CSnew}
\tilde\Omega_{\rm CS}^{(n,D+1)}\stackrel{def}=\Omega^{(n,D+2)}_{D+2}\,.
\ee
The departure of the new CS densities from the usual CS densities is stark, and these
differ in several essential respects from the usual ones described in the previous subsection. The most important
new features in question are
\begin{itemize}
\item
The field content of the new CS systems includes Higgs fields in addition to the Yang-Mills fields,
as a consequence of the dimensional reduction of gauge fields described in Section {\bf 4}. It should be
emphasised that the appearance of the Higgs field here is due to the imposition of symmetries in the descent mechanism,
in contrast with its presence in the models~\cite{Hong:1990yh,Jackiw:1990aw,NavarroLerida:2009dm} supporting $2+1$
dimensional CS vortices, where the Higgs field was introduced by hand with the expedient of satisfying the Derrick
scaling requirement.
\item
The usual dynamical CS densities defined with reference to the $n-$th CP density live in $2n-1$ dimensional Minkowski
space, $i.e.$, only in odd dimensional spacetime. By contrast, the new CS densities defined with reference to the $n-$th
CP densities live in $D+1$ dimensional Minkowski space, for all $D$ subject to
\be
\label{subject}
2n-2\ge D\ge 2\,,
\ee
 $i.e.$, in both odd, as well as even dimensions. Indeed, in any given $D$ there is an infinite tower of new CS
densities characterised by the integer $n$ subject to \re{subject}. This is perhaps the most important feature of
the new CS densities.
\item
The smallest simple group consistent with the nonvanishing of the $usual$ CS density in $2n-1$ dimensional spacetime
is $SO(2n)$, with the gauge connection taking its values in the $chiral$ Dirac representation. By contrast,
the gauge groups of the new CS densities in $D+1$ dimensional spacetime are fixed by the
prescription of the dimensional descent from which they result.
As $per$  the prescription of descent described in Section {\bf 4}, the gauge group now will be $SO(D+2)$,
independently of the integer $n$, while the Higgs field takes its values in the orthogonal complement of $SO(D+2)$ in
$SO(D+3)$. As such, it forms an iso-$(D+2)-$vector multiplet. 
\item
Certain properties of the new CS densities are remarkably different for $D$ even and $D$ odd.
\begin{itemize}
\item
Odd $D$: Unlike in the usual case \re{CS3}-\re{CS5}, the new CS terms are $gauge\ invariant$.
The gauge fields are $SO(D+2)$ and the Higgs are in $SO(D+3)$. $D$ being odd,
$D+3$ is even and hence the fields can be parametrised with respect to the $chiral$ (Dirac)
representations of $SO(D+3)$. An important consequence of this is the fact that now, both (electric) $A_0$
and (magnetic) $A_i$ fields lie in the same isotopic multiplets, in contrast to the $pseudo-$dyons described in the
previous section.
\item
Even  $D$: The new CS terms now consist of a $gauge\ variant$ part expressed only in terms of the gauge field, and a
$gauge\ invariant$ part expressed in terms of both gauge and Higgs fields. The leading, $gauge\ variant$, term differs
from the corresponding usual CS terms \re{CS3}-\re{CS5} only due to the presence of a (chiral) $\Ga_{D+3}$ matrix infront
of the Trace. The gauge and Higgs fields are again in $SO(D+2)$ and in $SO(D+3)$ respectively, but now, $D$ being even
$D+3$ is odd and hence the fields are parametrised with respect to the (chirally doubled up) full Dirac representations of
$SO(D+3)$. Hence the appearance of the chiral matrix infront of the Trace.
\end{itemize}
\end{itemize}
As in the usual CS models, the regular finite energy solutions of the new CS models are not topologically stable. These
solutions can be constructed numerically.

Before proceeding to display some typcal examples in the Subsection following, it is in order to make a small diversion
at this point to make a clarification. The new CS densities proposed are functionals of both the Yang--Mills, and, the
"isovector'' Higgs field. Thus, the systems to be described below are Chern-Simons--Yang-Mills-Higgs models in
a very specific sense, namely that the Higgs field is an intrinsic part of the new CS density. This is in contrast with
Yang-Mills--Higgs-Chern-Simons or Maxwell--Higgs-Chern-Simons models in $2+1$ dimensional spacetimes that
have appeared ubiquitously in the literature. It is important to emphasise that the latter are entirely different
from the systems introduced here, simply because the CS densities they employ are the $usual$ ones, namely \re{CS3} or
more often its Abelian~\footnote{There are, of course, Abelian CS densities in all odd spacetime dimensions but these do
not concern us here since in all $D+1$ dimensions with $D=2n\ge 4$, no regular solitons can be constructed.}  version
\[
\Omega_{\rm U(1)}^{(2)}=\vep_{\la\mu\nu}\,A_{\la}F_{\mu\nu}\,,
\]
while the CS densities employed here are $not$ simply functionals of the gauge field, but also of the (specific)
Higgs field. To put this in perspective, let us comment on the
well known $Abelian$ CS-Higgs solitons in $2+1$ dimensions constructed in \cite{Hong:1990yh,Jackiw:1990aw}
support self-dual vortices, which happen to be unique inasfar as
they are also topologically stable. (Their non-Abelian counterparts~\cite{NavarroLerida:2009dm} are not endowed with
topological stability.) The presence of the Higgs field in \cite{Hong:1990yh,Jackiw:1990aw,NavarroLerida:2009dm}
enables the Derrick scaling requirement to be satisfied by virtue of the presence of the Higgs self-interaction
potential. In the Abelian case in addition, it results in the topological stability of the vortices.
If it were not for the topological stability, it would not be
necessary to have a Higgs field merely to satisfy the Derrick scaling requirement. That can be achieved instead, $e.g.$,
by introducing a negative cosmological constant and/or gravity, as was done
in the $4+1$ dimensional case studied in \cite{Brihaye:2009cc}. Thus, the involvement of the Higgs field in
conventional ($usual$) Chern-Simons theories is not the only option. The reason for emphasising the optional status of
the Higgs field in the usual $2+1$ dimensional Chern-Simons--Higgs models is, that in the new models proposed here
the Higgs field is intrinsic to the definition of the (new) Chern-Simons density itself.

\subsection{Examples}
As discussed above, the new dynamical Chern-Simons densities
\[
\tilde\Omega_{\rm CS}^{(n,D+1)}[A_{\mu},\F]
\]
are characterised by the dimensionality of the space $D$ and the integer $n$ specifying the
dimension $2n$ of the bulk space from which the relevant residual system is arrived at. As in
Section {\bf 4} above, we restrict attention here to $n=2,3$ and $4$.

The case $n=2$ is empty, since according to \re{subject} the largest spacetime in which a new CS density can be
constructed is $2n-2$, $i.e.$, in $1+1$ dimensional Minkowsky space which we ignore.

The case $n=3$ is not empty, and affords us two nontrivial examples. The largest spacetime $2n-2$, in which a
new CS density can be constructed in this case is $3+1$ and the next in $2+1$ Minkowski space. These two densities
can be read off \re{3,5} and \re{CS34}, repectively,
\bea
\tilde\Omega_{\rm CS}^{(3,3+1)}&=&\vep_{\mu\nu\rho\si}\,\mbox{Tr}\ F_{\mu\nu}\,F_{\rho\si}\,\F
\label{3,3+1}\\
\tilde\Omega_{\rm CS}^{(3,2+1)}&=&
\vep_{\mu\nu\la}\,\mbox{Tr}\,\gamma_5\,\left[-2\eta^2A_{\la}\left(F_{\mu\nu}
-\frac23A_{\mu}A_{\nu}\right)+\left(\F\,D_{\la}\F-D_{\la}\F\,\F\right)\,F_{\mu\nu}
\right]\,.\label{3,2+1}
\eea
The case $n=4$ affords four nontrivial examples, those in $5+1$, $4+1$, $3+1$ and $2+1$ Minkowski space. These densities
can be read off \re{4,7}, \re{CS46}, \re{CS45} and \re{CS44}, repectively,
\bea
\tilde\Omega_{\rm CS}^{(4,5+1)}&=&
\vep_{\mu\nu\rho\si\tau\la}\,\mbox{Tr}\ F_{\mu\nu}\,F_{\rho\si}\,F_{\tau\la}\,\F\label{4,5+1}\\
\tilde\Omega_{\rm CS}^{(4,4+1)}&=&\vep_{\mu\nu\rho\si\la}\,\mbox{Tr}\,\Gamma_7
\bigg[A_{\la}\left(F_{\mu\nu}F_{\rho\si}-F_{\mu\nu}A_\rho{}A_{\si}+\frac25A_{\mu}A_{\nu}A_{\rho}A_{\si}\right)\nonumber\\
&&\qquad\qquad\qquad\qquad+
D_{\la}\F\left(\F F_{\mu\nu}F_{\rho\si}+F_{\mu\nu}\F F_{mn}+F_{\mu\nu}F_{mn}\F\right)\bigg]\label{4,4+1}\\
\tilde\Omega_{\rm CS}^{(4,3+1)}&=&
\vep_{\mu\nu\rho\si}\,\mbox{Tr}\bigg[
\F\left(\eta^2\,F_{\mu\nu}F_{\rho\si}+\frac29\,\F^2\,F_{\mu\nu}F_{\rho\si}
+\frac19\,F_{\mu\nu}\F^2F_{\rho\si}\right)\nonumber\\
&&\qquad\qquad\qquad\qquad
-\frac29\left(\F D_{\mu}\F D_{\nu}\F-D_{\mu}\F\F D_{\nu}\F+D_{\mu}\F D_{\nu}\F\F\right)F_{\rho\si}\bigg]\label{4,3+1}\\
\tilde\Omega_{\rm CS}^{(4,2+1)}
&=&\vep_{\mu\nu\la}\,\mbox{Tr}\,\Gamma_5\,\bigg\{6\eta^4\,A_{\la}\left(F_{\mu\nu}-\frac23\,A_{\mu}\,A_{\nu}
\right)\nonumber\\
&&\qquad\quad
-6\,\eta^2\left(\F\,D_{\la}\F-D_{\la}\F\,\F\right)\,F_{{\mu\nu}}\nonumber\\
&&\qquad\quad
+\left[\left(\F^2\,D_{\la}\F\,\F-\F\,D_{\la}\F\,\F^2\right)
-2\left(\F^3\,D_{\la}\F-D_{\la}\F\,\F^3\right)\right]F_{{\mu\nu}}
\bigg\}\,.\label{4,2+1}
\eea
It is clear that in any $D+1$ dimensional spacetime an infinite tower of CS densities
$\tilde\Omega_{\rm CS}^{(n,D+1)}$ can be defined, for all positive integers $n$. Of these,
those in even dimensional spacetimes are gauge invariant, $e.g.$, \re{3,3+1}, \re{4,5+1} and  \re{4,3+1},
while those in odd dimensional spacetimes are gauge variant, $e.g.$, \re{3,2+1}, \re{4,4+1} and \re{4,2+1},
the gauge variations in these cases being given formally by \re{gaugeCS3}
and \re{gaugeCS5}, with $g$ replaced by the appropriate gauge group here.

Static soliton solutions to models whose Lagrangians consist of the above introduced types of
CS terms together with Yang-Mills--Higgs (YMH) terms are currently under construction~\cite{Radu:2011zy}.
The only constraint in the choice of the detailed models employed is the requirement that the
Derrick scaling requirement be satisfied. Such solutions are constructed numerically. In
contrast to the monopole solutions, they are not endowed with topological stability
because the gauge group must be larger than $SO(D)$, for which the solutions to the
constituent YMH model is a stable monopole. Otherwise the CS term would vanish.

\subsection{Imposition of spherical symmetry}
Since the context of the present notes is one of constructing solitons, it may be helpful to state the Ans\"atze for the
fields subjected to spherical symmetry. The monopole and dyon-like solitons presented in Sections {\bf 7} and {\bf 8}
were also discussed in the context of spherical symmetry. One reason is that in that framework the Dirac-Yang nature of
the monopole fields becomes transparent, and, exposition of the overdetermination of some of the Bogomol'nyi equations
is natural. Also, dyonic
properties are easy to analyse subject to that symmetry. Less symmetric monopole and dyonic fields can also be studied
for the $SO(D)$ gauge fields and iso-$D-$vector Higgs fields of Sections {\bf 7} and {\bf 8}, with a manageable
additional effort.

In the context of the new Chern--Simons density however, where the gauge fields are $SO(D+2)$ and the Higgs multiplets are
iso-$(D+2)-$vectors, imposition of less stringent symmetry than spherical is impractical. Besides, the spherically
symmetric expressions in this case help to illustrate the gauge group and multiplet structure presented in the previous
subsection concretely.

In $D+1$ dimensional spacetime the
gauge connection $A_{\mu}=(A_i,A_0)$, $i=1,2,\dots D$, takes its values in $SO(D+2)$ and subject to spherical symmetry
is parametrised by the pair of triplets $\vec\xi(r)$, $\vec\chi(r)$ and the triplet~\footnote{This triplet,
$\vec A_r(r)$, plays the role of a connection in the residual one dimensional system after the imposition of symmetry,
and encodes the $SO(3)$ arbitrariness of this Ansatz. In one dimension there is no curvature hence it can be
gauged~\cite{Brihaye:2009cc} away in practice.}
$\vec A_r(r)$. The Higgs field $\F$ takes its values in the orthogonal complement of $SO(D+2)$
in $SO(D+3)$ and is parametrised by the triplet $\vec\f(r)$. The explicit expression of this Ansatz for the Higgs
field and the gauge connection are,
\bea
\F&=&\f^M\,\ga_{M,D+3}+\f^{D+3}\,\hat x_j\,\ga_{j,D+3}\label{higgs}\\
A_0&=&-(\vep\chi)^M\,\hat x_j\,\ga_{jM}-
\chi^{D+3}\,\ga_{D+1,D+2}\label{a0p}\\
A_i&=&\left(\frac{\xi^{D+3}+1}{r}\right)\ga_{ij}\hat x_j+
\left[\left(\frac{\xi^M}{r}\right)\left(\delta_{ij}-\hat x_i\hat x_j\right)+
(\vep A_r)^M\,\hat x_i\hat x_j\right]\ga_{jM}+\nonumber\\
&&\qquad\qquad\qquad\qquad\qquad\qquad\qquad\qquad +A_r^{D+3}\,
\hat x_i\,\ga_{D+1,D+2}\,,\label{aip}
\eea
where  $\vep_{MN}$ is the two dimensional Levi-Civita symbol.
In \re{higgs}, \re{a0p} and \re{aip}, the index $M$ runs over two values $M=D+1,\, D+2$, such that 
$\xi\equiv(\xi^M,\xi^{D+3})$, $\vec\chi\equiv(\chi^M,\chi^{D+3})$, $\vec A_r\equiv(A_r^M,A_r^{D+3})$, and
$\vec\f\equiv(\f^M,\f^{D+3})$. The spin matrices $\ga_{ab}=(\ga_{ij},\,\ga_{iM},\,\ga_{i,D+3},\,,\ga_{MN},\ga_{M,D+3})$ are
the generators of $SO(D+3)$, in a unified notation to cover both the Dirac, $\Ga_{ab}$, and the chiral, $\Si_{ab}$ as
the case may be.

It can now clearly be seen from \re{higgs}, \re{a0p} and \re{aip}, that all components of the gauge connection
$A_{\mu}=(A_i,A_0)$ take their values in the algebra of $SO(D+2)$, with the Higgs field in the orthogonal complement
of $SO(D+2)$ in $SO(D+3)$. This is in essential contrast with the Julia--Zee type dyons, for which the electric
component of the gauge connection $A_0$ does not have
the same multiplet structure as the magnetic component $A_i$. It has instead precisely the same
multiplet structure as the Higgs field as seen in \re{dysphodd}-\re{dyspheven}.

The above choice of the multiplet structure of $A_0$ is only one of two possibilities, namely the one that differs from
the Julia-Zee case. In the latter case, to support solutions regular at the origin in $\R^D$ with $D\ge 4$, $A_0$ has to
take its values outside the algebra of $SO(D)$, and like the Higgs field $\F$, in the orthogonal complement in $SO(D+1)$.
Here however, the gauge group is not $SO(D)$, but rather, $SO(D+2)$ permitting nonvanishing (new-)CS density. As a
consequence it was possible to let $A_0$ take its values in the algebra of $SO(D+2)$ like the magnetic component of
the connection $A_i$, and unlike $\F$. It is however just as legitimate to choose instead that $A_0$ take its values
in the orthogonal complement of $SO(D+2)$ in $SO(D+3)$. In that case, the spherically symmetric Ansatz \re{a0p}
for $A_0$ would be replaced by
\be
\label{a0p1}
A_0=\chi^M\,\ga_{M,D+3}+\chi^{D+3}\,\hat x_j\,\ga_{j,D+3}\,.
\ee
Solitons of such CS-Higgs models in $3+1$ and $2+1$ dimensions are now under active consideration~\cite{NRT}.

\section{Outlook}
Solitons in higher dimensions, both in flat space and their gravitating versions, have the potential to be
employed in the construction of string theory solitons. (The presentation in these notes is restricted to considerations
in flat space. The gravitating versions follow systematically, and are deferred.) Solitons of gauge
fields play a very special role in this context, because both Abelian and non-Abelian matter feature prominently in
heterotic string theory and in Supergravities.

The present notes are restricted to non-Abelian gauge fields, and more specifically to Yang-Mills--Higgs systems.
Solitons of the Yang-Mills systems, the instantons, are topologically stable in even
spacelike dimensions, stabilised by Chern-Pontryagin charges. The gauge connection of these solutions
are asymptotically pure gauge, resulting in vanishing curvature on
the boundary. Solitons of the Yang-Mills--Higgs systems of the type considered in these notes
(namely with the Higgs field taking values of an isotopic $D-$vector) on the other hand, are
topologically stable monopoles which can be defined in both even and odd spacelike dimensions.
These are asymptotically Dirac--Yang Monopoles.
The major difference of monopoles and instantons is, that the gauge connection of a monopole is
asymptotically one-half pure-gauge and not pure-gauge, resulting in nonvanishing curvature on
the boundary. This difference can be important in certain applications,
in particular to some examples in the AdS-CFT correspondence.

Another important feature of Yang-Mills--Higgs systems generally, versus Yang-Mills systems is,
that the solutions to the former exhibit symmetry breaking, and the presence of the dimensionful
vacuum expectation value of the Higgs field breaks the scale. In the presence of a Higgs self-interaction potential,
the resulting soliton is exponentially localised.

The main limitation of monopoles is that they are
asymptotically Dirac--Yang fields. As such the Yang--Mills curvature decays as $r^{-2}$
asymptotically, and hence in all spacelike dimensions higher than $three$
the (energy) integral of the $usual$ Yang--Mills density is divergent. Only high enough order YM curvature terms
decay appropriately to yield finite energy in any given (higher) dimension. Thus, the requirement of finite energy
in higher dimensions restricts the models to be employed to consist only of higher order YM curvature terms, in the
absence of the usual quadratic Yang-Mills term~\footnote{It may be worth to digress here to mention that pure
Yang-Mills systems in the absence of a Higgs field are not subject to this limitation. In the
absence of a Higgs field, the solitons of pure Yang-Mills systems are instantons~\cite{Belavin:1975fg,Tchrakian:1984gq},
which are {\it pure--gauge} at infinity. This faster decay enables the inclusion of the usual (quadratic) Yang-Mills
term in higher dimensiions.} as the case may be.

This limitation is somewhat mitigated in the case of Yang-Mills--Higgs models derived from the dimensional descent over
codimensions $N$, with $N\ge 2$. In all those cases the quadratic Yang-Mills density can appear in a specific guise,
namely $via$ the term
\[
\mbox{Tr}\left(\{S,F_{ij}\}+[D_i\F,D_j\F]\right)^2\,,\qquad S=(\eta^2+\F^2)\,,
\]
in the Hamiltonian density,
$e.g.$, in \re{H53}, \re{H44}, \re{H35}, \re{H33} and \re{H26}. Clearly, the term $\eta^4\mbox{Tr}\,F_{ij}^2$ is present
here, albeit in a rather couched manner. In the cases arising from the descent over unit codimension $N=1$, such
terms are absent since the presence of a Higgs potential in the residual system is predicated on the presence of
nontrivial components of the curvature on the codimension. Unfortunately, these models on $\R^{4p-1}$ are interesting
as they are the only ones for which the monopole saturates the Bogomol'nyi lower bound.

This limitation need not necessarily be a disadvantage since such applications have been usefully made in the literature:
in Strings from five branes~\cite{Duff:1990wu}, closed strings from instantons~\cite{Minasian:2001ib}
and cosmic strings from open heterotic string~\cite{Polchinski:2005bg}, where the only Yang-Mills term appearing is
$F^4$. The actual field configurations employed in
\cite{Duff:1990wu,Minasian:2001ib,Polchinski:2005bg}, as in the earliest prototype~\cite{Strominger:1990et} for the
heterotic string are those of $instantons$ on $\R^8$ and $\R^4$ respectively, and not of $monopoles$.
But the usual BPS $monopole$ on $\R^3$ is also
used~\cite{Harvey:1991jr} in this context. Obviously, the BPS $monopole$ on $\R^7$ of Section {\bf 7.3} above can
likewise be used analogously to the use of the $8-$dimensional instanton in \cite{Duff:1990wu}
This limitation applies both to the construction of monopoles, dyons, and the solitons of the new
Chern-Simons terms (which are functionals of both gauge and Higgs fields, defined in all dimensions)
presented in Sections {\bf 7}, {\bf 8} and {\bf 9} respectively.

Another avenue of applications occurs the construction of supersymmetric selfgravitating solitons~\cite{Gibbons:1993xt},
where both $instantonic$ (in the absence of Higgs fields) and $monopolic$ field configurations are employed.
In the case where monopole field configuations are employed in \cite{Gibbons:1993xt}, the relation of the
latter with the mechanism applied in
\cite{Harvey:1991jr} has been explored in \cite{Gibbons:1995zt}. In the case where
instanton field configuations are employed in \cite{Gibbons:1993xt},
this has been extended to cover the case where instead of the instanton,
the {\it dyonic instanton}~\cite{Lambert:1999ua}
is employed~\cite{Eyras:2000dg}. Unfortunately this latter scheme cannot be extended to higher dimensions,
due to the absence of higher dimensional dyonic instantons. The higher
dimensional extensions of such models, with the exception of the last, \cite{Eyras:2000dg},
can proceed systematically, either using the higher dimensional $instantons$
alluded to in Section {\bf 2}, or the $monopoles$ in Section {\bf 7}. Like in the
generalistions~\cite{Duff:1990wu,Polchinski:2005bg} of
\cite{Strominger:1990et}, here too higher order Yang-Mills and Yang-Mills--Higgs terms will apprear in the absence of
the usual $quadratic$ YM and YMH densities.

Concerning the new Chern-Simons densities introduced here, these can be applied to probems of gravitating gauge fields.
Indeed Chern-Simons densities play a central role in \cite{Gibbons:1993xt} and \cite{Eyras:2000dg}, the latter being
related to the new CS terms. (This analysis is pending.) One interesting outcome of the application of CS densities is
their effect on the thermodynamic stability of black holes, as discovered in \cite{Brihaye:2010wp}.

In all putative further generalisations of
\cite{Strominger:1990et,Harvey:1991jr}\cite{Duff:1990wu,Minasian:2001ib,Polchinski:2005bg}, $e.g.$,
\cite{Duff:1990wu,Minasian:2001ib,Polchinski:2005bg}, and of \cite{Gibbons:1993xt}, one $caveat$ is
that the field configurations exploited must satisfy self--duality or Bogomol'nyi equations. 
In the absence of a Higgs field, $e.g.$ for \cite{Duff:1990wu,Minasian:2001ib,Polchinski:2005bg} and some of the systems
considered in \cite{Gibbons:1993xt}, this criterion is satified if one employs the instantons on $\R^{4p}$.
Alternatively, this can be achieved by using instead the instantons on $S^{2n}$, $CP^n$
and generally on symmetric compact coset spaces may also be exploited. In the presence of a Higgs field these
configurations are the monopoles in Section {\bf 7}, and most of these do not saturate the Bogomol'nyi bound. The only
exceptions, whose Bogomol'nyi equations are not overdetermined, are the monopoles on $\R^{4p-1}$ and the
(generalised) vortices on $\R^{2}$.

Looking forward, it may be worthwhile to consider question of the limitation of Yang-Mills--Higgs solitons in dimensions
higher than $3+1$, where the usual (quadratic) YM and YMH terms are absent. This is a problem of Higgs theories. The
presence of a Higgs field, along with a symmetry breaking Higgs self-interaction potential, is necessary only
when exponential localisation is required, and the scale is broken by the vacuum expectation value of the Higgs field.
Should the presence of (usual quadratic) YM term in dimensions higher than $3+1$ be required, there are two alternatives
to using a Higgs model. With both these choices the solitons employed are instantonic and asymptotically there is power
decay, exponential localisation being thereby sacrificed. These two options are:
\begin{itemize}
\item
In the first variant the Higgs field is simply absent. The resulting
configurations are instantons, which are asymptotically $pure$-$gauge$ and decay fast enough. As a consequence
finite energy solitons are consistent with the presence of (usual) quadratic YM terms.
In all dimensions higher than $4+1$, these
theories are necessarily scale breaking and the instantons display power decay (See Section {\bf 2}.).
In the appropriate (necessarily even spacelike) dimensions, these solitons are stabilised by a Pontryagin charge,
$e.g.$, the example in \cite{Burzlaff:1993kf} on $\R^6$.
\item
The second variant involves gauged, $S^n$-valued sigma models~\cite{Faddeev}, which also exhibit instantonic
configurations. These are the  solitons of the $SO(N)$ gauged $O(D+1)$ sigma
models on $\R^D$, with $2\le N\le D$. The $SO(2)$ gauged solitons (vortices) of the $O(3)$ sigma model on $\R^2$ were
given in \cite{Schroers:1995ns}, the $SO(3)$ gauged solitons of the $O(4)$ sigma model on $\R^3$
in \cite{Arthur:1996wy,yves}, and the $SO(4)$ gauged solitons of the $O(5)$ sigma model on $\R^4$
in \cite{yves}. The generic case of $SO(D)$ gauged $O(D+1)$ sigma models on $\R^D$ is
discussed in \cite{SODOD+1}. But what is peculiar to gauged $O(D+1)$ sigma models on $\R^D$ is that they can be $SO(N)$
gauged, with $2\le N\le D$. Such an exmple is given in \cite{Piette:1997ny,Radu:2005jp} for the $SO(2)$ gauged
$O(4)$ sigma models on $\R^3$. This is in contrast to (gauged) Higgs models on $\R^D$ of the type described in
Section {\bf 7}, with $iso-D-$vector Higgs which support monopoles. In the latter case the gauge group must
be $SO(D)$ since the covariant derivative of the hedghog field at infinity must vanish.
\end{itemize}
Using either of these variants for the purpose of constructing solitons falls outside the remit of the present notes.
What is perhaps more interesting and hence may deserve a comment is the corresponding definitions of the new
Chern-Simons terms. As seen in Section {\bf 9} the crucial quantity there are the dimensionally reduced Chern-Pontryagin
densities, that are {\it total divergence}. The topological charge densities of $S^n$-valued sigma models, and, of the
gauged $S^n$-valued sigma models, are not expressed explicitly as {\it total divergences}. They are however
essentially total divergence in the sense that when subjected to the variational calculus, the resulting
Euler-lagrange equations turn out to be trivial. Equally, these topological charge densities can be cast in explicit
total divergence form by suitable parametrising the $S^n$-valued fields such that the sigma model constraint is
implicitly imposed. As a consequence on can define new(er) Chern-Simons densities which are functionals of both the gauge
fields, and the $S^n$-valued fields. Like the usual Chern-Simons densities in odd dimensional spacetimes, the
solitons of such theories will feature instantonic gauge fields, allowing the presence of the usual Yang-Mills density
in the Lagrangian. But now, these new(er) Chern-Simons densities are defined in both odd {\bf and} even
dimensional spacetimes.

\newpage
\begin{small}
 
\end{small}


\begin{thebibliography}{99}
\bibitem{Dirac:1931kp}
  P.~A.~M.~Dirac,
  ``Quantised singularities in the electromagnetic field,''
  Proc.\ Roy.\ Soc.\ Lond.\  A {\bf 133} (1931) 60.
\bibitem{Yang:1977qv}
  C.~N.~Yang,
  ``Generalization Of Dirac's Monopole To SU(2) Gauge Fields,''
  J.\ Math.\ Phys.\  {\bf 19} (1978) 320.
\bibitem{Tchrakian:2008zz}
  T.~Tchrakian,
  ``Dirac-Yang monopoles in all dimensions and their regular counterparts,''
  Phys.\ Atom.\ Nucl.\  {\bf 71} (2008) 1116.
\bibitem{'tHooft:1974qc}
  G.~'t Hooft,
  ``Magnetic monopoles in unified gauge theories,''
  Nucl.\ Phys.\  B {\bf 79} (1974) 276.
\bibitem{Polyakov:1974ek}
  A.~M.~Polyakov,
  ``Particle spectrum in quantum field theory,''
  JETP Lett.\  {\bf 20} (1974) 194
  [Pisma Zh.\ Eksp.\ Teor.\ Fiz.\  {\bf 20} (1974) 430].
\bibitem{Abrikosov:1956sx}
  A.~A.~Abrikosov,
  ``On the Magnetic properties of superconductors of the second group,''
  Sov.\ Phys.\ JETP {\bf 5} (1957) 1174
  [Zh.\ Eksp.\ Teor.\ Fiz.\  {\bf 32} (1957) 1442].
\bibitem{Nielsen:1973cs}
  H.~B.~Nielsen and P.~Olesen,
  ``Voretex-line models for dual strings,''
  Nucl.\ Phys.\  B {\bf 61} (1973) 45.
\bibitem{Tchrakian:1984gq}
  D.~H.~Tchrakian,
  ``Spherically Symmetric Gauge Field Configurations With Finite Action In 4P-Dimensions (P = Integer),''
  Phys.\ Lett.\  B {\bf 150} (1985) 360.
\bibitem{Tchrakian:2002ti}
  T.~Tchrakian,
  ``Winding number versus Chern-Pontryagin charge,''
  arXiv:hep-th/0204040.
\bibitem{Julia:1975ff}
  B.~Julia and A.~Zee,
  ``Poles With Both Magnetic And Electric Charges In Nonabelian Gauge Theory,''
  Phys.\ Rev.\  D {\bf 11} (1975) 2227.
\bibitem{Radu:2005rf}
  E.~Radu and D.~H.~Tchrakian,
  ``Static BPS 'monopoles' in all even spacetime dimensions,''
  Phys.\ Rev.\  D {\bf 71} (2005) 125013
  [arXiv:hep-th/0502025].
\bibitem{Tchrakian:1978sf}
  D.~H.~Tchrakian,
  ``N-Dimensional Instantons And Monopoles,''
  J.\ Math.\ Phys.\  {\bf 21} (1980) 166.
\bibitem{Grossman:1984pi}
  B.~Grossman, T.~W.~Kephart and J.~D.~Stasheff,
  ``Solutions To Yang-Mills Field Equations In Eight-Dimensions And The Last Hopf Map,''
  Commun.\ Math.\ Phys.\  {\bf 96} (1984) 431
  [Erratum-ibid.\  {\bf 100} (1985) 311].
\bibitem{Burzlaff:1993kf}
  J.~Burzlaff and D.~H.~Tchrakian,
  ``Nonselfdual solutions of gauge field models in 2n-dimensions,''
  J.\ Phys.\ A  {\bf 26} (1993) L1053.
\bibitem{Saclioglu:1986qn}
  C.~Saclioglu,
  ``Scale invariant gauge theories and selfuality in higher dimensions,''
  Nucl.\ Phys.\  B {\bf 277} (1986) 487.
\bibitem{Fujii:1986ty}
  K.~Fujii,
  ``Extended Yang-Mills Models On Even Dimensional Spaces,''
  Lett.\ Math.\ Phys.\  {\bf 12} (1986) 363.
Supplement to the paper ``Extended Yang-Mills Models on even-dimensional spaces'',
{ibid.} {\bf 12} (1896) 371.
\bibitem{Corrigan:1982th}
  E.~Corrigan, C.~Devchand, D.~B.~Fairlie and J.~Nuyts,
  ``First Order Equations For Gauge Fields In Spaces Of Dimension Greater Than Four,''
  Nucl.\ Phys.\  B {\bf 214} (1983) 452.
\bibitem{Fairlie:1984mp}
  D.~B.~Fairlie and J.~Nuyts,
  ``Spherically Symmetric Solutions Of Gauge Theories In Eight-Dimensions,''
  J.\ Phys.\ A  {\bf 17} (1984) 2867.
\bibitem{Fubini:1985jm}
  S.~Fubini and H.~Nicolai,
  ``The Octonionic Instanton,''
  Phys.\ Lett.\  B {\bf 155} (1985) 369.
\bibitem{Belavin:1975fg}
  A.~A.~Belavin, A.~M.~Polyakov, A.~S.~Schwartz and Yu.~S.~Tyupkin,
  ``Pseudoparticle solutions of the Yang-Mills equations,''
  Phys.\ Lett.\  B {\bf 59} (1975) 85.
\bibitem{Witten:1976ck}
  E.~Witten,
  ``Some exact multipseudoparticle solutions of classical Yang-Mills  theory,''
  Phys.\ Rev.\ Lett.\  {\bf 38} (1977) 121.
\bibitem{Chakrabarti:1985qj}
  A.~Chakrabarti, T.~N.~Sherry and D.~H.~Tchrakian,
  ``On Axially Symmetric Selfdual Gauge Field Configurations In 4p Dimensions,''
  Phys.\ Lett.\  B {\bf 162} (1985) 340.
\bibitem{Spruck:1997eb}
  J.~Spruck, D.~H.~Tchrakian and Y.~Yang,
  ``Multiple instantons representing higher-order Chern-Pontryagin classes,''
  Commun.\ Math.\ Phys.\  {\bf 188} (1997) 737.
\bibitem{Tchrakian:1996bc}
  D.~H.~Tchrakian and Y.~S.~Yang,
  ``The existence of generalised selfdual Chern-Simons vortices,''
  Lett.\ Math.\ Phys.\  {\bf 36} (1996) 403.
\bibitem{Tchrakian:1990gc}
  D.~H.~Tchrakian and A.~Chakrabarti,
  ``How overdetermined are the generalized selfduality relations?,''
  J.\ Math.\ Phys.\  {\bf 32} (1991) 2532.
\bibitem{O'Se:1987fx}
  D.~O'Se and D.~H.~Tchrakian,
  ``Conformal properties of the BPST instantons of the generalised Yang-Mills system,''
  Lett.\ Math.\ Phys.\  {\bf 13} (1987) 211.
\bibitem{O'Brien:1988rs}
  G.~M.~O'Brien and D.~H.~Tchrakian,
  ``Spin connection generalised Yang-Mills fields on double dual generalised Einstein backgrounds,''
  J.\ Math.\ Phys.\  {\bf 29} (1988) 1212.
\bibitem{Kihara:2007di}
  H.~Kihara and M.~Nitta,
  ``A Classical Solution in Six-dimensional Gauge Theory with Higher Derivative Coupling,''
  Phys.\ Rev.\  D {\bf 77} (2008) 047702
  [arXiv:hep-th/0703166].
\bibitem{Radu:2007az}
  E.~Radu, D.~H.~Tchrakian and Y.~Yang, ``Spherically symmetric selfdual Yang-Mills instantons on curved backgrounds
  in all even dimensions,''
  Phys.\ Rev.\  D {\bf 77} (2008) 044017
  [arXiv:0707.1270 [hep-th]].
\bibitem{Ma:1990ja}
  Z.~Ma and D.~H.~Tchrakian,
  ``Gauge Field Systems On CP(N),''
  J.\ Math.\ Phys.\  {\bf 31} (1990) 1506.
\bibitem{Kihara:2008zg}
  H.~Kihara and M.~Nitta,
  ``Generalized Instantons on Complex Projective Spaces,''
  J.\ Math.\ Phys.\  {\bf 50} (2009) 012301
  [arXiv:0807.1259 [hep-th]].
\bibitem{Kihara:2007vz}
  H.~Kihara and M.~Nitta,
  ``Exact Solutions of Einstein-Yang-Mills Theory with Higher-Derivative Coupling,''
  Phys.\ Rev.\  D {\bf 76} (2007) 085001
  [arXiv:0704.0505 [hep-th]].
\bibitem{Kihara:2009ea}
  H.~Kihara, M.~Nitta, M.~Sasaki, C.~M.~Yoo and I.~Zaballa,
  ``Dynamical Compactification and Inflation in Einstein-Yang-Mills Theory with Higher Derivative Coupling,''
  Phys.\ Rev.\  D {\bf 80} (2009) 066004
  [arXiv:0906.4493 [hep-th]].
\bibitem{Chingangbam:2009jy}
  P.~Chingangbam, H.~Kihara and M.~Nitta,
  ``Gauge symmetry breaking in ten-dimensional Yang-Mills theory dynamically compactified on $S^6$,''
  Phys.\ Rev.\  D {\bf 81} (2010) 085008
  [arXiv:0912.3128 [hep-th]].
\bibitem{de Alfaro:1976qz}
  V.~de Alfaro, S.~Fubini and G.~Furlan,
  ``A New Classical Solution Of The Yang-Mills Field Equations,''
  Phys.\ Lett.\  B {\bf 65} (1976) 163.
\bibitem{O'Brien:1987jy}
  G.~M.~O'Brien and D.~H.~Tchrakian,
  ``Meron Field Configurations In Every Even Dimension,''
  Phys.\ Lett.\  B {\bf 194} (1987) 411.
\bibitem{Schwarz:1977ix}
  A.~S.~Schwarz,
  ``On Symmetric Gauge Fields,''
  Commun.\ Math.\ Phys.\  {\bf 56} (1977) 79.
\bibitem{Romanov:1977rr}
  V.~N.~Romanov, A.~S.~Schwarz and Yu.~S.~Tyupkin,
  ``On Spherically Symmetric Fields In Gauge Theories,''
  Nucl.\ Phys.\  B {\bf 130} (1977) 209.
\bibitem{Schwarz:1981mb}
  A.~S.~Schwarz and Yu.~S.~Tyupkin,
  ``Dimensional Reduction Of The Gauge Field Theory,''
  Nucl.\ Phys.\  B {\bf 187} (1981) 321.
\bibitem{Forgacs:1979zs}
  P.~Forgacs and N.~S.~Manton,
  ``Space-Time Symmetries In Gauge Theories,''
  Commun.\ Math.\ Phys.\  {\bf 72} (1980) 15.
\bibitem{Sherry:1982fd}
  T.~N.~Sherry and D.~H.~Tchrakian,
  ``Dimensional Reduction Of A Six-Dimensional Selfdual Gauge Field Theory,''
\bibitem{Ma:1986pu}
  Z.~Q.~Ma, G.~M.~O'Brien and D.~H.~Tchrakian,
  ``Dimensional reduction of higher order topological invariants: Descent by even steps and applications,''
  Phys.\ Rev.\  D {\bf 33} (1986) 1177.
\bibitem{Ma:1988um}
  Z.~Q.~Ma and D.~H.~Tchrakian,
  ``Dimensional reduction of higher order topological invariants: The case $CP^n$,''
  Phys.\ Rev.\  D {\bf 38} (1988) 3827.
\bibitem{O'Brien:1988xr}
  G.~M.~O'Brien and D.~H.~Tchrakian,
  ``Spherically symmetric SO(4) instanton of a nonabelian Higgs model in four-dimensions,''
  Mod.\ Phys.\ Lett.\  A {\bf 4} (1989) 1389.
\bibitem{Arafune:1974uy}
  J.~Arafune, P.~G.~O.~Freund and C.~J.~Goebel,
  ``Topology of Higgs Fields,''
  J.\ Math.\ Phys.\  {\bf 16} (1975) 433.
\bibitem{Yisong1}
Yisong~Yang,
"Existence of Solutions for a Generalized Yang-Mills Theory,"
Lett.\ Math.\ Phys.\  {\bf 19} (1990) 257.
\bibitem{Yisong2}
Yisong~Yang,
"Self-Dual Monopoles in a Seven-DimensionalGauge Theory,"
Lett.\ Math.\ Phys.\  {\bf 20} (1990) 285.
\bibitem{Yisong3}
Yisong~Yang,
"Existence of Static BPS Monopoles and Dyons in Arbitrary (4 p − 1)-Dimensional Spaces,"
Lett.\ Math.\ Phys.\  {\bf 77} (2006) 249.
\bibitem{ST}
T.N. Sherry and D.H. Tchrakian, "Dimensional reduction of Chern--Pontryagin densities over codimensions $CP^N$'',
in preparation.
\bibitem{Kleihaus:1998kd}
  B.~Kleihaus, D.~O'Keeffe and D.~H.~Tchrakian,
  ``Calculation of the mass of a generalised monopole,''
  Phys.\ Lett.\  B {\bf 427} (1998) 327.
\bibitem{Tchrakian:1999me}
  D.~H.~Tchrakian and F.~Zimmerschied,
  ``'t~Hooft tensors as Kalb-Ramond fields of generalised monopoles in all  odd
  Phys.\ Rev.\  D {\bf 62} (2000) 045002
  [arXiv:hep-th/9912056].
\bibitem{Ma:1992zy}
  Z.~Q.~Ma and D.~H.~Tchrakian,
  ``Wu-Yang fields,''
  Lett.\ Math.\ Phys.\  {\bf 26} (1992) 179.
\bibitem{Skyrme:1962vh}
  T.~H.~R.~Skyrme,
  ``A Unified Field Theory Of Mesons And Baryons,''
  Nucl.\ Phys.\  {\bf 31} (1962) 556.
\bibitem{Jacobs:1978ch}
  L.~Jacobs and C.~Rebbi,
  ``Interaction Energy Of Superconducting Vortices,''
  Phys.\ Rev.\  B {\bf 19} (1979) 4486.
\bibitem{Kleihaus:1998gy}
  B.~Kleihaus, D.~O'Keeffe and D.~H.~Tchrakian,
  ``Interaction energies of generalised monopoles,''
  Nucl.\ Phys.\  B {\bf 536} (1998) 381
  [arXiv:hep-th/9806088].
\bibitem{Burzlaff:1994tf}
  J.~Burzlaff, A.~Chakrabarti and D.~H.~Tchrakian,
  ``Generalized Abelian Higgs models with selfdual vortices,''
  J.\ Phys.\ A  {\bf 27} (1994) 1617.
\bibitem{Arthur:1998nh}
  K.~Arthur, Y.~Brihaye and D.~H.~Tchrakian,
  ``Interaction energy of generalized Abelian Higgs vortices,''
  J.\ Math.\ Phys.\  {\bf 39} (1998) 3031.
\bibitem{Breitenlohner:2009zi}
  P.~Breitenlohner and D.~H.~Tchrakian,
  ``Gravitating BPS Monopoles in all d=4p Spacetime Dimensions,''
  Class.\ Quant.\ Grav.\  {\bf 26} (2009) 145008
  [arXiv:0903.3505 [gr-qc]].
\bibitem{MullerKirsten:1990qw}
  H.~J.~W.~Muller-Kirsten and D.~H.~Tchrakian,
  J.\ Phys.\ A  {\bf 23} (1990) L363.
\bibitem{Tchrakian:1990qx}
  D.~H.~Tchrakian,
  Phys.\ Lett.\  B {\bf 244} (1990) 458.
\bibitem{Paturyan:2005ik}
  V.~Paturyan, E.~Radu and D.~H.~Tchrakian,
  ``Solitons and soliton antisoliton pairs of a Goldstone model in 3+1 dimensions,''
  J.\ Phys.\ A  {\bf 39} (2006) 3817
  [arXiv:hep-th/0509056].
\bibitem{Radu:2007zz}
  E.~Radu and D.~H.~Tchrakian,
  ``Goldstone models in D + 1 dimensions, D = 3, 4, 5, supporting stable and
  J.\ Phys.\ A  {\bf 40}, 10129 (2007).
\bibitem{Arthur:1995ws}
  K.~Arthur and D.~H.~Tchrakian,
  ``Vortex Solutions In A Class Of 2-D Symmetry Breaking Skyrme-Like Models,''
  J.\ Math.\ Phys.\  {\bf 36} (1995) 6566.
\bibitem{Barriola:1989hx}
  M.~Barriola and A.~Vilenkin,
  ``Gravitational Field of a Global Monopole,''
  Phys.\ Rev.\ Lett.\  {\bf 63} (1989) 341.
\bibitem{Harari:1990cz}
  D.~Harari and C.~Lousto,
  ``Repulsive gravitational effects of global monopoles,''
  Phys.\ Rev.\  D {\bf 42} (1990) 2626.
\bibitem{Maison:1999pi}
  D.~Maison and S.~L.~Liebling,
  ``Some remarks on gravitational global monopoles,''
  Phys.\ Rev.\ Lett.\  {\bf 83} (1999) 5218
  [arXiv:gr-qc/9908038].
\bibitem{Cohen:1988sg}
  A.~G.~Cohen and D.~B.~Kaplan,
  ``The exact metric about global cosmic strings,''
  Phys.\ Lett.\  B {\bf 215} (1988) 67.
\bibitem{Gregory:1988xc}
  R.~Gregory,
  ``Global String Singularities,''
  Phys.\ Lett.\  B {\bf 215} (1988) 663.
\bibitem{Lambert:1999ua}
  N.~D.~Lambert and D.~Tong,
  ``Dyonic instantons in five-dimensional gauge theories,''
  Phys.\ Lett.\  B {\bf 462} (1999) 89
  [arXiv:hep-th/9907014].
\bibitem{Schechter:1980cz}
  M.~Schechter and R.~Weder,
  ``A Theorem On The Existence Of Dyon Solutions,''
  Annals Phys.\  {\bf 132} (1981) 292.
\bibitem{Brihaye:1998vr}
  Y.~Brihaye, B.~Kleihaus and D.~H.~Tchrakian,
  ``Dyon-Skyrmion lumps,''
  J.\ Math.\ Phys.\  {\bf 40} (1999) 1136
  [arXiv:hep-th/9805059].
\bibitem{Prasad:1975kr}
  M.~K.~Prasad and C.~M.~Sommerfield,
  ``An Exact Classical Solution For The 'T Hooft Monopole And The Julia-Zee Dyon,''
  Phys.\ Rev.\ Lett.\  {\bf 35} (1975) 760.
\bibitem{Deser:1982vy}
  S.~Deser, R.~Jackiw and S.~Templeton,
  ``Three-Dimensional Massive Gauge Theories,''
  Phys.\ Rev.\ Lett.\  {\bf 48} (1982) 975.
\bibitem{Deser:1981wh}
  S.~Deser, R.~Jackiw and S.~Templeton,
  ``Topologically massive gauge theories,''
  Annals Phys.\  {\bf 140} (1982) 372
  [Erratum-ibid.\  {\bf 185} (1988) 406]
  [Annals Phys.\  {\bf 185} (1988) 406]
  [Annals Phys.\  {\bf 281} (2000) 409].
\bibitem{Jackiw:2003pm}
  R.~Jackiw and S.~Y.~Pi,
  ``Chern-Simons modification of general relativity,''
  Phys.\ Rev.\  D {\bf 68} (2003) 104012
  [arXiv:gr-qc/0308071].
\bibitem{Jackiw:1985}
R. Jackiw, "Chern-Simons terms and cocycles in physics and mathematics",
in E.S. Fradkin $Festschrift$, Adam Hilger, Bristol (1985)
\bibitem{Rubakov:1986am}
  V.~A.~Rubakov and A.~N.~Tavkhelidze,
  ``Stable anomalous states of superdense matter in gauge theories,''
  Phys.\ Lett.\  B {\bf 165} (1985) 109.
\bibitem{Brihaye:2009cc}
  Y.~Brihaye, E.~Radu and D.~H.~Tchrakian,
  ``$AdS_5$ solutions in Einstein--Yang-Mills--Chern-Simons theory,''
  Phys.\ Rev.\  D {\bf 81} (2010) 064005
  [arXiv:0911.0153 [hep-th]].
\bibitem{Hong:1990yh}
  J.~Hong, Y.~Kim and P.~Y.~Pac,
  ``On The Multivortex Solutions Of The Abelian Chern-Simons-Higgs Theory,''
  Phys.\ Rev.\ Lett.\  {\bf 64} (1990) 2230.
\bibitem{Jackiw:1990aw}
  R.~Jackiw and E.~J.~Weinberg,
  ``Selfdual Chern-Simons Vortices,''
  Phys.\ Rev.\ Lett.\  {\bf 64} (1990) 2234.
\bibitem{NavarroLerida:2009dm}
  F.~Navarro-Lerida and D.~H.~Tchrakian,
  ``Non-Abelian Yang-Mills-Higgs vortices,''
  Phys.\ Rev.\  D {\bf 81} (2010) 127702
  [arXiv:0909.4220 [hep-th]].
\bibitem{Radu:2011zy}
Eugen Radu and D.H. Tchrakian, "New Chern-Simons densities in both odd and even dimensions",
to appear in  S.G. Matinyan {\it Festschrift}, eds. V. Gurzadyan and A. Sedrakian, World Scientific (2011),
 arXiv:1101.5068 [hep-th].
\bibitem{NRT}
  F.~Navarro-Lerida, E.~Radu and D.~H.~Tchrakian,
  ``Non-Abelian Chern-Simons--Higgs solitons in $3+1$ and $2+1$ dimensions,''
   $work\ in\ progress.$
\bibitem{Duff:1990wu}
  M.~J.~Duff and J.~X.~Lu,
  ``Strings from five-branes,''
  Phys.\ Rev.\ Lett.\  {\bf 66} (1991) 1402.
\bibitem{Strominger:1990et}
  A.~Strominger,
  ``Heterotic solitons,''
  Nucl.\ Phys.\  B {\bf 343} (1990) 167
  [Erratum-ibid.\  B {\bf 353} (1991) 565].
\bibitem{Minasian:2001ib}
  R.~Minasian, S.~L.~Shatashvili and P.~Vanhove,
  ``Closed strings from SO(8) Yang-Mills instantons,''
  Nucl.\ Phys.\  B {\bf 613} (2001) 87
  [arXiv:hep-th/0106096].
\bibitem{Polchinski:2005bg}
  J.~Polchinski,
  ``Open heterotic strings,''
  JHEP {\bf 0609} (2006) 082
  [arXiv:hep-th/0510033].
\bibitem{Harvey:1991jr}
  J.~A.~Harvey and J.~Liu,
  ``Magnetic monopoles in N=4 supersymmetric low-energy superstring theory,''
  Phys.\ Lett.\  B {\bf 268} (1991) 40.
\bibitem{Gibbons:1993xt}
  G.~W.~Gibbons, D.~Kastor, L.~A.~J.~London, P.~K.~Townsend and J.~H.~Traschen,
  ``Supersymmetric selfgravitating solitons,''
  Nucl.\ Phys.\  B {\bf 416} (1994) 850
  [arXiv:hep-th/9310118].
\bibitem{Gibbons:1995zt}
  G.~W.~Gibbons and P.~K.~Townsend,
  ``Antigravitating BPS monopoles and dyons,''
  Phys.\ Lett.\  B {\bf 356} (1995) 472
  [arXiv:hep-th/9506131].
\bibitem{Eyras:2000dg}
  E.~Eyras, P.~K.~Townsend and M.~Zamaklar,
  ``The heterotic dyonic instanton,''
  JHEP {\bf 0105} (2001) 046
  [arXiv:hep-th/0012016].
\bibitem{Brihaye:2010wp}
  Y.~Brihaye, E.~Radu and D.~H.~Tchrakian,
  Phys.\ Rev.\ Lett.\  {\bf 106} (2011) 071101
  [arXiv:1011.1624 [hep-th]].
\bibitem{Faddeev}
The proposal of such a construction is one of the results in:
L.D. Fadde'ev, Lett. Math. Phys. {\bf 1} (1976) 289.
\bibitem{Schroers:1995ns}
  B.~J.~Schroers,
  ``Bogomolny solitons in a gauged O(3) sigma model,''
  Phys.\ Lett.\  B {\bf 356} (1995) 291
  [arXiv:hep-th/9506004].
\bibitem{Arthur:1996wy}
  K.~Arthur and D.~H.~Tchrakian,
  ``SO(3) Gauged Soliton of an O(4) Sigma Model on $R_3$,''
  Phys.\ Lett.\  B {\bf 378} (1996) 187
  [arXiv:hep-th/9601053].
\bibitem{yves}
Y.~Brihaye and D.~H.~Tchrakian,
``Solitons/Instantons in d-dimensional SO(d) gauged O(d+1) Skyrmemodels'',
Nonlinearity, {\bf 11} (1998) 891-912.
\bibitem{SODOD+1}
D.H. Tchrakian, ``Topologically stable lumps in $SO(d)$ gauged
$O(d+1)$ sigma models in $d$ dimensions
: $d=2,3,4$'' Lett. Math. Phys. {\bf 40} (1997) 191-201
\bibitem{Piette:1997ny}
  B.~M.~A.~Piette and D.~H.~Tchrakian,
  ``Topologically stable soliton in the U(1) gauged Skyrme model,''
  Phys.\ Rev.\  D {\bf 62} (2000) 025020
  [arXiv:hep-th/9709189].
\bibitem{Radu:2005jp}
  E.~Radu and D.~H.~Tchrakian,
  ``Spinning U(1) gauged skyrmions,''
  Phys.\ Lett.\  B {\bf 632} (2006) 109
  [arXiv:hep-th/0509014].


\end{thebibliography}
\end{document}